\documentclass[moor]{informs1}  
 \usepackage{natbib}
 \NatBibNumeric
 \def\bibfont{\small}%
 \bibpunct[, ]{[}{]}{,}{n}{}{,}%

\usepackage[colorlinks=true,breaklinks=true,bookmarks=true,urlcolor=blue,
     citecolor=blue,linkcolor=blue,bookmarksopen=false,draft=false]{hyperref}

\def\EMAIL#1{\href{mailto:#1}{#1}}
\def\URL#1{\href{#1}{#1}}         

\TheoremsNumberedThrough     

\EquationsNumberedThrough    
\usepackage{amssymb, bm, bbm, subfigure}
\usepackage{amsmath}
\usepackage{amsfonts}
\usepackage{latexsym}
\usepackage{hyperref}
\usepackage{graphicx}
\usepackage{color}

\newcommand{\ds}{\displaystyle}

\newcommand{\ba}{\begin{array}}
\newcommand{\ea}{\end{array}}

\renewcommand{\l}{\left}
\renewcommand{\r}{\right}
\renewcommand{\be}{\begin{equation}}
\renewcommand{\ee}{\end{equation}}
\newcommand{\eps}{\varepsilon}
\newcommand{\ups}{\upsilon}

\renewcommand{\mc}{\mathcal}

\newcommand{\ov}{\overline}

\newcommand{\Z}{\mathbb{Z}}

\newcommand{\E}{\mathbb{E}}

\newcommand{\R}{\mathbb{R}}
\newcommand{\N}{\mathbb{N}}

\newcommand{\summ}{\sum\limits}
\renewcommand{\P}{\mathbb{P}}

\newcommand{\de}{\mathrm{d}}
\newcommand{\ora}{\overrightarrow}\newcommand{\ola}{\overleftarrow}

\newcommand{\se}{\text{ if }}

\DeclareMathOperator{\Var}{Var}

\def\1{\mathbbm{1}}
\def\qed{\hfill \vrule height 7pt width 7pt depth 0pt\medskip}
\def\Z{\mathbb{Z}}
\def\N{\mathbb{N}}
\def\E{\mathbb{E}}
\def\R{\mathbb{R}}

\def\P{\mathbb{P}}

\begin{document}


 \RUNAUTHOR{Acemoglu, Como, Fagnani, and Ozdaglar}

\RUNTITLE{Opinion fluctuations and disagreement}

\TITLE{Opinion fluctuations and disagreement in social networks}

\ARTICLEAUTHORS{%
\AUTHOR{Daron Acemoglu}
\AFF{Department of Economics, Massachusetts Institute of Technology, 77 Massachusetts Avenue, Cambridge, MA, 02139, \EMAIL{darn@mit.edu}, \URL{http://econ-www.mit.edu/faculty/acemoglu/}}
\AUTHOR{Giacomo Como}
\AFF{Department of Automatic Control, Lund University, BOX 118,  22100,  Lund, Sweden, \EMAIL{giacomo.como@control.lth.se}, \URL{http://www.control.lth.se/Staff/GiacomoComo/}}
\AUTHOR{Fabio Fagnani}
\AFF{Dipartimento di Scienze Matematiche, Politecnico di Torino, corso Stati Uniti 24, 10129, Torino, Italy, \EMAIL{fabio.fagnani@polito.it}, \URL{http://calvino.polito.it/~fagnani/indexeng.html}}
\AUTHOR{Asuman Ozdaglar}
\AFF{Laboratory for Information and Decision Systems and Department of Electrical Engineering, Massachusetts Institute of Technology, 77 Massachusetts Avenue, Cambridge, MA, 02139, \EMAIL{asuman@mit.edu}, 
\URL{http://web.mit.edu/asuman/www/}}

} 

\ABSTRACT{%
We study a tractable opinion dynamics model that generates long-run disagreements and persistent opinion fluctuations. Our model involves an inhomogeneous stochastic gossip process of continuous opinion dynamics in a society consisting of two types of agents: \emph{regular agents}, who update their beliefs according to information that they receive from their social neighbors; and \emph{stubborn agents}, who never update their opinions and might represent leaders, political parties or media sources attempting to influence the beliefs in the rest of the society. When the society contains stubborn agents with different opinions, the belief dynamics never lead to a consensus (among the regular agents). Instead, beliefs in the society fail to converge almost surely, the belief profile keeps on fluctuating in an ergodic fashion, and it converges in law to a non-degenerate random vector.

The structure of the graph describing the social network and the location of the stubborn agents within it shape the opinion dynamics. The expected belief vector is proved to evolve according to an ordinary differential equation coinciding with the Kolmogorov backward equation of a continuous-time Markov chain on the graph with absorbing states corresponding to the stubborn agents, and hence to converge to a harmonic vector, with every regular agent's value being the weighted average of its neighbors' values, and boundary conditions corresponding to the stubborn agents' beliefs. Expected cross-products of the agents' beliefs allow for a similar characterization in terms of coupled Markov chains on the graph describing the social network.

We prove that, in large-scale societies which are \emph{highly fluid}, meaning that the product of the mixing time of the Markov chain on the graph describing the social network and the relative size of the linkages to stubborn agents vanishes as the population size grows large, a condition of \emph{homogeneous influence} emerges, whereby the stationary beliefs' marginal distributions of most of the regular agents have approximately equal first and second moment. 
}%



\MSCCLASS{Primary: 91D30, 60K35; Secondary: 93A15 
}
\ORMSCLASS{Primary: Games/group decisions: stochastic; Secondary: Markov processes,
Random walk 
}
\HISTORY{}

\maketitle

\section{Introduction}

Disagreement among individuals in a society, even on central
questions that have been debated for centuries, is the norm;
agreement is the rare exception. How can disagreement of this sort
persist for so long? Notably, such disagreement is not a consequence
of lack of communication or some other factors leading to fixed opinions.
Disagreement remains even as individuals communicate and sometimes change their opinions.

Existing models of communication and learning,
based on Bayesian or non-Bayesian updating mechanisms, typically
lead to consensus provided that communication takes place over a
strongly connected network (e.g., Smith and Sorensen \cite{SS},
Banerjee and Fudenberg \cite{ban-fud}, Acemoglu, Dahleh, Lobel and
Ozdaglar \cite{ilan}, Bala and Goyal \cite{bala-goyal}, Gale and
Kariv \cite{gale-kariv}, DeMarzo, Vayanos and Zwiebel
\cite{demarzo}, Golub and Jackson \cite{golub}, Acemoglu, Ozdaglar
and ParandehGheibi \cite{influence}, Acemoglu, Bimpikis and Ozdaglar \cite{Bimpikis}), and are thus unable to explain
persistent disagreements. One notable exception is provided by models that incorporate a form of \emph{homophily} mechanism in communication, whereby individuals are more likely to exchange opinions or communicate with others that have similar beliefs, and fail to interact with agents whose beliefs differ from theirs by more than some given confidence threshold. This mechanism was first proposed by Axelrod \cite{Axelrod} in the discrete opinion dynamics setting, and then by Krause \cite{Krause}, and Deffuant and Weisbuch \cite{Deffuant}, in the continuous opinion dynamics framework. Such belief dynamics typically lead to the emergence of different asymptotic opinion clusters (see, e.g., \cite{Lorenz05,BlondelHendrickxTsitsiklis09,ComoFagnani10}), but fail to explain persistent opinion fluctuations in the society, as well as the role of influential agents in the opinion formation process. In fact, the latter phenomena have been empirically observed and reported in the social science literature, see, e.g., the stream of work originated with Kramer's paper \cite{Kramer:71} documenting large swings in voting behavior within short periods, and the sizable literature in social psychology (e.g., Cohen \cite{Cohen:03}) documenting changes in political beliefs as a result of parties or other influential organizations.

In this paper, we investigate a tractable opinion dynamics model
that generates  both long-run disagreement and opinion
fluctuations. We consider an \emph{inhomogeneous stochastic gossip
model} of communication wherein there is a fraction of
\emph{stubborn agents} in the society who never change their
opinions. We show that the presence of stubborn agents with
competing opinions leads to persistent opinion fluctuations and
disagreement among the rest of the society.

More specifically, we consider a society envisaged as a social
network of $n$ interacting agents (or individuals), communicating
and exchanging information. Each agent $a$ starts with an opinion
(or belief) $X_{a}(0)\in \mathbb{R}$ and is then activated according
to a Poisson process in continuous time. Following this event, she
meets one of the individuals in her \emph{social neighborhood}
according to a pre-specified stochastic process. This process represents an
underlying \emph{social network}. We distinguish between two types
of individuals, stubborn and regular. Stubborn agents,
which are typically few in number, never change their opinions: they
might thus correspond to media sources, opinion leaders, or political parties wishing to influence the rest of the society, and, in a first approximation, not getting any feedback from it. In contrast, regular agents,
which make up the great majority of the agents in the social
network, update their beliefs to some weighted average of their
pre-meeting belief and the belief of the agent they met. The opinions
generated through this information exchange process form a
Markov process whose long-run behavior is the focus of our analysis.

First, we show that, under general conditions, these opinion dynamics never lead to a consensus (among the regular agents). In fact, regular agents' beliefs fail to converge almost surely, and keep on fluctuating in an ergodic fashion. Instead, the belief of each regular agent converges in law to a non-degenerate stationary random variable, and, similarly, the vector of beliefs of all agents jointly converge to a non-degenerate stationary random vector. This model therefore provides a new approach to understanding persistent disagreements and opinion fluctuations.

Second, we investigate how the structure of the graph describing the social network and the location of the stubborn agents within it shape the behavior of the opinion dynamics. The expected belief vector is proved to evolve according to an ordinary differential equation coinciding with the Kolmogorov backward equation of a continuous-time Markov chain on the graph with absorbing states corresponding to the stubborn agents, and hence to converge to a harmonic vector, with every regular agent's value being the weighted average of its neighbors' values, and boundary conditions corresponding to the stubborn agents' beliefs. Expected cross-products of the agents' beliefs allow for a similar characterization in terms of coupled Markov chains on the graph describing the social network. The characterization of the expected stationary beliefs as harmonic functions is then used in order to find explicit solutions for some social networks with particular structure or symmetries.

Third, in what we consider the most novel contribution of our analysis, we study the behavior of the stationary beliefs in large-scale \emph{highly fluid} social networks, defined as networks where the product between the fraction of links incoming in the stubborn agent set times the mixing time of the associated Markov chain is small. 
We show that in highly fluid social networks, the expected value and variance of the stationary beliefs of most of the agents concentrate around certain values  as the population size grows large. We refer to this result as  \emph{homogeneous influence} of stubborn agents on the rest of the society---meaning that their influence on most of the agents in the society is approximately the same.
The applicability of this result is then proved by providing several examples of large-scale random networks, including the Erd\"os--R\'enyi graph in the connected regime, power law networks, and small-world networks.  We wish to emphasize that homogeneous influence in highly fluid societies needs not imply approximate consensus among the agents, whose beliefs may well fluctuate in an almost uncorrelated way. Ongoing work of the authors is aimed at a deeper understanding of this topic.

Our main contribution partly stems from novel applications of
several techniques of applied probability in the study of opinion
dynamics. In particular, convergence in law and ergodicity of the
agents' beliefs is established by first rewriting the dynamics in
the form of an iterated affine function system and then proving almost sure convergence of the time-reversed process \cite{DiaconisFreedman99}. On
the other hand, our estimates of the behavior of the expected
values and variances of the stationary beliefs in large-scale
highly fluid networks are based on techniques from the theory of
Markov chains and mixing times \cite{AldousFillbook,LevinPeresWilmer}, as well as on results in
modern random graph theory \cite{Durrettbook}.


In addition to the aforementioned works on learning and opinion
dynamics, this paper is related to some of the literature in the statistical physics of
social dynamics: see \cite{CastellanoFortunatoLoreto} and
references therein for an overview of such research line. More
specifically, our model is closely related to some work by Mobilia
and co-authors \cite{Mobilia03,Mobilia05,Mobilia07}, who study a
variation of the discrete opinion dynamics model, also called the
\emph{voter model}, with inhomogeneities, there referred to as
\emph{zealots}: such zealots are agents which tend to favor one
opinion in \cite{Mobilia03,Mobilia05}, or are in fact equivalent
to our stubborn agents in \cite{Mobilia07}. These works generally
present analytical results for some regular graphical structures
(such as regular lattices \cite{Mobilia03,Mobilia05}, or complete
graphs \cite{Mobilia07}), and are then complemented by numerical
simulations. In contrast, we prove convergence in distribution and
characterize the properties of the limiting distribution for
general finite graphs. Even though our model involves continuous
belief dynamics, we will also show that the voter model with
zealots of \cite{Mobilia07} can be recovered as a special case of
our general framework.


Our work is also related to work on consensus and gossip
algorithms, which is motivated by different problems, but
typically leads to a similar mathematical formulation (Tsitsiklis \cite%
{johnthes}, Tsitsiklis, Bertsekas and Athans \cite{distasyn},
Jadbabaie, Lin and Morse \cite{ali}, Olfati-Saber and Murray
\cite{reza}, Olshevsky and Tsitsiklis \cite{alexlong}, Fagnani and
Zampieri \cite{fabio-sandro}, Nedi\'c and Ozdaglar
\cite{distpaper}). In consensus problems, the focus is on whether
the beliefs or the values held by different units (which might
correspond to individuals, sensors, or distributed processors)
converge to a common value. Our analysis here does not focus on
limiting consensus of values, but in contrast, characterizes the
stationary fluctuations in values.


The rest of this paper is organized as follows: In Section
\ref{model}, we introduce our model of interaction between the
agents, describing the resulting evolution of individual beliefs, and
we discuss two special cases, in which the arguments simplify particularly, and some fundamental features of the general case are highlighted.
Section \ref{sectconvprop} presents convergence results on the evolution of agent beliefs over time, for a given social network: the beliefs are shown to converge in distribution, and to be an ergodic process, while in general they do not  converge almost surely.
Section \ref{sect1st2ndmoment} presents a characterization of the first and second moments of the stationary beliefs in terms of the hitting probabilities of two coupled Markov chains on the graph describing the social network.
Section \ref{sect:reversible} presents explicit computations of the expected stationary beliefs and variances for some special network topologies.
Section \ref{sectestimates} provides bounds on the level of dispersion of the first two moments of the stationary beliefs: it is shown that, in highly fluid networks, most of the agents have almost the same stationary expected belief and variance.
Section \ref{conclusions} presents some concluding remarks.

\vskip 1pc

\noindent{\bf Basic Notation and Terminology}
We will typically label the entries of vectors by elements of
finite alphabets, rather than non-negative integers, hence
$\R^{\mc I}$ will stand for the set of vectors with entries
labeled by elements of the finite alphabet $\mc I$. An index
denoted by a lower-case letter will implicitly  be assumed to run
over the finite alphabet denoted by the corresponding calligraphic
upper-case letter (e.g. $\sum_i$ will stand for $\sum_{i\in\mc
I}$). For any finite set $\mc J$, we use the notation $\1_{\mc J}$
to denote the indicator function over the set $\mc J$, i.e.,
$\1_{\mc J}(j)$ is equal to 1 if $j\in \mc J$, and equal to $0$
otherwise. For a matrix $M\in\R^{\mc I\times\mc J}$, $M^{\rm
T}\in\R^{\mc J\times\mc I}$ will stand for its transpose,
$||M||$ for its $2$-norm. For a probability distribution $\mu$
over a finite set $\mc I$, and a subset $\mc J\subseteq\mc I$ we
will write $\mu(\mc J):=\sum_{j}\mu_j$. If $\nu$ is another
probability distribution on $\mc I$, we will  use the notation
$||\mu-\nu||_{TV}:=\frac12\sum\nolimits_i|\mu_i-\nu_i|=\sup\l\{|\mu(\mc J)-\nu(\mc J)|:\,\mc J\subseteq\mc I\r\}\,,$
for the total variation distance between $\mu$ and $\nu$. The
probability law (or distribution) of a random variable $Z$ will be
denoted by $\mc L(Z)$. Continuous-time
Markov chains on a finite set $\mc V$ will be characterized by their transition rate matrix $M\in\R^{\mc V\times\mc V}$, which has zero row sums, and whose non-diagonal elements are nonnegative and correspond to the rates at which the chain jumps from a state to another (see \cite[Ch.s 2-3]{Norris}). If $V(t)$ and $V'(t)$ are Markov chains on $\mc V$, defined on the same probability space, we will
use the notation $\P_v(\,\cdot\,)$, and $\P_{vv'}(\,\cdot\,)$, for
the conditional probability measures given the events $V(0)=v$,
and, respectively, $(V(0),V'(0))=(v,v')$. Similarly, for some
probability distribution $\pi$ over $\mc V$ (possibly the
stationary one),
$\P_{\pi}(\,\cdot\,):=\sum_{v,v'}\pi_v\pi_{v'}\P_{vv'}(\,\cdot\,)$
will denote the conditional probability measure of the Markov chain
with initial distribution $\pi$, while $\E_v[\,\cdot\,]$,
$\E_{v,v'}[\,\cdot\,]$, and $\E_{\pi}[\,\cdot\,]$ will denote the
corresponding conditional expectations.
For two non-negative real-valued sequences $\{a_n:\,n\in\N\}$, $\{b_n:\,n\in\N\}$, we
will write $a_n=O(b_n)$ if for some positive constant $K$, $a_n\le Kb_n$ for
all sufficiently large $n$, $a_n=\Theta(b_n)$ if $b_n=O(a_n)$, $a_n=o(b_n)$ if $\lim_{n}a_n/b_n=0$.

\section{Belief evolution model} \label{model}

\begin{figure}
\begin{center}
\includegraphics[height=6cm,width=8cm]{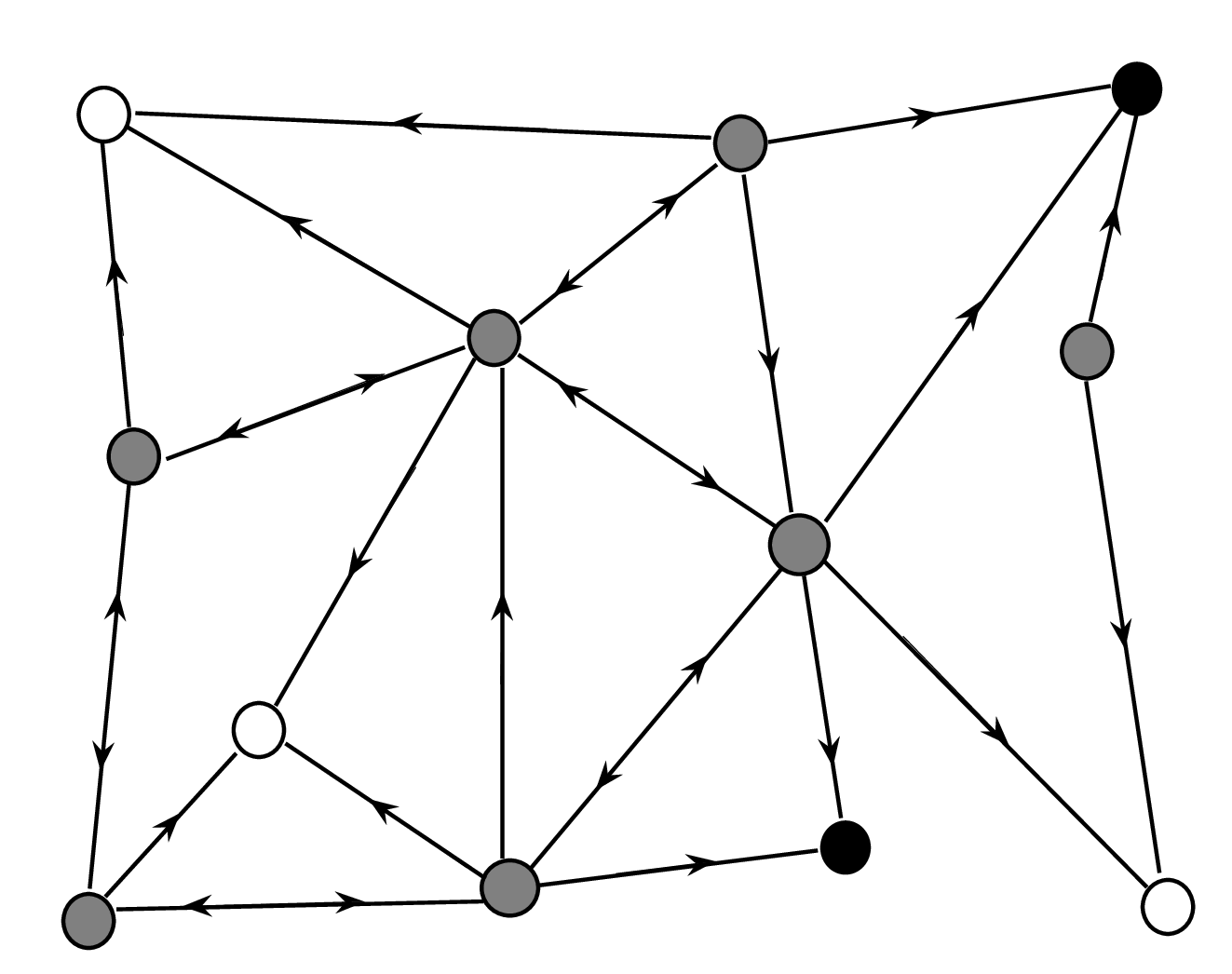}
\caption{\label{stubborn} A social network with seven regular agents (colored in grey), and five stubborn agents (colored in white, and black, respectively). The presence of a directed link $(v,v')$ indicates that agent $v$ is influenced by the opinion of agent $v'$. Therefore, links are only incoming to the stubborn agents, while links between pairs of regular agents may be uni- or bi-directional.}
\end{center}
\end{figure}

We consider a finite population $\mc V$ 
of interacting agents, of possibly very large size $n:=|\mc V|$. The
connectivity among the agents is described by a simple directed
graph $\ora{\mc G}=\big(\mc V,\ora{\mc E}\big)$, whose node set is
identified with the agent population, and where $\ora{\mc
E}\subseteq\mc V\times\mc V\setminus\mc D$, with $\mc D:=\{(v,v):v\in\mc V\}$, stands for the set of directed links  among the agents.\footnote{Notice that we don't allow for parallel links or loops.}

At time $t\ge0$, each agent $v\in\mc V$ holds a \emph{belief} (or
\emph{opinion}) about an underlying state of the world, denoted by
$X_v(t)\in \R$. The full vector of beliefs at time $t$ will be
denoted by $X(t)\:=\{X_v(t):\,v\in\mc V\}$. We distinguish between
two types of agents: regular and stubborn. Regular agents
repeatedly update their own beliefs, based on the observation of
the beliefs of their out-neighbors in $\ora{\mc G} $.
\emph{Stubborn agents} never change their opinions, i.e., they do
not have any out-neighbors. Agents which are not stubborn are
called \emph{regular}. We will denote the set of regular agents by
$\mc A$, the set of stubborn agents by $\mc S$, so that the set of
all agents is $\mc V=\mc A\cup\mc S$ (see Figure \ref{stubborn}).

More specifically, the agents' beliefs evolve according to the
following stochastic update process. At time $t=0$, each agent
$v\in\mc V$ starts with an initial belief $X_v(0)$. The beliefs of the stubborn agents stay constant in time:
$$X_s(t)=X_s(0)=:x_s\,,\qquad s\in\mc S\,,\ t\ge0\,.$$
In contrast, the beliefs of the regular agents are updated as
follows. To every directed link in $\ora{\mc E}$ of the form
$(a,v)$, where necessarily $a\in\mc A$, and $v\in\mc V$, a clock
is associated, ticking at the times of an independent Poisson
process of rate $r_{av}>0$. If the $(a,v)$-th clock ticks at time
$t$, agent $a$ meets agent $v$ and updates her belief to a convex
combination of her own current belief and the current belief of
agent $v$:
\be\label{beliefupdate}X_a(t)=(1-\theta_{av})X_a(t^-)+\theta_{av}X_v(t^-)\,,\ee
where $X_v(t^-)$ stands for the left limit $\lim_{u\uparrow
t}X_v(u)$. Here, the scalar $\theta_{av}\in(0,1]$ is a {\it trust
parameter} that represents the confidence that the regular agent
$a\in\mc A$ puts on agent $v$'s belief.\footnote{We have imposed
that at each meeting instance, only one agent updates her belief.
The model can be easily extended to the case where both agents
update their beliefs simultaneously, without significantly
affecting any of our general results.} That $r_{av}$ and
$\theta_{av}$ are strictly positive for all $(a,v)\in\ora{\mc E}$
is simply a convention (since if $r_{av}\theta_{av}=0$, one can
always consider the subgraph of $\ora{\mc G} $ obtained by
removing the link $(a,v)$ from $\ora{\mc E}$). Similarly, we also
adopt the convention that $r_{vv'}=\theta_{vv'}=0$ for all
$v,v'\in\mc V$ such that $(v,v')\notin\ora{\mc E}$ (hence,
including loops $v'=v$). For every regular agent $a\in\mc A$,
let $\mc S_a\subseteq\mc S$ be the subset of stubborn agents which
are reachable from $a$ by a directed path in $\ora{\mc G} $. We
refer to $\mc S_a$ as the set of stubborn agents \emph{
influencing} $a$. For every stubborn agent $s\in\mc S$,  $\mc
A_s:=\{a:\,s\in\mc S_a\}\subseteq\mc A$ will stand for the set of
regular agents \emph{influenced} by $s$.

The tuple $\mc N=\l(\ora{\mc G} , \{\theta_e\}, \{r_e\}\r)$ contains
the entire information about patterns of interaction among the
agents, and will be referred to as the {\it social network}.
Together with an assignment of a probability law for the initial
belief vector, $\mc L(X(0))$,
the social network designates a {\it society}.
 Throughout the paper, we make the following assumptions regarding the underlying social network.

\begin{assumption}\label{assumptionconnected}
Every regular agent is influenced by some stubborn agent, i.e., $\mc S_a$ is non-empty for every $a$ in $\mc A$.
\end{assumption}


Assumption \ref{assumptionconnected} may be easily removed.
If there are some regular agents which are not influenced by any
stubborn agent, then there is no link in $\mc E$ connecting the set $\mc R$ of such regular agents to $\mc V\setminus\mc R$. Then, one may decompose the subgraph obtained by restricting $\mc G$ to $\mc R$ into its communicating classes, and apply the results in \cite{fabio-sandro} (see Example 3.5 therein), showing that, with probability one, a consensus on a random belief is achieved on every such communicating class.


We denote the {\it total meeting
rate of agent $v\in \mc V$} by $r_v$, i.e., $r_v:=\sum_{v'}r_{vv'}$, and the {\it total meeting rate of all agents} by $r$,
i.e., $r:=\sum_{v}r_v$. We use $N(t)$ to denote the {\it
total number of agent meetings (or link activations) up to time
$t\ge0$}, which is simply a Poisson arrival process of rate $r$. We
also use the notation $T_{(k)}$ to denote the {\it time of the $k$-th
belief update}, i.e., $T_{(k)}:=\inf\{t\ge0:\,N(t)\ge k\}$.

To a given social network, we associate 
the matrix $Q\in\R^{\mc V\times\mc V}$, with entries
\be\label{Qdef}Q_{vw}:=\theta_{vw}r_{vw}\qquad
Q_{vv}:=-\sum\nolimits_{v'\ne v}Q_{vv'}\,,\qquad v\ne w\in\mc V\,.\ee
In the rest of the paper, we will often consider a continuous-time Markov chain $V(t)$ on $\mc V$ with transition rate matrix $Q$. 

The following example describes the canonical construction of a
social network from an undirected graph, and will be used often in
the rest of the paper.
\begin{example}\label{examplestandardRW} Let $\mc G=(\mc V,\mc E)$ be a connected multigraph,\footnote{I.e., $\mc E$ is a multi-set of unordered pairs of elements of $\mc V$. This allows for the possibility of considering parallel links and loops.} and $\mc S\subseteq\mc V$, $\mc A=\mc
V\setminus\mc S$. Define the directed graph $\ora{\mc G}=(\mc
V,\ora{\mc E})$, where $(a,v)\in\ora{\mc E}$ if and only if $a\in\mc
A$, $v\in\mc V\setminus\{a\}$, and $\{a,v\}\in\mc E$, i.e., $\ora{\mc G}$ is the
directed graph obtained by making all links in $\mc E$ bidirectional
except links between a regular and a stubborn agent, which are
unidirectional (pointing from the regular agent to the stubborn
agent). For $v,w\in\mc V$, let $\kappa_{v,w}$ denote the multiplicity of the link $\{v,w\}$ in $\mc E$ (each self-loop contributing as $2$), and let $d_v=\sum_w\kappa_{v,w}$ be the degree of node $v$ in $\mc G$. (In particular, $\kappa_{a,v}=\1_{\mc E}(\{a,v\})$ if $\mc G$ is a simple graph, i.e., it does not contain neither loops nor parallel links.) Let the trust parameter be constant, i.e.,
$\theta_{av}=\theta\in(0,1]$ for all $(a,v)\in\ora{\mc E}$. Define
\be\label{rav1} r_{av}=d_a^{-1}\kappa_{a,v}\1_{\mc V\setminus\{a\}}(v)\,,\qquad a\in\mc A\,,\ v\in\mc V\,.\ee 
This concludes the construction of the social
network $\mc N=\l(\ora{\mc G} , \{\theta_e\}, \{r_e\}\r)$. In particular, one has
$$Q_{av}=\theta\kappa_{a,v}/d_a\,,\qquad \forall (a,v)\in\ora{\mc E}\,.$$
Observe that connectedness of $\mc G$ implies that Assumption \ref{assumptionconnected} holds. Finally, notice that nothing prevents the multigraph $\mc G$ from having (possibly parallel) links between two nodes both in $\mc S$. However, such links do not have any correspondence in the directed graph $\ora {\mc G}$, and in fact they are irrelevant for the belief dynamics, since stubborn agents do not  update their beliefs. 
\end{example}\medskip
We conclude this section by discussing in some detail two special
cases whose simple structure sheds light on the main features of the
general model. In particular, we consider a social network with a
single regular agent and a social network where the trust parameter
satisfies $\theta_{av}=1$ for all $a\in {\cal A}$ and $v\in {\cal
V}$. We
show that in both of these cases agent beliefs fail to converge almost surely.

\subsection{Single regular agent}\label{noasconv}
\begin{figure}
\begin{center}
\subfigure[]{\includegraphics[height=5cm,width=2cm]{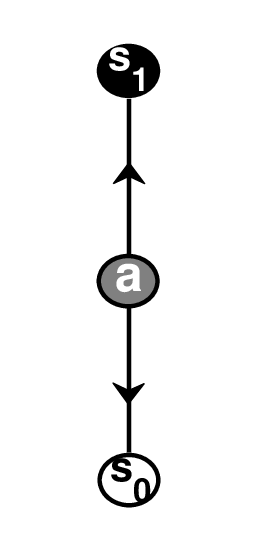}}\hspace{0cm}
\subfigure[]{\includegraphics[height=5cm,width=6.5cm]{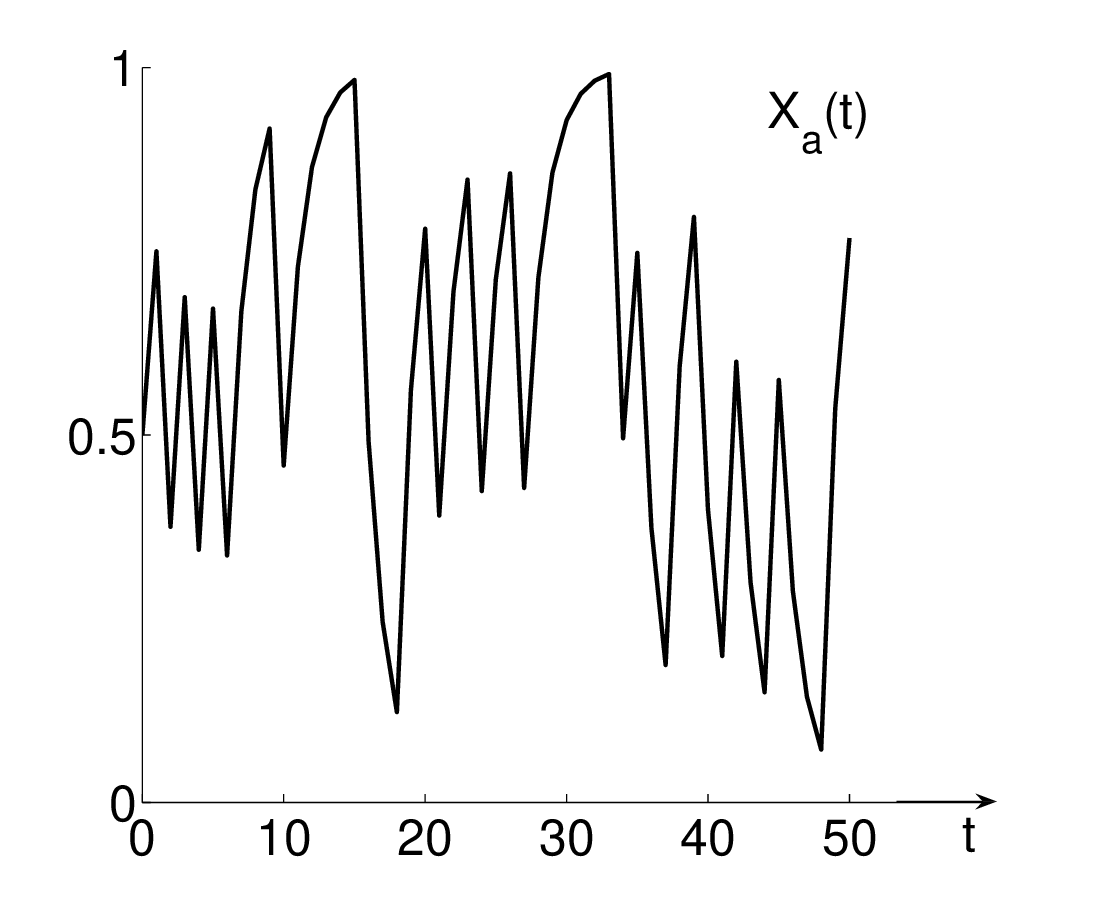}}\hspace{0cm}
\subfigure[]{\includegraphics[height=5cm,width=6cm]{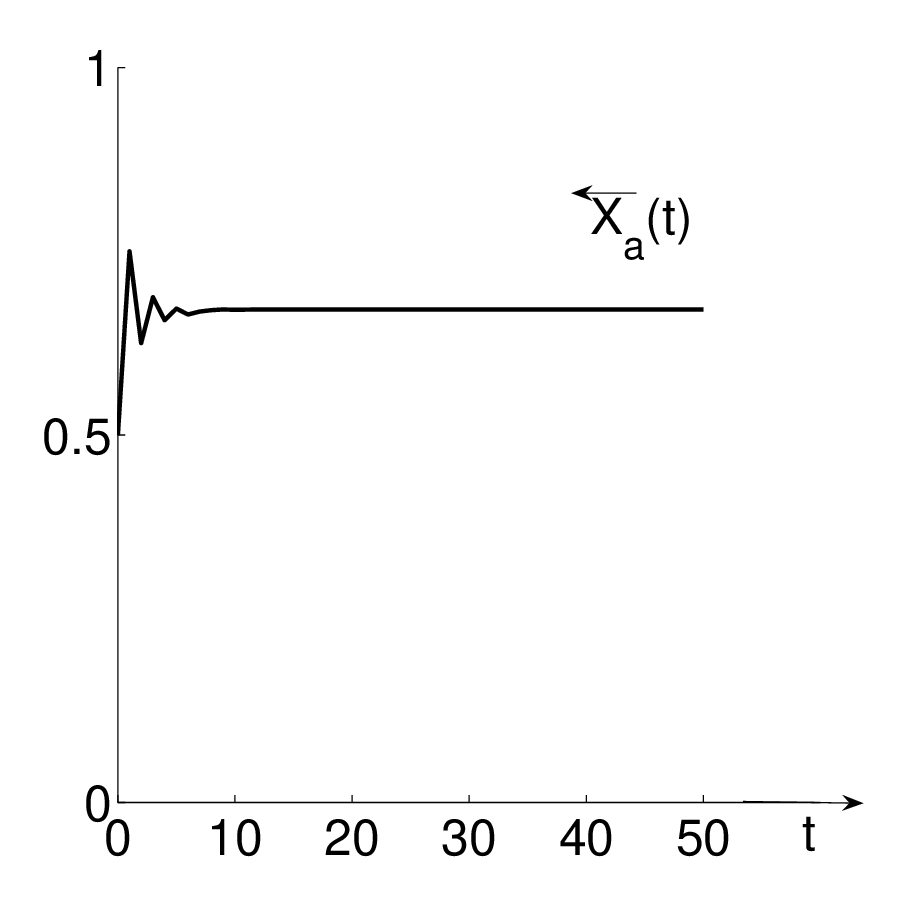}}
\caption{\label{figFWBW}
Typical sample-path behavior of the belief of the regular agent in the simple social network topology depicted in (a). In (b) the actual belief process $X_a(t)$, fluctuating ergodically on the interval $[0,1]$; in (c), the time-reversed process, rapidly converging to a stationary random belief $X_a$.}
\end{center}
\end{figure}
Consider a society consisting of a single regular agent, i.e., $\mc
A=\{a\}$, and two stubborn agents, $\mc S=\{s_0,s_1\}$ (see Fig.~\ref{figFWBW}(a)). Assume that
$r_{as_0}=r_{as_1}=1/2$, $\theta_{as_0}=\theta_{as_1}=1/2$, $x_{s_0}=0$,
$x_{s_1}=1$, and $X_a(0)=0$. Then, one has for all $t\ge 0$,
$$X_a(t)=\sum_{1\le k\le N(t)}2^{k-N(t)-1}B(k)\,,$$
where $N(t)$ is the total number of agent meetings up to time $t$
(or number of arrivals up to time $t$ of a rate-$1$ Poisson
process), and $\{B(k):\,k\in\N\}$ is a sequence of Bernoulli($1/2$)
random variables, independent mutually and from the process $N(t)$.
Observe that, almost surely, arbitrarily long strings of contiguous zeros and ones appear in the sequence $\{B(k)\}$, while the number of meetings $N(t)$ grows unbounded. It follows that, with probability one
$$\liminf_{t\to\infty} X_a(t)=0\,,\qquad\limsup_{t\to\infty}X_a(t)=1\,,$$
so that the belief $X_a(t)$ does not converge almost surely.


On the other hand, observe that, since
$\sum_{k>n}2^{-k}|B(k)|\le2^{-n}$, the series
$X_a:=\sum_{k\ge1}2^{-k}B(k)$ is sample-wise converging. It
follows that,  as $t$ grows large, the time-reversed process
$$\ola{X_a}(t):=\sum_{1\le k\le N(t)}2^{-k}B(k)$$ converges to
$X_a$, with probability one, and, a fortiori, in distribution.
Notice that, for all positive integer $k$, the binary $k$-tuples
$\{B(1),\ldots,B(k)\}$ and $\{B(k),\ldots,B(1)\}$ are uniformly distributed over $\{0,1\}^k$, and independent from the
Poisson arrival process $N(t)$. It follows that,  for all $t\ge0$,
the random variable $\ola X_a(t)$ has the same distribution as
$X_a(t)$. Therefore, $X_a(t)$ converges in distribution to $X_a$ as
$t$ grows large. Moreover, it is a standard fact (see
e.g.~\cite[pag.92]{RaoSwift06}) that $X_a$ is uniformly distributed
over the interval $[0,1]$. Hence, the probability distribution of
$X_a(t)$ is asymptotically uniform on $[0,1]$.

The analysis can in fact be extended to any trust parameter
$\theta_{is}=\theta_{is'}=\theta\in(0,1)$. In this case, one gets
that $$X_a(t)=\theta\sum_{1\le k\le N(t)}(1-\theta)^{N(t)-k}B(k)$$
converges in law to the stationary belief
\be\label{Xadef}X_a:=\theta(1-\theta)^{-1}\sum_{k\ge1}(1-\theta)^kB(k)\,.\ee
As explained in \cite[Section 2.6]{DiaconisFreedman99}, for every
value of $\theta$ in $(1/2,1)$, the probability law of $X_a$ is
singular, and in fact supported on a Cantor set. In contrast, for
almost all values of $\theta\in(0,1/2)$, the probability law of
$X_a$ is absolutely continuous with respect to Lebesgue's
measure.\footnote{See \cite{PeresSolomyak96}. In fact, explicit
counterexamples of values of $\theta\in(0,1/2)$ for which the
asymptotic measure is singular are known. For example, Erd\"os
\cite{Erdos39,Erdos40} showed that, if $\theta=(3-\sqrt5)/2$, then
the probability law of $X_a $ is singular.}
In the extreme case $\theta=1$, it is not hard to see that $X_a(t)=B(N(t))$ converges in distribution to a random variable $X_a$ with Bernoulli($1/2$) distribution.
On the other hand, observe that, regardless of the fine structure of
the probability law of the stationary belief $X_a$, i.e.,~on whether
it is absolutely continuous or singular, its moments can be characterized for all values of $\theta\in(0,1]$. In
fact, it follows from (\ref{Xadef}) that the expected value of $X_a$ is given by
$$\E[X_a]=\theta(1-\theta)^{-1}\sum_{k\ge1}(1-\theta)^k\E[B(k)]=\theta\sum_{k\ge0}(1-\theta)^k\frac12=\frac12\,,$$
and, using the mutual independence of the $B(k)$'s, the variance of
$X_a$ is given by
$$\Var[X_a]=\theta^2(1-\theta)^{-2}\sum_{k\ge1}(1-\theta)^{2k}\Var[B(k)]=\theta^2\sum_{k\ge0}(1-\theta)^{2k}\frac14=\frac{\theta}{4(2-\theta)}\,.$$
Observe that the expected value of $X_a$ is independent from $\theta$, while its variance increases from $0$ to a maximum of $1/2$ as $\theta$ is increased from $0$ to $1$. 


\subsection{Voter model with zealots}\label{sect:voter}
\begin{figure}
\begin{center}
\includegraphics[height=6cm,width=10cm]{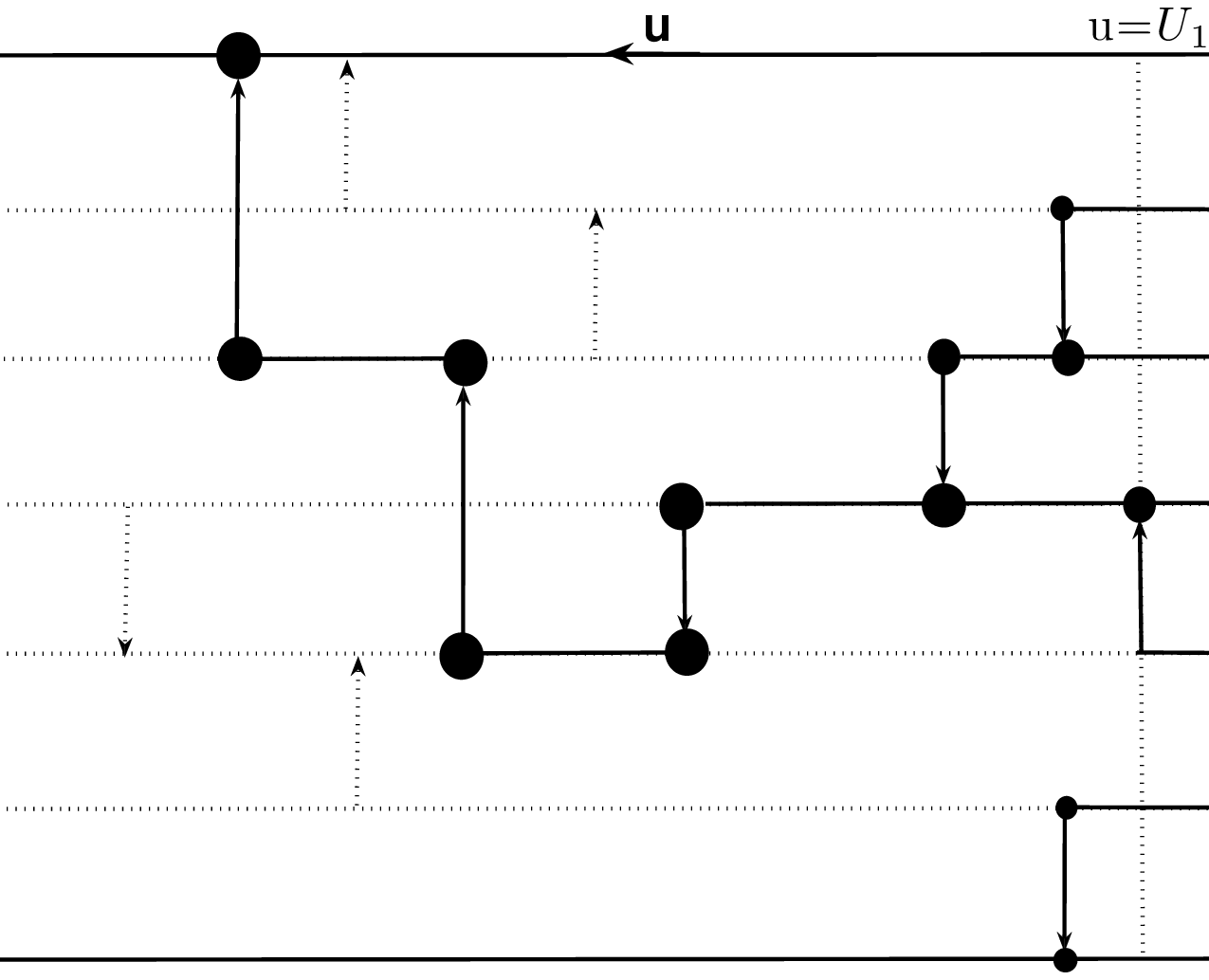}
\end{center}
\caption{\label{figdualRW} Duality between the voter model with zealots and the coalescing Markov chains process with absorbing states. The network topology is a line with five regular agents and two stubborn agents placed in the two extremities. The time index for the opinion dynamics, $t$, runs from left to right, whereas the time index for the coalescing Markov chains process, $u$, runs from right to left. Both dotted and solid arrows represent meeting instances. Fixing a time horizon $t>0$, in order to trace the beliefs $X(t)$, one has to follow coalescing Markov chains starting at $u=0$ in the different nodes of the network, and jumping from a state to another one in correspondence to the solid arrows. The particles are represented by bullets at times of their jumps. Clusters of coalesced particles are represented by bullets of increasing size.}
\end{figure}
We now consider the special case when the social network topology $\ora{\mc G}$ is arbitrary, and $\theta_{av}=1$ for all $(a,v)\in\mc E$. In this case, whenever a link $(a,v)\in\ora{\mc E}$ is activated, the regular agent $a$ adopts agent $v$'s current opinion as such, completely disregarding her own current opinion.

This opinion dynamics, known as the \emph{voter model}, was introduced independently by Clifford and Sudbury
\cite{CliffordSudbury73}, and Holley and Liggett \cite{HolleyLiggett75}. It has been extensively studied in the framework of
interacting particle systems \cite{Liggettbook,Liggettbook2}. While
most of the research focus has been on the case when the graph is an infinite lattice, the voter model on finite graphs,
and without stubborn agents, was considered, e.g., in
\cite{ConnelyWelsh82,Cox89}, \cite[Ch.~14]{AldousFillbook}, and
\cite[Ch.~6.9]{Durrettbook}: in this case, consensus is achieved in
some finite random time, whose distribution
depends on the graph topology only.

In some recent work \cite{Mobilia07} a variant with one or more
stubborn agents (there referred to as \emph{zealots}) has been
proposed and analyzed on the complete graph. We wish to
emphasize that such voter model with zealots can be recovered as a
special case of our model, and hence our general results, to be
proven in the next sections, apply to it as well. However, we briefly discuss this special case here, since proofs are much more intuitive, and allow one to anticipate some of the general results.

The main tool in the analysis of the voter model is the \emph{dual
process}, which runs backward in time and allows one to identify the
source of the opinion of each agent at any time instant.
Specifically, let us focus on the belief of a regular agent $a$ at
time $t>0$. Then, in order to trace $X_a(t)$, one has to look at the
last meeting instance of agent $a$ that occurred no later than time
$t$. If such a meeting instance occurred at some time
$t-U_1\in[0,t]$ and the agent met was $v\in\mc V$, then the belief
of agent $a$ at time $t$ coincides with the one of agent $v$ at time
$t-U_1$, i.e., $X_a(t)=X_v(t-U_1)$. The next step is to look at the
last meeting instance of agent $v$ occurred no later than time
$t-U_1$; if such an instance occurred at time $t-U_2\in[0,t-U_1]$,
and the agent met was $w$, then $X_a(t)=X_v(t-U_1)=X_w(t-U_2)$.
Clearly, one can iterate this argument, going backward in time,
until reaching time $0$. In this way, one implicitly defines a
continuous-time Markov chain $V_a(u)$ with state space $\mc V$, which starts at $V_a(0)=a$ and stays put there until time $U_1$, when it jumps to node $v$ and stays put there in the time interval $[U_1,U_2)$, then jumps at time
$U_2$ to node $w$, and so on. It is not hard to see that, thanks to
the fact that the meeting instances are independent Poisson
processes, the Markov chain $V_a(u)$ has transition rate matrix $Q$. In particular, it halts when it hits some state $s\in\mc S$. This shows that
$\mc L(X_a(t))=\mc L(X_{V_a(t)}(0))\,.$
More generally, if one is interested in the joint probability
distribution of the belief vector $X(t)$, then one needs to consider
$n-|\mc S|$ continuous-time Markov chains, $\{V_a(t):\,a\in\mc A\}$ each one starting from
a different node in $\mc A$ (specifically, $V_a(0)=a$ for all $a\in\mc A$), and run simultaneously on $\mc V$ (see
Figure \ref{figdualRW}). These Markov chains move independently with
transition rate matrix $Q$, until the first time that they either
meet, or they hit the set $\mc S$: in the former case, they stick
together and continue moving on $\mc V$ as a single particle, with
transition rate matrix $Q$; in the second case, they halt. This
process is known as the \emph{coalescing Markov chains process} with
absorbing set $\mc S$. Then, one gets that \be\label{Xtdist}\mc
L(\{X_a(t):\,a\in\mc A\})=\mc L(\{X_{V_a(t)}(0):\,a\in\mc A\})\,.\ee
Equation (\ref{Xtdist}) establishes a \emph{duality} between the
voter model with zealots and the coalescing Markov chains process
with absorbing states. In particular, Assumption
\ref{assumptionconnected} implies that, with probability one, each $V_a(u)$ will hit the set $\mc S$ in some finite random time $T_{\mc S}^a$, so that in particular the vector
$\{V_a(u):\,a\in\mc A\}$ converges in distribution, as $u$ grows large, to an
$\mc S^{\mc A}$-valued random vector $\{V_a(T^a_{\mc S}):\,a\in\mc
A\}$. It then follows from (\ref{Xtdist}) that $X(t)$ converges in
distribution to a stationary belief vector $X$ whose entries are
given by $X_s=x_s$ for every stubborn agent $s\in\mc S$, and
$X_a=x_{V_a(T^a_{\mc S})}$ for every regular agent $a\in\mc A$.

\section{Convergence in distribution and ergodicity of the beliefs}\label{sectconvprop}
\begin{figure}
\begin{center}
\subfigure[]{\vspace{1cm}\includegraphics[height=5cm,width=1.8cm]{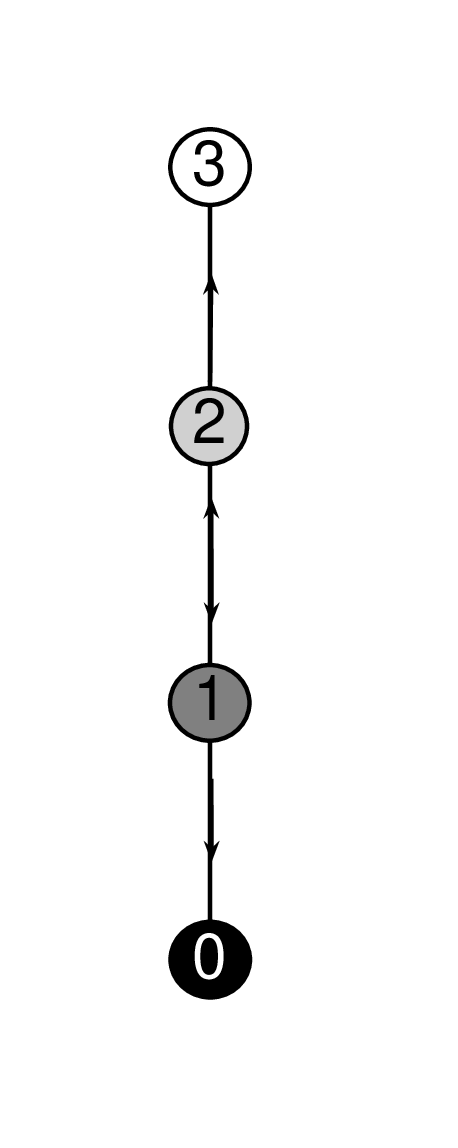}}
\subfigure[]{\includegraphics[height=5cm,width=6cm]{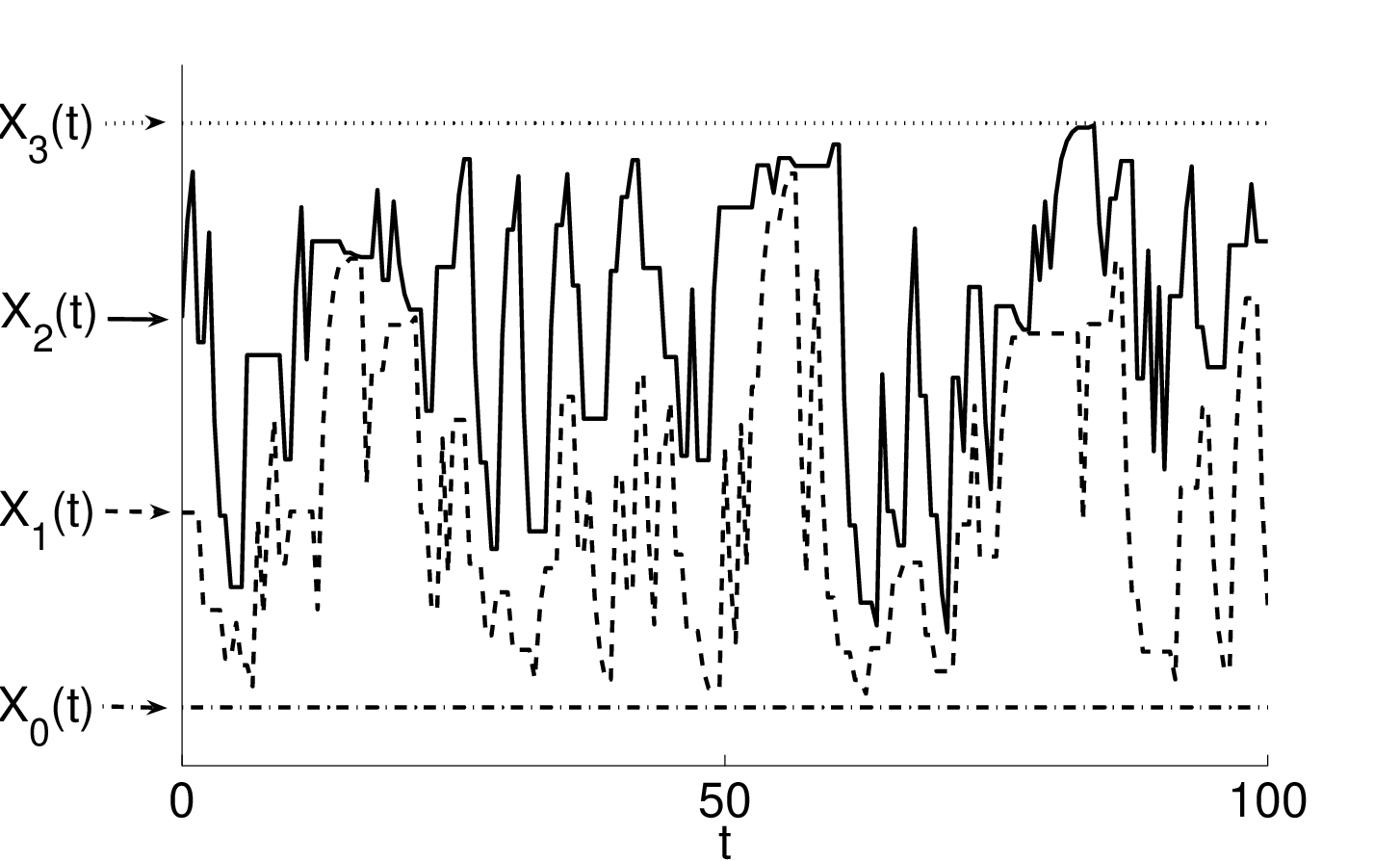}}
\subfigure[]{\includegraphics[height=5cm,width=6cm]{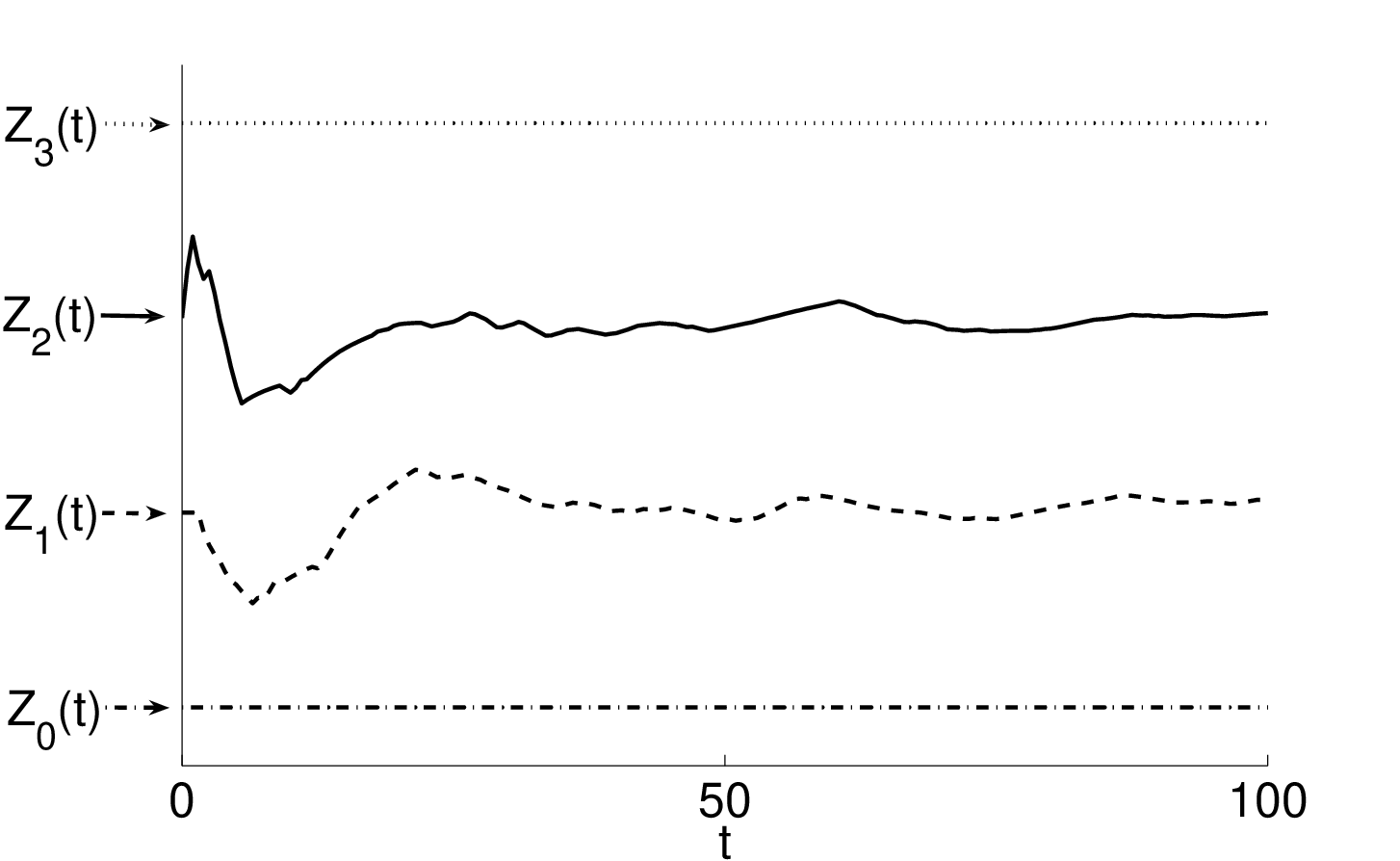}}
\caption{\label{Figergplot}Typical sample-path behavior of the
beliefs, and their ergodic averages for a social network with
population size $n=4$. The topology is a line graph, displayed in
(a). The stubborn agents correspond to the two extremes of the
line, $\mc S=\{0,3\}$, while their constant opinions are $x_0=0$,
and $x_3=1$. The regular agent set is $\mc A=\{1,2\}$. The
confidence parameters, and the interaction rates are chosen to be
$\theta_{av}=1/2$, and $r_{av}=1/3$, for all $a=1,2$, and $v=a\pm1$.
In picture (b), the trajectories of the actual beliefs $X_v(t)$, for
$v=0,1,2,3$, are reported, whereas picture (c) reports the
trajectories of their ergodic averages
$Z_v(t):=t^{-1}\int_0^tX_v(u)\de u$.}
\end{center}
\end{figure}

This section is devoted to studying the convergence properties of
the random belief vector $X(t)$ for the general update model
described in Section \ref{model}. Figure \ref{Figergplot} reports the
typical sample-path behavior of the agents' beliefs for a simple
social network with population size $n=4$, and line graph topology,
in which the two stubborn agents are positioned in the extremes and
hold beliefs $x_0<x_3$. As shown in Fig.~\ref{Figergplot}(b), the beliefs of the two
regular agents, $X_1(t)$, and $X_2(t)$, fluctuate persistently in the interval
$[x_0,x_3]$. On the other hand, the
time averages of the two regular agents' beliefs rapidly approach a
limit value, of $2x_0/3+x_3/3$ for agent $1$, and $x_0/3+2x_3/3$ for
agent $2$.

As we will  see below, such behavior is rather general. In our
model of social network with at least two stubborn agents having
non-coincident constant beliefs, the regular agent beliefs fail to converge almost
surely: we have seen this in the special cases of
Section \ref{noasconv}, while a general result in this sense will
be stated as Theorem \ref{coroNOconvP1}. On the other hand, we
will prove that, regardless of the initial regular agents'
beliefs, the belief vector $X(t)$ is convergent in distribution to
a random stationary belief vector $X$ (see Theorem
\ref{theoconvergence}), and in fact it is an ergodic process (see
Corollary \ref{coroergodic}).

In order to prove Theorem \ref{theoconvergence}, we will  rewrite
$X(t)$ in the form of an iterated affine function system \cite{DiaconisFreedman99}. 
Then, we will consider the so-called time-reversed belief process. This is a stochastic process whose marginal probability
distribution, at any time $t\ge0$, coincides with the one of the
actual belief process, $X(t)$. In contrast to $X(t)$, the
time-reversed belief process is in general not Markov, whereas it
can be shown to converge to a random stationary belief vector with
probability one. From this, we recover convergence in distribution
of the actual belief vector $X(t)$.

Formally, for any time instant $t\ge0$, let us introduce the projected belief vector $Y(t)\in\R^{\mc A}$,
where $Y_a(t)=X_a(t)$ for all $a\in\mc A$. Let $I_{\mc A}\in\R^{\mc
A\times\mc A}$ be the identity matrix, and for $a\in\mc A$, let
$e_{(a)}\in\R^{\mc A}$ be the vector whose entries are all zero, but
for the $a$-th which equals $1$. For every
positive integer $k$, consider the random matrix $A(k)\in\R^{\mc
A\times\mc A}$, and the random vector $B(k)\in\R^{\mc A}$, defined
by $$A(k)=I_{\mc A}+\theta_{aa'}\l(e_{(a)}e_{(a')}^{\rm T}-e_{(a)}e_{(a)}^{\rm T}\r)\,\qquad
B(k)=0\,,$$ if the $k$-th activated link is $(a,a')\in\ora{\mc E}$, with
$a,a'\in\mc A$, and
$$A(k)=I_{\mc A}-\theta_{as}e_{(a)}e_{(a)}^{\rm T}\,\qquad
B(k)=e_{(a)}\theta_{as}x_s\,,$$ if the $k$-th activated link is
$(a,s)\in\ora{\mc E}$, with $a\in\mc A$, and $s\in\mc S$. Define the
matrix product \be\label{oraA}\ora A(k,l):=A(l)A(l-1)\ldots A(k+1)A(k)\,,\qquad
1\le k\le l\,,\ee with the convention that $\ora A(k,l)=I_{\mc A}$
for $k>l$. Then, at the time $T_{(l)}$ of the $l$-th belief update, one has
$$Y(T_{(l)})=A(l)Y(T_{(l)}^-)+B(l)=A(l)Y(T_{(l-1)})+B(l)\,,\qquad l\ge1\,,$$
so that, for all $t\ge 0$,
\be\label{IFS1} Y(t)= \ora A(1,N(t))Y(0)+\summ_{1\le k\le N(t)}\ora A(k+1,N(t))B(k)\,,\ee
where we recall that $N(t)$ is the total number of agents' meetings up to time $t$.
Now, define the \emph{time-reversed belief process}
\be\label{TRdef}\ola Y(t):= \ola A(1,N(t))Y(0)+\summ_{1\le k\le N(t)}\ola A(1,k-1)B(k)\,,\ee
where
$$\ola A(k,l):=A(k)A(k+1)\ldots A(l-1)A(l)\,,\qquad k\le l\,,$$
with the convention that $\ola A(k,l)=I_{\mc A}$ for $k>l$.
The following is a fundamental observation (cf.~\cite{DiaconisFreedman99}):
\begin{lemma}\label{TRlemma1}
For all $t\ge0$, $Y(t)$ and $\ola Y(t)$ have the same probability distribution.
\end{lemma}
\proof{Proof.}
Notice that $\l\{\l(A(k),B(k)\r):\,k\in\N\r\}$ is a sequence of independent and identically distributed random variables, independent from the process $N(t)$. This, in particular, implies that, the $l$-tuple $\{(A(k),B(k)):\,1\le k\le l\}$ has the same distribution as the $l$-tuple $\{(A(l-k+1),B(l-k+1)):\,1\le k\le l\}$, for all $l\in\N$. From this, and the identities (\ref{IFS1}) and (\ref{TRdef}), it follows that the belief vector $Y(t)$ has the same distribution as $\ola Y(t)$, for all $t\ge0$.
\qed\endproof

The second fundamental result is that, in contrast to the actual
regular agents' belief vector $Y(t)$, the time-reversed belief
process $\ola Y(t)$ converges almost surely. 

\begin{lemma}\label{TRlemma2}
Let Assumption \ref{assumptionconnected} hold. Then, for every value of the
stubborn agents' beliefs $\{x_s\}\in\R^{\mc S}$, there exists an
$\R^{\mc A}$-valued random variable $Y$, such that,
$$\P\l(\lim_{t\to\infty}\ola Y(t)=Y\r)=1\,,$$
for every initial distribution $\mc L(Y(0))$ of the regular agents' beliefs.
\end{lemma}
\proof{Proof.}
Observe that the expected entries of $A(k)$,  and $B(k)$, are given by
$$\E[A_{aa'}(k)]=\frac{Q_{aa'}}r\,,\qquad\E[A_{aa}(k)]=1+\frac{Q_{aa}}{r}\,,
\qquad \E[B_a(k)]=\frac1r\sum\nolimits_{s}Q_{as}x_s\,,$$ for all
$a\ne a'\in\mc A$. In particular, $\E[A(k)]$ is a substochastic matrix.
It follows from Perron-Frobenius' theory that the spectrum of $\E[A(k)]$ is contained in the disk centered in $0$ of radius $\rho$,
where $\rho\in[0,+\infty)$ is its largest in module eigenvalue, with corresponding left eigenvector $y$ with nonnegative entries.
 Moreover, Assumption \ref{assumptionconnected} implies that, for all nonempty subsets $\mc J\subseteq\mc A$, there exists some $j\in\mc J$ and $v\in\mc V\setminus\mc J$ such that $(j,v)\in\mc E$ (otherwise $\mc S_j=\emptyset$ for all $j\in\mc J$). Therefore $\sum_{a}\E[A_{ja}]\le1-r^{-1}Q_{jv}<1$. Choosing $\mc J$ as the support of the eigenvector $y$ gives
$\rho\sum\nolimits_ay_a
=\sum_a(\E[A(k)]^{\rm T}y)_a
=\sum\nolimits_jy_j\sum\nolimits_a\E[A_{ja}(k)]<\sum\nolimits_jy_j\,,$
so that $\rho<1$. Then, using the Jordan canonical decomposition, one can show that
$$\l|\l|\E\l[\ola A(1,k)\r]\r|\r|_{\infty}\le Ck^{n-1}\rho^k\,,\qquad\forall k\ge0\,,$$
where $C$ is a constant depending on $\E[A(1)]$ only. 
Upon observing that the $\ola A(1,k)$ has non-negative entries, and using the inequality $\E[\max\{Z,W\}]\le\E[Z]+\E[W]$ valid for all nonnegative-valued random variables $Z$ and $W$, one gets that
\be\label{estimateQ}
\E\l[\l|\l|\ola A(1,k)\r|\r|_{1}\r]
=\E\l[\max\limits_{a'}\summ\nolimits_{a}\ola A_{aa'}(1,k)\r]
\le\summ_{a,a'}\E\l[\ola A_{a,a'}(1,k)\r]
\le n\l|\l|\E\l[\ola A(1,k)\r]\r|\r|_{\infty}
\le Cnk^{n-1}\rho^k\,.\ee
Now, fix some $\ups\in(\rho,1)$. 
From thÄe independence of $\ola A(1,k-1)$ and $B(k)$ it follows that, for all $k\ge1$,
$$
\ba{rcl}
\P\l(\l|\l|\ola A(1,k-1)B(k)\r|\r|_1\ge\ups^{k-1}\r)
&\le&\ups^{-k+1}\E\l[||\ola A(1,k-1)B(k)||_1\r]\\
&\le&\ups^{-k+1}\E\l[||\ola A(1,k-1)||_1||B(k)||_1\r]\\
&=&\ups^{-k+1}\E\l[||\ola A(1,k-1)||_1\r]\E\l[||B(k)||_1\r]\\
&\le&  \beta_{k-1}\,,\ea
$$
where $\beta_k:=Cnk^{n-1}(\rho/\ups)^k\E[||B(1)||_1]$. 
Since $\sum_{k\ge0}\beta_{k}<\infty$, the above bound and the Borel-Cantelli lemma imply that, with probability one,  $||\ola A(1,k-1)B(k)||_1<\ups^{k-1}$ for all but finitely many values of $k\ge1$.
Hence, almost surely, 
the series $$Y:=\sum_{k\ge1}\ola A(1,k-1)B(k)$$ is absolutely convergent. An analogous argument shows almost sure convergence of $\ola A(1,k)Y(0)$ to $0$, as $k$ grows large. 
Since, with probability one, $N(t)$ goes to infinity as $t$ grows large, one has that
$$\lim\limits_{t\to\infty}\ola Y(t)=\lim\limits_{t\to\infty}\ola A(1,N(t))Y(0)+\lim\limits_{t\to\infty}\summ_{1\le j\le N(t)}\ola A(1,j-1)B(j)=Y\,,$$
with probability one. This completes the proof.
\qed\endproof

Lemma \ref{TRlemma1} and Lemma \ref{TRlemma2} allow one to prove convergence in distribution of $X(t)$ to a random belief vector $X$, as stated in the following result.

\begin{theorem}\label{theoconvergence}
Let Assumption \ref{assumptionconnected} hold. Then, for every value of the
stubborn agents' beliefs $\{x_s\}\in\R^{\mc S}$, there exists an
$\R^{\mc V}$-valued random variable $X $, such that, for every initial distribution $\mc L(X(0))$ satisfying $\P(X_s(0)=x_s)=1$ for every $s\in\mc S$, 
$$\lim_{t\to\infty}\E[\varphi(X(t))]=\E[\varphi(X)]\,,$$
for all bounded and continuous test functions $\varphi:\R^{\mc V}\to\R$. Moreover, the probability law of the stationary belief vector $X$ is invariant for the system, i.e., if $\mc L(X(0))=\mc L(X)$, then $\mc L(X(t))=\mc L(X)$ for all $t\ge0$.
\end{theorem}
\proof{Proof.}
It follows from Lemma \ref{TRlemma2} $\ola Y(t)$ converges to $Y$ with probability one, and a
fortiori in distribution. By Lemma \ref{TRlemma1}, $\ola Y(t)$ and $Y(t)$ are identically distributed.
Therefore,  $Y(t)$ converges to $Y$ in distribution, and the first part of the claim follows by defining $X_a=Y_a$ for all $a\in\mc A$, and $X_s=x_s$ for all $s\in\mc S$. For the second part of the claim, it is sufficient to observe that the distribution of $Y=\sum_{k\ge1}\ola A(1,k-1)B(k)$ is the same as the one of $Y':=A(0)Y+B(0)$, where $A(0)$, and $B(0)$, are independent copies of $A(1)$, and $B(1)$, respectively.
\qed\endproof

Motivated by Theorem \ref{theoconvergence}, for any agent $v\in \mc
V$, we refer to the random variable $X_v$ as the {\it stationary
belief of agent $v$}. 
Using standard ergodic theorems for Markov processes, an immediate
implication of Theorem \ref{theoconvergence} is the following
corollary, which shows that time averages of continuous
functions of agent beliefs with bounded expectation are given by their expectation over the
limiting distribution. Choosing the relevant function properly, this
enables us to express the empirical averages of, and correlations
across, agent beliefs in terms of expectations over the limiting
distribution, highlighting the ergodicity of agent beliefs.

\begin{corollary}\label{coroergodic}

Let Assumption \ref{assumptionconnected} hold. Then, for every value of the
stubborn agents' beliefs $\{x_s\}\in\R^{\mc S}$, with probability one,
$$\lim_{t\to\infty}\frac1t\int_0^t\varphi(X(u))\de u=\E[\varphi(X)]\,,$$
where $X$ is the stationary belief vector and $\varphi:\R^{\mc V}\to\R$ is any continuous test function such that $\E[\varphi(X)]$ exists and is finite.
\end{corollary}
\proof{Proof.}
Let $Y(t)$ and $Y$ be the projections of the belief vector at time $t\ge0$, and of the stationary belief vector $A$, respectively, to the regular agents set $\mc A$. Let $\tilde Y(0)$ be an $\R^{\mc A}$-valued random vector, independent from $Y(0)$ and such that $\mc L(\tilde Y(0))=\mc L(Y)$. Let $Y(t)$ be as in (\ref{IFS1}), and
$$\tilde Y(t)=\ora A(1,N(t))\tilde Y(0)+\summ_{1\le k\le N(t)}\ora A(k+1,N(t))B(k)\,,$$
where $N(t)$ is the total number of agents' meetings up to time $t$, and $\ora A(k,l)$ is defined as in (\ref{oraA}). Then,
$\tilde Y(t)-Y(t)=\ora A(1,N(t))(\tilde Y(0)-Y(0))\,.$
Arguing as in the proof of Lemma \ref{TRlemma2}, one shows that $\lim_{t\to\infty}||\tilde Y(t)-Y(t)||=0\,,$ with probability one. Now, for $t>0$, let the vectors $\tilde X(t)$ and $X(t)$ be defined by $\tilde X_a(t)=\tilde Y_a(t)$, $X_a(t)=Y_a(t)$ for $a\in\mc A$, and $\tilde X_s(t)=X_s(t)=x_s$ for $s\in\mc S$, and observe that, with probability one, $\sup_{t\ge0}|X(t)|\le\max_v|X_v(0)|<\infty$, $\sup_{t\ge0}|\tilde X(t)|\le\max_v|X_v|<\infty$. Then, for every continuous $\varphi:\R^{\mc V}\to\R$, one has that
$$\lim_{t\to\infty}|\varphi(\tilde X(t))-\varphi(X(t))|=0\,,$$
with probability one. 
On the other hand, stationarity of the process $\tilde X(t)$ allows one to apply the ergodic theorem (see, e.g., \cite[Theorem 6.2.12]{Strook}), showing that, if $\E[\varphi(X)]$ exists and is finite, then
$$\lim_{t\to\infty}\frac1t\int_0^t\varphi(\tilde X(s))\de s=\E[\varphi(X)]\,,$$
with probability one. Then, for any continuous $\varphi$ such that $\E[\varphi(X)]$ exists and is finite, one has that 
$$\l|\frac1t\int_0^t\varphi(X(s))\de s-\E[\varphi(X)]\r|\le 
\l|\frac1t\int_0^t\varphi(\tilde X(s))\de s-\E[\varphi(X)]\r|+ 
\frac1t\int_0^t\l|\varphi(X(s))-\varphi(\tilde X(s))\r|\de s\stackrel{t\to\infty}{\longrightarrow}0\,,
$$
with probability one. 
\qed\endproof

Theorem \ref{theoconvergence}, and Corollary \ref{coroergodic},
respectively, show that the beliefs of all the agents converge in
distribution, and that their empirical distributions converge almost
surely, to a random stationary belief vector $X$. In contrast, the
following theorem shows that the stationary belief of a regular
agent which is connected to at least two stubborn agents with
different beliefs is a non-degenerate random variable. As a
consequence, the belief of every such regular agent keeps on
fluctuating with probability one. Moreover, the theorem shows that
the difference between the belief of a regular agent influenced by at least two stubborn agents with
different beliefs, and the belief of any other agent does not converge to zero with probability one, so that disagreement between them
persists in time. For $a\in\mc A$, let $\mc X_a=\{x_s:\,s\in\mc S_a\}$
denote the set of stubborn agents' belief values influencing agent
$a$. 

\begin{theorem}\label{coroNOconvP1}
Let Assumption \ref{assumptionconnected} hold, and let $a\in\mc A$ be such that $|\mc X_a|\ge2$. Then, the stationary belief $X_a$ is a non-degenerate random variable. Furthermore, $\P(X_a\ne X_{v})>0$ for all $v\in\mc V\setminus\{a\}$.
\end{theorem}
\proof{Proof.}
With no loss of generality, since the distribution of the stationary belief vector $X$ does not depend on the probability law of the initial beliefs of the regular agents, we can assume that such a law is the stationary one, i.e., that $\mc L(X(0))=\mc L(X)$. Then, Theorem \ref{theoconvergence} implies that $\mc L(X(t))=\mc L(X)$ for all  $t\ge0$.

Let $a\in\mc A$ be such that $X_a$ is degenerate. Then, almost
surely, $X_a(t)=x_a$ for almost all $t$, for some constant $x_a$.
Then, as we will  show below, all the out-neighbors of $a$ will
have their beliefs constantly equal to $x_a$ with probability one.
Iterating the argument until reaching the set $\mc S_a$, one
eventually finds that $x_{s}=x_a$ for all $s\in\mc S$, so that
$|\mc X_a|=1$. This proves the first part of the claim. For the
second part, assume that $X_a=X_{a'}$ almost surely for some $a\ne
a'$. Then, one can prove that, with probability one, every
out-neighbor of $a$ or $a'$ agrees with $a$ or $a'$ at any time.
Iterating the argument until reaching the set $\mc S_a\cup\mc
S_{a'}$, one eventually finds that $|\mc X_s\cup\mc X_{s'}|=1$.

One can reason as follows in order to see that, if $v$ is an out-neighbor of $a$, and $X_a=x_a$ is degenerate, then $X_v(t)=x_a$ for all $t$. Let $T^{av}_{(k)}$ be the $k$-th activation of the link $(a,v)$. Then, Equation (\ref{beliefupdate}) implies that \be\label{contradict0}X_v(T^{av}_{(k)})=x_a\,,\qquad \forall n\ge1\,.\ee
Now, define $T^*:=\inf\{t\ge0:\,X_v(t)\ne x_a\}$, and assume by contradiction that $\P(T^*<\infty)>0$. By the strong Markov property, and the property that link activations are independent Poisson processes, this would imply that \be\label{contr}\P(\text{first link activated after }T^*\text{ is }(a,v)|\,\mc F_{T^*})>0\qquad\text{ on }\qquad\{T^*<\infty\}\,,\ee which would contradict (\ref{contradict0}). Then, necessarily $T^*=\infty$, and hence $X_v=X_a$, with probability one.

On the other hand, assume that $\P(X_a=X_{v})=1$ for some $v\in\mc V\setminus\{a\}$. Then, with probability one $X_a(t)=X_v(t)$ for all rational $t\ge0$. Since, as proved above, with probability one $X_a(t)$ is not constant in $t$, both $X_a(t)$ and $X_v(t)$ should jump simultaneously. However, the probability of this to occur is zero since link activations are independent Poisson processes. Therefore, necessarily $\P(X_a=X_{v})<1$.
\qed\endproof

Even though, by Theorem \ref{theoconvergence}, the belief of any agent always converges in distribution, Theorem \ref{coroNOconvP1} shows that, if a regular agent $a$ is influenced by stubborn agents with different beliefs, then her stationary belief $X_a$ is non-degenerate. By Corollary \ref{coroergodic}, this implies that, with probability one, her belief $X_a(t)$ keeps on fluctuating and does not stabilize on a limit. Similarly, Theorem \ref{coroNOconvP1} and Corollary \ref{coroergodic} imply that, if a regular agents is influenced by stubborn agents with different beliefs, then, with probability one, her belief will not achieve a consensus asymptotically with any other agent in the society. 

\section{Expected beliefs and belief crossproducts}
\label{sect1st2ndmoment}
\begin{figure}
\begin{center}
\hspace{-1cm}\subfigure[]{\includegraphics[height=4.5cm,width=5cm]{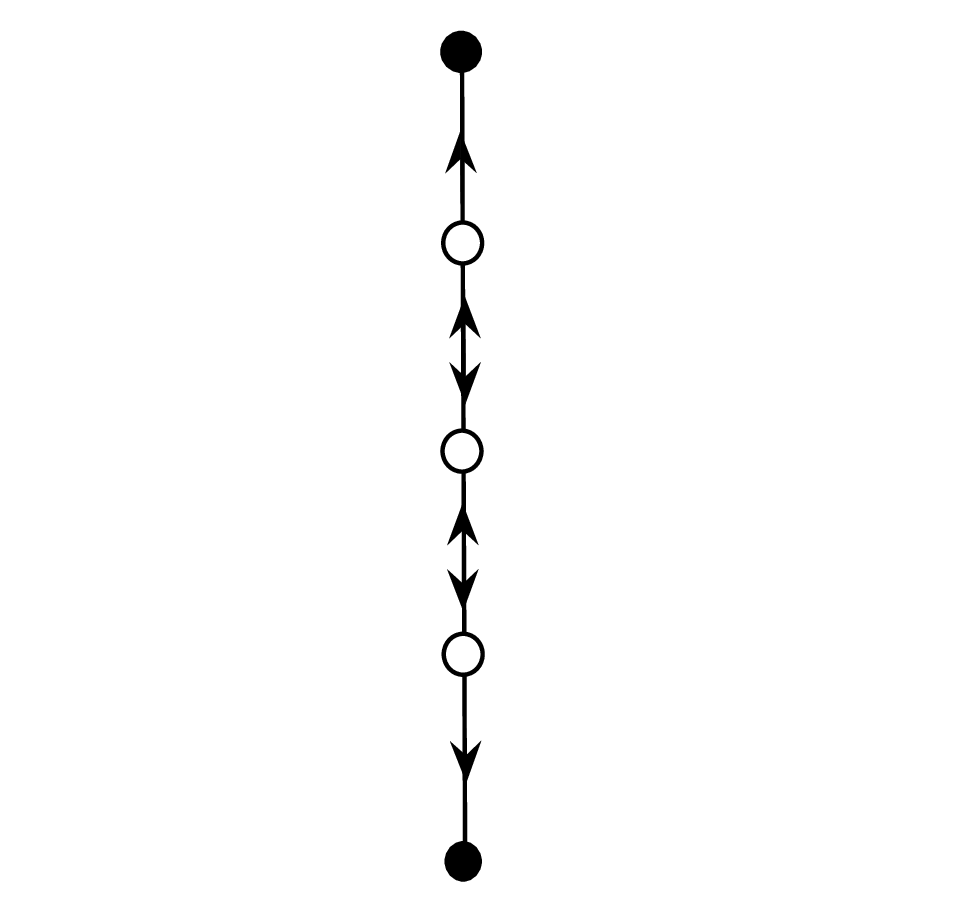}}\hspace{-1cm}
\subfigure[]{\includegraphics[height=4.5cm,width=5cm]{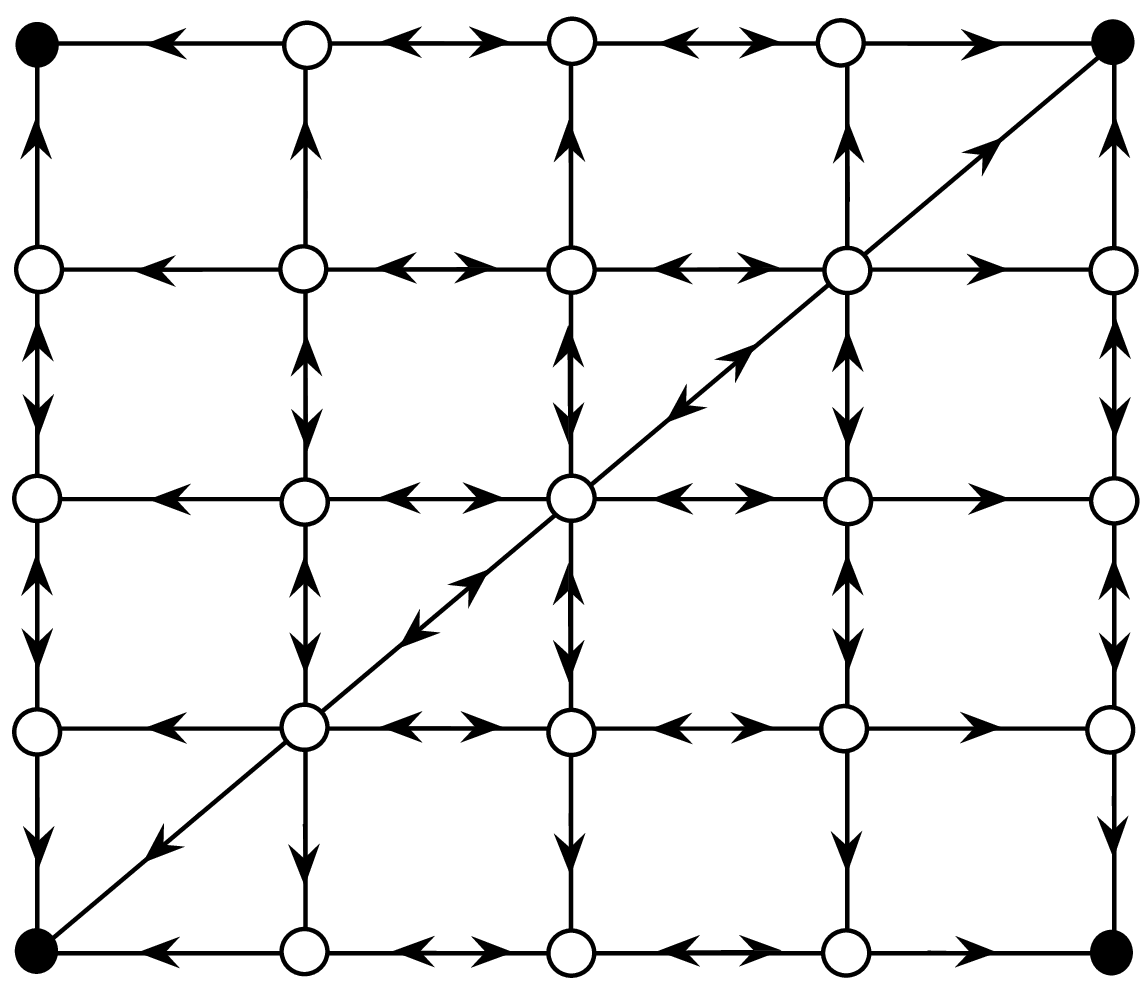}}\hspace{1cm}
\subfigure[]{\includegraphics[height=4.5cm,width=5cm]{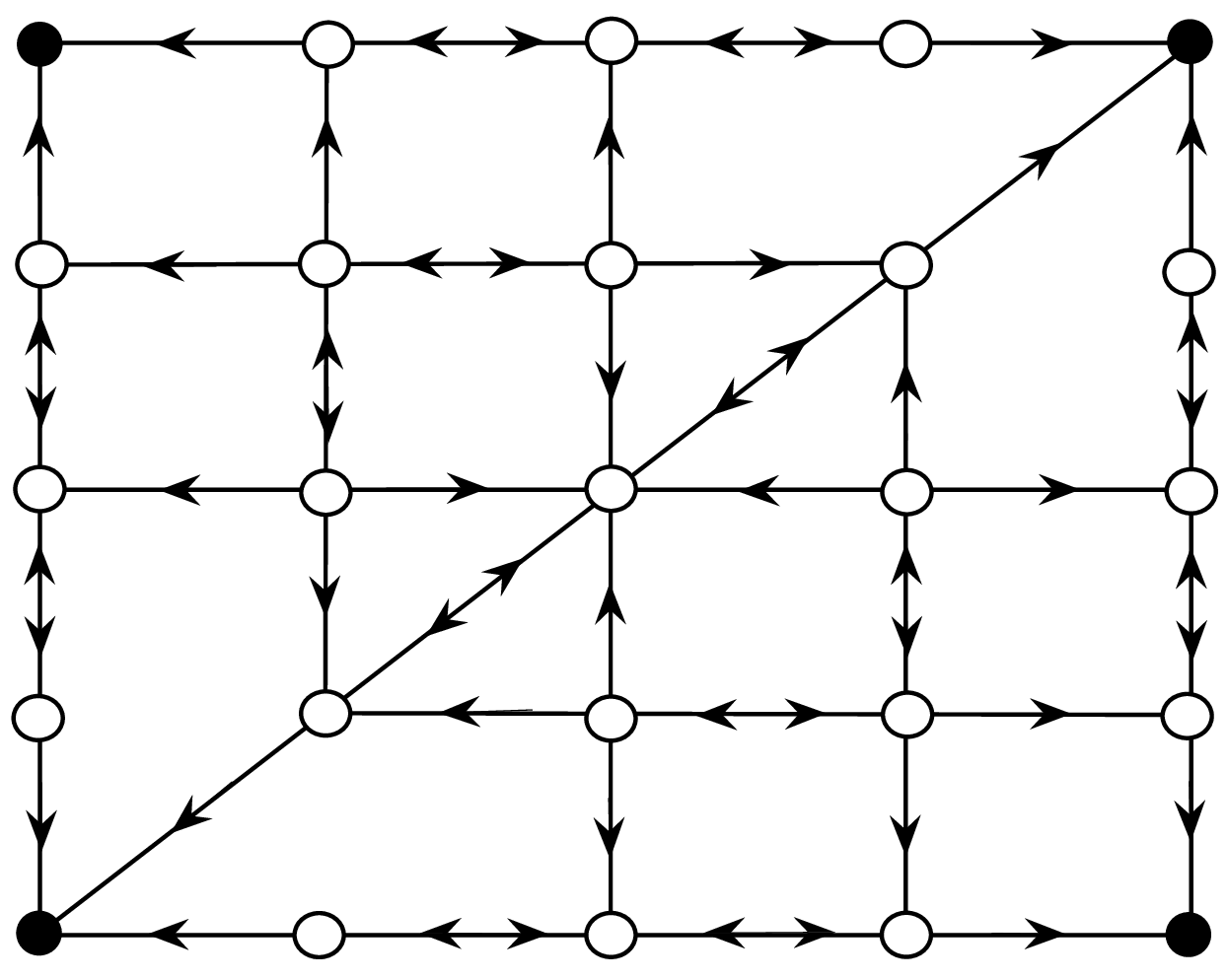}}
\caption{\label{figlineproduct}In (a), a network topology consisting
of a line with three regular agents and two stubborn agents placed
in the extremes. In (b), the directed graph representing the possible state transitions of the corresponding coupled Markov chains pair $(V(t),V'(t))$ when $\theta_e\in(0,1)$ for all $e\in\stackrel{\rightarrow}{\mc E}$. 
Such coupled chains pair has $25$ states, four of which are absorbing states. The components of $(V(t),V'(t))$ jump independently to neighbor states, unless they are either on the diagonal, or one of them is in
$\mc S$: in the former case, there is some chance that the two
components jump as a unique one, thus inducing a direct connection
along the diagonal; in the latter case, the only component that can
keep moving is the one which has not hit $\mc S$, while the one who
hit $\mc S$ is bound to remain constant from that point on. In (c),
the state transition graph is reported for the Markov chains pair $(V(t),V'(t))$ in the extreme
case when $\theta_{e}=1$ for all $e\in\stackrel{\rightarrow}{\mc E}$.
 In this case, the coupled Markov chains are
coalescing: once they meet, they stick together, moving as a single
particle, and never separating from each other. This reflects the
fact that there are no outgoing links from the diagonal set.
}
\end{center}
\end{figure}
In this section, we provide a characterization of the expected beliefs and belief crossproducts of the agents. 
In particular we will provide explicit characterizations of the stationary expected beliefs and belief crossproducts in terms
of hitting probabilities of a pair of coupled Markov chains on
$\ora{\mc G}=(\mc V,\ora{\mc E})$.\footnote{Note that the set of
states for such Markov chain corresponds to the set of agents,
therefore we use the terms ``state" and ``agent" interchangeably in
the sequel.} Specifically, we consider a coupling
$(V(t),V'(t))$ of continuous-time Markov chains with state space $\mc V$, such that
both $V(t)$ and $V'(t)$ have transition rate matrix
$Q$, as defined in (\ref{Qdef}). The pair $(V(t),V'(t))$ is a Markov chain on the state space $\mc V\times\mc V$ with transition rate matrix $K$ whose entries are given by
\be\label{Mdef}K_{(v,v')(w,w')}:=\l\{\ba{ll}
Q_{vw}&\se\ v\ne v'\,,\ v\ne w\,,\ v'=w'\\[3pt]
Q_{v'w'}&\se\ v\ne v'\,,\  v=w\,,\ v'\ne w'\\[3pt]
0&\se\ v\ne v'\,,\ v\ne w\,,\ v'\ne w'\\[3pt]
Q_{vv}+Q_{v'v'}&\se\ v\ne v'\,,\ v=w\,,\ v'=w'\\[3pt]
\theta_{vw}Q_{vw}&\se\ v=v'\,,\ w=w'\,,\ v\ne w\\[3pt]
(1-\theta_{vw})Q_{vw}&\se\ v=v'\,,\ v\ne w\,,\ v'= w'\\[3pt]
(1-\theta_{vw'})Q_{vw'}&\se\ v=v'\,,\ v= w\,,\ v'\ne w'\\[3pt]
0&\se\ v=v'\,,\ w\ne w'\,,\ w\ne v\,,\ w'\ne v'\\[3pt]
2Q_{vv}+\sum_{v''\ne v}\theta_{vv''}Q_{vv''}&\se\ v=v'\,,\ v=w\,,\ v'=w'\,.\ea\r.\ee
The first four lines of (\ref{Mdef}) state that, conditioned on
$(V(t),V'(t))$ being on a pair of non-coincident nodes $(v,v')$,
each of the two components, $V(t)$ (respectively, $V'(t)$), jumps to
a neighbor node $w$,  with transition rate $Q_{vw}$ (respectively,
to a neighbor node $w'$ with transition rate $Q_{v'w'}$), whereas
the probability that both components jump at the same time is zero.
On the other hand, the last five lines of (\ref{Mdef}) state that,
once the two components have met, i.e., conditioned on
$V(t)=V'(t)=v$, they have some chance to stick together and jump as
a single particle to a neighbor node $w$, with rate
$\theta_{vw}Q_{vw}$, while each of the components $V(t)$
(respectively, $V'(t)$) has still some chance to jump alone to a
neighbor node $w$ with rate $(1-\theta_{vw})Q_{vw}$ (resp., to $w'$
with rate $(1-\theta_{vw'})Q_{vw'}$). In the extreme case when
$\theta_{vw}=1$ for all $(v,w)$ in $\mc E$, the sixth and seventh line of the righthand
side of (\ref{Mdef}) equal $0$, and in fact one recovers the
expression for the transition rates of two coalescing Markov chains:
once $V(t)$ and $V'(t)$ have met, they stick together and move as a
single particle, never separating from each other. See Figure \ref{figlineproduct} for a visualization of the possible state transitions of $(V(t),V'(t))$.

For $v,w,v',w'\in\mc V$, and $t\ge0$, we will denote by
\be\label{gammatetatdef}
\gamma^v_{w}(t):=\P_v(V(t)=w)=\P_v(V'(t)=w)\,,\qquad
\eta^{vv'}_{ww'}(t)=\P_{vv'}(V(t)=w,V'(t)=w')\,,\ee the marginal and
joint transition probabilities of the two Markov chains at time
$t$. It is a standard fact (see, e.g., \cite[Theorem
2.8.3]{Norris}) that such transition probabilities satisfy the \emph{Kolmogorov
backward equations} \be\label{backward} \frac{\de}{\de
t}\gamma^v_w(t)=\sum_{\tilde v}Q_{v\tilde v}\gamma^{\tilde
v}_w(t)\,,\qquad \frac{\de}{\de t}\eta^{vv'}_{ww'}(t)=\sum_{\tilde
v,\tilde v'}K_{(v,v')(\tilde v,\tilde v')}\eta^{\tilde v\tilde v
'}_{ww'}(t)\,,\qquad v,v',w,w'\in\mc V\,,
 \ee
with initial condition \be\label{initialcond}\gamma^v_w(0)=\l\{\ba{rcl}1&\se&v=w\\0&\se&v\ne w\,,\ea\r.\qquad
\eta^{vv'}_{ww'}(0)=\l\{\ba{rcl}1&\se&(v,v')=(w,w')\\0&\se&(v,v')\ne(w,w')\,.\ea\r.\ee

The next simple result provides a fundamental link between the belief evolution process introduced in Section \ref{model} and the coupled Markov chains, by showing that the expected values and expected cross-products of the agents' beliefs satisfy the same linear system (\ref{backward}) of ordinary differential equations as the transition probabilities of $(V(t),V'(t))$.

\begin{lemma}\label{lemmaODE}
For all $v,v'\in\mc V$, and $t\ge0$, it holds
\be\label{heateqn}\frac{\de}{\de t}\E[X_v(t)]=\sum_{w}Q_{vw}\E[X_w(t)]\,,\qquad \frac{\de}{\de t}\E[X_v(t)X_{v'}(t)]=\sum_{w,w'}K_{(v,v')(ww')}\E[X_w(t)X_{w'}(t)]\,,\ee
so that 
\be\label{dualityfinitetime}
\E[X_v(t)]=\sum_w\gamma^v_w\E[X_w(0)]\,,\qquad \E[X_v(t)X_{v'}(t)]=\sum_{w,w'}\eta^{vv'}_{ww'}\E[X_w(0)X_{w'}(0)]\,.
\ee
\end{lemma}
\proof{Proof.} Recall that, for the belief update model introduced in Section \ref{model}, arbitrary agents' interactions occur at the ticking instants $T_{(k)}$ of a Poisson clock of rate $r$. Moreover, with conditional probability $r_{vw}/r$, any such interaction involves agent $v$ updating her opinion to a convex combination of her current belief and the one of agent $w$, with weight $\theta_{vw}$ on the latter. It follows that, for all $k\ge0$, and $v\in\mc V$,
$$\E[X_v(T_{(k+1)})|\mc F_{T_{(k)}}]-X_v(T_{(k)})=\frac1r\sum_{w}r_{vw}\theta_{vw}(X_w(T_{(k)})-X_v(T_{(k)}))=\frac1r\sum_wQ_{vw}X_w(T_{(k)})\,.$$
Then, the above and the fact that the Poisson clock has rate $r$ imply the left-most equation in (\ref{heateqn}).

Similarly, or all $v\ne v'\in\mc V$, one gets that
$$
\ba{l}\ds\E[X_v(T_{(k+1)})X_{v'}(T_{(k+1)})|\mc F_{T_{(k)}}]-X_v(T_{(k)})X_{v'}(T_{(k)})\\[7pt]
\qquad\qquad\qquad\qquad\qquad\qquad\qquad=\frac1r\sum_wr_{vw}\theta_{vw}\l(X_w(T_{(k)})X_{v'}(T_{(k)})-X_v(T_{(k)})X_{v'}(T_{(k)})\r)\\[7pt]
\qquad\qquad\qquad\qquad\qquad\qquad\qquad\quad+\frac{1}r\sum_{w'}r_{v'w'}\theta_{v'w'}\l(X_{v}(T_{(k)})X_{w'}(T_{(k)})-X_v(T_{(k)})X_{v'}(T_{(k)})\r)\\[10pt]
\qquad\qquad\qquad\qquad\qquad\qquad\qquad=\frac1r\sum_{w,w'}K_{(v,v')(w,w')}X_wX_{w'}\,,\ea
$$
as well as
$$\ba{rcl}\E[X_v^2(T_{(k+1)})|\mc F_{T_{(k)}}]-X_v^2(T_{(k)})&=&
\frac{1}r\sum_wr_{vw}\E\l[\l((1-\theta_{vw})X_v+\theta_{vw}X_w\r)^2-X_v^2\r]\\[7pt]
&=&\frac{1}r\sum_wr_{vw}\theta_{vw}\l(\theta_{vw}X_w^2+2(1-\theta_{vw})X_vX_w-(2-\theta_{vw})X_v^2\r)\\[7pt]
&=&\frac{1}r\sum_wQ_{vw}(1-\theta_{av})\l(X_aX_v-X_a^2\r)\\[5pt]
&&+\frac{1}r\sum_{w'}Q_{vw'}(1-\theta_{vw'})\l(X_vX_{w'}-X_v^2\r)\\[5pt]
&&+\frac{1}r\sum_{w}Q_{vw}\theta_{vw}\l(X_w^2-X_v^2\r)\\[7pt]
&=&\frac{1}r\sum_{w,w'}K_{(v,v)(w,w')}X_wX_{w'}\,.
\ea$$
Then, the two identities above and the fact that the Poisson clock has rate $r$ imply the right-most equation in (\ref{heateqn}).
It follows from (\ref{backward}), (\ref{initialcond}), and (\ref{heateqn}) that $\sum_w\gamma^v_w(t)\E[X_w(0)]$ and $\E[X_v(t)]$ satisfy the same linear system of differential equations, and the same holds true for $\E[X_v(t)X_{v'}(t)]$ and $\sum_{w,w'}\eta^{vv'}_{ww'}(t)\E[X_w(0)X_{w'}(0)]$. This readily implies (\ref{dualityfinitetime}).
\qed\endproof

We are now in a position to prove the main result of this section characterizing the expected values and expected cross-products of the agents' stationary beliefs in terms of the hitting probabilities of the coupled Markov chains. Let us denote by $T_{\mc S}$ and $T'_{\mc S}$ the
{\it hitting times} of the Markov chains $V(t)$, and respectively
$V'(t)$, on the set of stubborn agents $\mc S$, i.e., $$T_{\mc
S}:=\inf\{t\ge0:\,V(t)\in\mc S\}\,,\qquad T'_{\mc
S}:=\inf\{t\ge0:\,V'(t)\in\mc S\}\,.$$
Observe that Assumption \ref{assumptionconnected} implies that both $T_{\mc S}$ and $T'_{\mc S}$ are finite with probability one for every initial distribution of the pair $(V(0),V'(0))$.
Hence, for all $v,v'\in\mc V$, we can define the {\it hitting probability distributions}
$\gamma^v$ over $\mc S$, and $\eta^{vv'}$ over $\mc S^2$, whose
entries are respectively given by
\be\label{gammadef}\ba{l}\gamma^v_s:=\P_v(V(T_{\mc S})=s)\,,\qquad s\in\mc S\,,\\[5pt]
\eta^{vv'}_{ss'}:=\P_{vv'}(V(T_{\mc S})=s,V'(T'_{\mc S})=s')\,,\qquad s,s'\in\mc S\,.\ea\ee
Then, we have the following:
\begin{theorem}\label{theoasymptoticmeanvariance}
Let Assumption \ref{assumptionconnected} hold.
Then, for every value of the
stubborn agents' beliefs $\{x_s\}\in\R^{\mc S}$,
\be\label{EXv}\E[X_v]=\summ_{s}\gamma_s^vx_s\,,\qquad\E[X_vX_{v'}]=\summ_{s,s'}\eta_{ss'}^{vv'}x_sx_{s'}\,, \qquad v,v'\in\mc V\,.\ee
 Moreover, $\{\E[X_v]:\,v\in\mc V\}$ and $\{\E[X_vX_{v'}]:\,v,v'\in\mc V\}$ are the unique vectors in $\R^{\mc V}$, and $\R^{\mc V\times\mc V}$ respectively, satisfying
\be\label{harmonic}\sum_vQ_{av}\E[X_v]=0\,,\qquad\E[X_s]=x_s\,,\qquad \forall a\in\mc A\,,\quad \forall s\in\mc S
\,,\ee
\be\label{harmonic2}\sum_{w,w'}K_{(a,a'),(w,w')}\E[X_{w}X_{w'}]=0\,,\qquad\E[X_vX_{v'}]=\E[X_v]\E[X_{v'}]\,,\qquad \forall a,a'\in\mc A\,,\ \forall (v,v')\in\mc V^2\setminus\mc A^2\,.\ee
\end{theorem}
\proof{Proof.}
Assumption \ref{assumptionconnected} implies that
$\lim_{t\to\infty}\gamma^v_s(t)=\gamma^v_s$ for every $s\in\mc S$, and $\lim_{t\to\infty}\gamma^v_a(t)=0$ for every $a\in\mc A$.
Therefore, (\ref{dualityfinitetime}) implies that
\be\label{eq3}\lim_{t\to\infty}\E[X_v(t)]=\sum_s\gamma^v_sx_s\,,\qquad \forall v\in\mc V\,.\ee
Now, if the initial belief distribution $\mc L(X(0))$ coincides with the stationary one $\mc L(X)$, one has that $\mc L(X(t))=\mc L(X)$ for all $t\ge0$, so that in particular $\E[X_v(t)]=\E[X_v]$, and hence $\lim_{t\to\infty}\E[X_v(t)]=\E[X_v]$ for all $v$. Substituting in the righthand side of (\ref{eq3}), this proves the leftmost identity in (\ref{EXv}). The rightmost identity in (\ref{EXv}) follows from an analogous argument.

In order to prove the second part of the claim, observe that the expected stationary beliefs and belief cross-products necessarily satisfy (\ref{harmonic}) and (\ref{harmonic2}), since, by Lemma \ref{lemmaODE}, they evolve according to the autonomous differential equations (\ref{heateqn}), and are convergent by the arguments above. On the other hand,  uniqueness of the solutions of (\ref{harmonic}) and (\ref{harmonic2}) follows from \cite[Ch.~2, Lemma 27]{AldousFillbook}.
\qed\endproof

\begin{remark}
Since the stationary beliefs $X_v$ take values in the interval $[\min_s x_s,\max_sx_s]$, one has that both $\E[X_v]$ and $\E[X_{v}X_{v'}]$ exist and are finite. Hence, Corollary \ref{coroergodic} implies that the asymptotic empirical averages of the agents' beliefs and their cross-products, i.e., of the almost surely constant limits
\[\lim_{t\to\infty}\frac1t\int_0^tX_v(u)\de u,\qquad \lim_{t\to\infty}\frac1t\int_0^tX_{v}(u)X_{v'}(u)\de u\,,\qquad v,v'\in\mc V\]
coincide with the expected stationary beliefs and belief crossproducts, i.e., $\E[X_v]$ and
$\E[X_{v}X_{v'}]$, respectively, independently of the distribution of initial regular agents' beliefs.
\end{remark}

\begin{remark}\label{remarkx*} As a consequence of Theorem \ref{theoasymptoticmeanvariance}, one gets that, if $\mc X_a=\{x_*\}$, then $X_a=x_*$, and, by Corollary \ref{coroergodic}, $X_a(t)$ converges to $x_*$ with probability one. Hence, in particular, $\E[X_a]=x_*$, and $\Var[X_a]=0$. This can be thought of as a sort of complement to Theorem \ref{coroNOconvP1}.
\end{remark}

\section{Explicit computations of stationary expected beliefs and variances}\label{sect:reversible}
We present now a few examples of explicit computations of the
stationary expected beliefs and variances for social
networks obtained using the construction in Example
\ref{examplestandardRW}, starting from a simple undirected graph $\mc G=(\mc V,\mc E)$. Recall that, in this case, $Q_{av}=\theta/d_a$ for all $a\in\mc A$, and $v\in\mc V$ such that $\{a,v\}\in\mc E$. It then follows from Theorem \ref{theoasymptoticmeanvariance} that the expected stationary beliefs can be characterized as the unique vectors in $\R^{\mc V}$ satisfying
\be\label{harmonic3}\E[X_a]=\frac1{d_a}\sum_{v:\{v,a\}\in\mc E}\E[X_v]\,,\qquad\E[X_s]=x_s\,,\qquad\forall a\in\mc A\,,\ \forall s\in\mc S\,.\ee
Moreover, in the special case when $\theta=1$, the second moments of the stationary beliefs are the unique solutions of
\be\label{harmonic4}\E[X_a^2]=\frac1{d_a}\sum_{v:\{v,a\}\in\mc E}\E[X_v^2]\,,\qquad\E[X_s^2]=x^2_s\,,\qquad\forall a\in\mc A\,,\ \forall s\in\mc S\,.\ee
\begin{figure}
\begin{center}
\subfigure{\includegraphics[height=4.5cm,width=6.5cm]{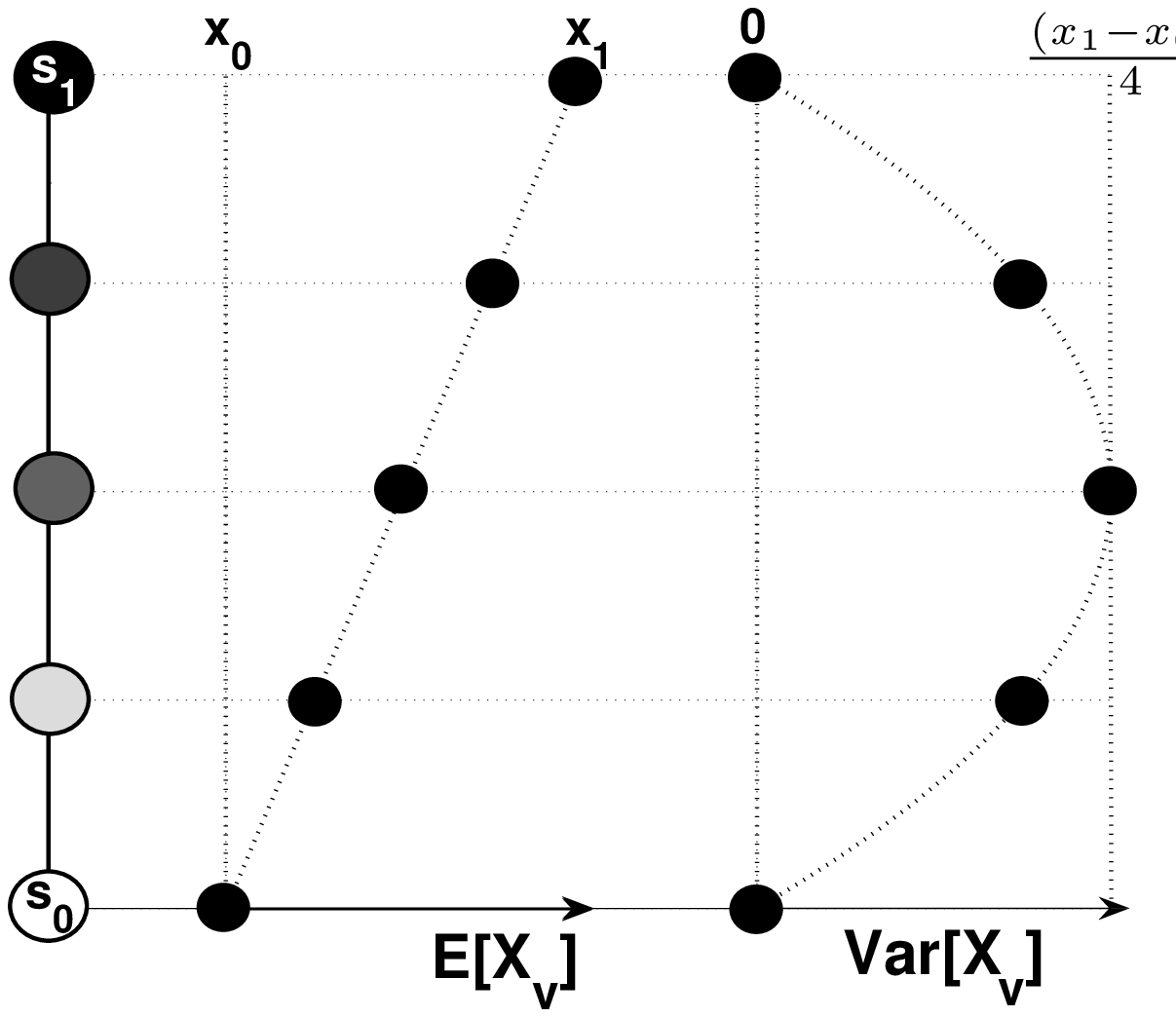}}\hspace{1cm}
\subfigure{\includegraphics[height=4.5cm,width=5.5cm]{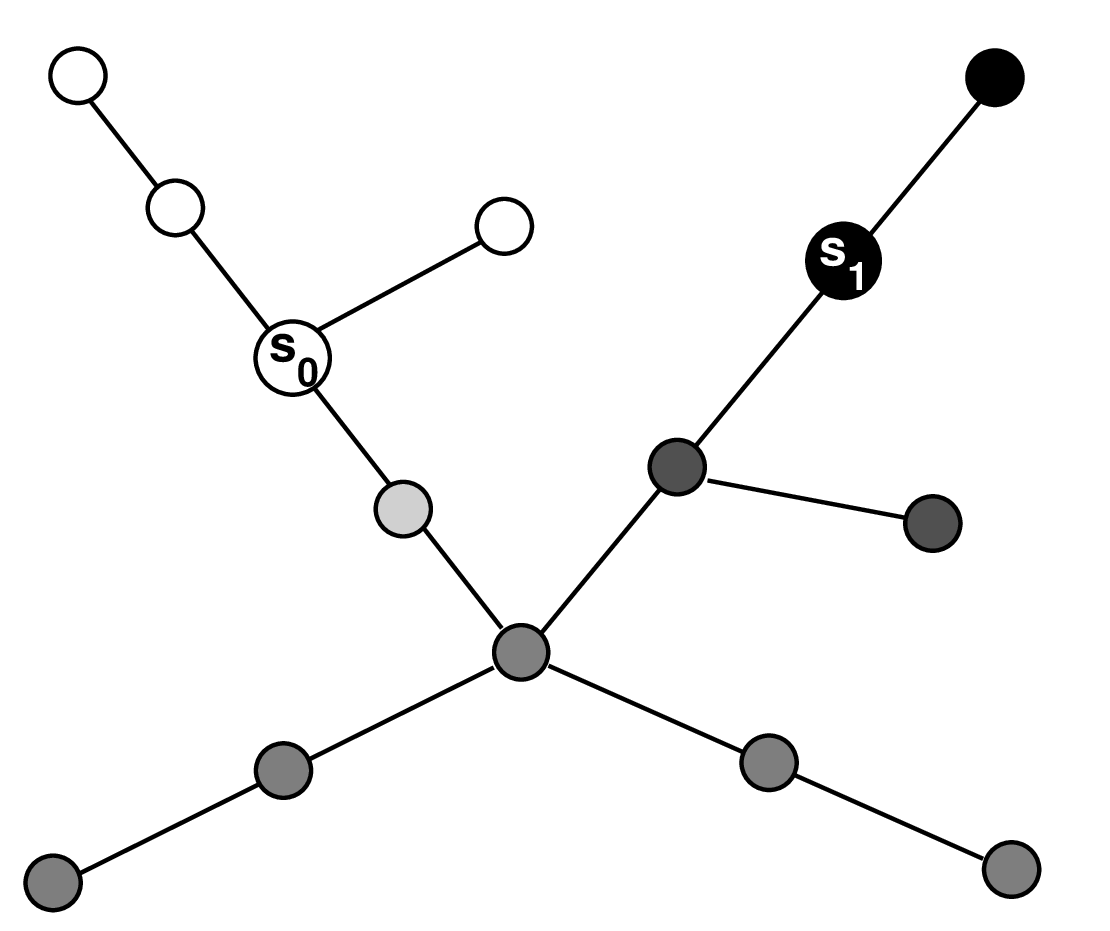}}
\caption{\label{linefig}In the left-most figure, expected stationary beliefs and variances (the latter valid for the special case when $\theta=1$) in a social network with a line graph topology with $n=5$, and stubborn agents positioned in the two extremities. The expected stationary beliefs are linear interpolations of the two stubborn agents' beliefs, while their variances follow a parabolic profile with maximum in the central agent, and zero variance for the two stubborn agents $s_0$, and $s_1$.
In the right-most figure, expected stationary beliefs in a social network with a tree-like topology, represented by different levels of gray. The solution is obtained by linearly interpolating between the two stubborn agents' beliefs, $x_{0}$ (white), and $x_{1}$ (black),  on the vertices lying on the path between $s_0$ and $s_1$, and then extended by putting it constant on each of the connected components of the subgraph obtained by removing the links of such path.}
\end{center}
\end{figure}

\begin{example}(\textbf{Tree})\label{exampletree}
Let us consider the case when $\mc G=(\mc V,\mc E)$ is a tree and the stubborn agent set $\mc S$ consists of only two elements, $s_0$ and $s_1$, with beliefs $x_0$, and $x_1$, respectively.
%
For $v,w\in\mc V$, let $d(v,w)$ denote their distance, i.e., the length of the shortest path connecting them in $\mc G$.
Let $m:=d(s_0,s_1)$, and $\mc W=\{s_0=w_0,w_1,\ldots,w_{m-1},w_m=s_1\}$, where $\{w_{i-1},w_i\}\in\mc E$ for all $1\le i\le m$, be the unique path connecting $s_0$ to $s_1$ in $\mc G$. 
Then, we can partition the rest of the node set as $\mc V\setminus\mc W=\bigcup_{0\le i\le m}\mc V_w$, where $\mc V_i$ is the set of nodes $v\in\mc V\setminus\mc W$ such that the unique paths from $v$ to $s_0$ and $s_1$ both pass through $w_i$. Since the set of neighbors of every $v\in\mc V_i$ is contained in $\mc V_i\cup\{w_i\}$, (\ref{harmonic3}) implies that 
\be\label{eq:tree1}\E[X_v]=\E[X_{w_i}]\,,\qquad\forall v\in\mc V_i\,,\ 1\le i\le m\,.\ee
Hence, one is left with determining the values of $\E[X_{w_i}]$, for $0\le i\le m$. 
Observe that clearly \be\label{eq:tree2}\E[X_{w_0}]=x_{s_0}\,,\qquad\E[X_{w_m}]=x_{s_1}\,.\ee 
On the other hand, for all $0<i<m$, the neighborhood of $w_i$ consists of $w_{i-1}$, $w_{i+1}$, and possibly some elements of $\mc V_i$. Then, (\ref{harmonic3}) and (\ref{eq:tree1}) imply that  
\be\label{eq:tree3}\E[X_{w_i}]=\frac12\E[X_{w_{i-1}}]+\frac12\E[X_{w_{i+1}}]\,,\qquad 0<i<m\,.\ee
Now, observe that, since 
$$\frac imx_{s_1}+\frac{m-i}mx_{s_0}=\frac12(i-1)x_{s_1}+\frac12(m-i+1)x_{s_0}+\frac12(i+1)x_{s_1}+\frac12(m-i-1)x_{s_1}\,,$$
then the unique solution of (\ref{eq:tree2}) and (\ref{eq:tree3}) is given by
\be\label{eq:tree4}\E[X_{w_i}]=\frac imx_{s_1}+\frac{m-i}mx_{s_0}\,,\qquad 0\le i\le m\,.\ee
Upon observing that $d(v,s_j)=d(w_i,s_j)$ for all $w\in\mc V_i$, $0\le i\le m$, and $j=0,1$, we may rewrite (\ref{eq:tree1}) and (\ref{eq:tree4}) as 
\be\label{eq:trees}\E[X_v]=\l\{\ba{lcl}
x_i&\se&|\mc S_{v}|=\{x_i\}\,,\,\,i=0,1\,,\\[10pt]
\ds\frac{d(v,s_0)x_1+d(v,s_1)x_0}{d(v,s_0)+d(v,s_1)}\,,&\se&|\mc S_v|=2\,.\ea\r.\ee
In other words, the stationary  expected beliefs are linear interpolations of the beliefs of the stubborn agents. 
A totally analogous argument shows that, if the confidence parameter satisfies $\theta=1$, then (\ref{harmonic4}) is satisfied by
$$\E[X_a^2]=\frac{d(a,s_0)x_1^2+d(a,s_1)x_0^2}{d(a,s_0)+d(a,s_1)}\,,$$
so that the stationary variance of agent $a$'s belief is given by
$$\Var[X_a]=\E[X_a^2]-\E[X_a]^2=\frac{d(a,s_0)d(a,s_1)}{(d(a,s_0)+d(a,s_1))^2}\l(x_0-x_1\r)^2\,.$$
The two equations above show that the belief of each regular agent keeps on fluctuating ergodically around a value which depends on the relative distance of the agent from the two stubborn agents. The amplitude of such fluctuations is maximal for central nodes, i.e., those which are homogeneously distant from both stubborn agents. This can be given the intuitive explanation that, the closer a regular agent is to a stubborn agent $s$ with respect to the other stubborn agent $s'$, the more frequent her, possibly indirect, interactions are with agent $s$ and the less frequent her interactions are with $s'$, and hence the stronger the influence is from $s$ rather than from $s'$. Moreover, the more equidistant a regular agent $a$ is from $s_0$ and $s_1$, the higher the uncertainty is on whether, in the recent past, agent $a$ has been influenced by either $s_0$, or $s_1$.

\begin{figure}\begin{center}
\subfigure{\includegraphics[height=4.5cm,width=5cm]{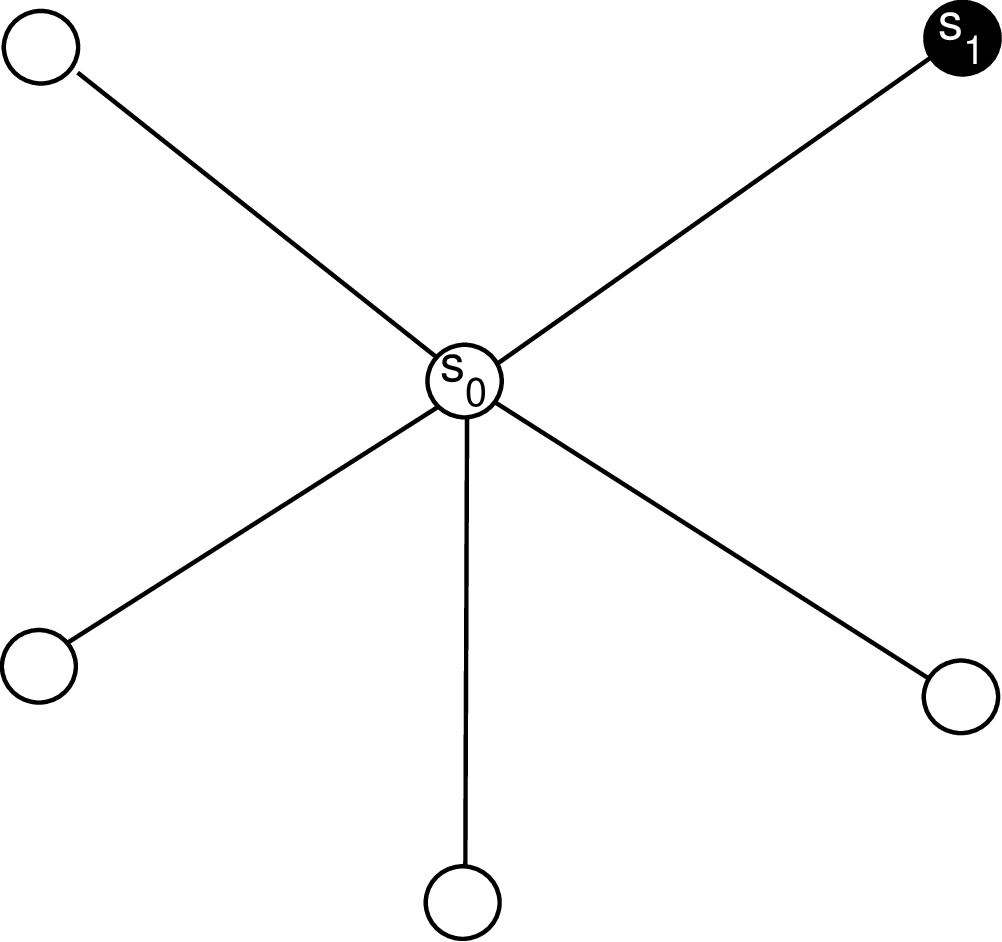}}\hspace{1.5cm}
\subfigure{\includegraphics[height=4.5cm,width=5cm]{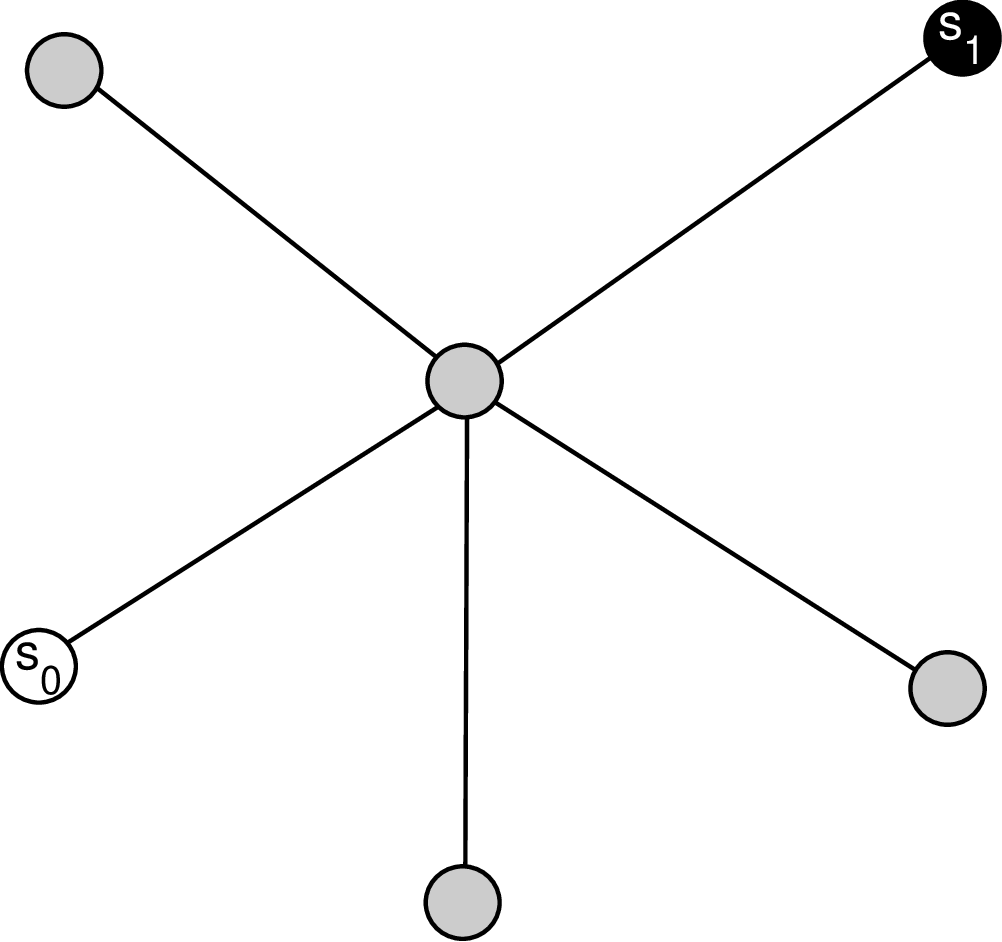}}
\caption{\label{starfig}Two social network with a special case of tree-like topology, known as star graph, and two stubborn agents. In social network depicted in left-most figure one of the stubborn agents, $s_0$, occupies the center, while the other one, $s_1$, occupies one of the leaves. There, all regular agents' stationary beliefs coincide with the belief $x_{0}$ of $s_0$, represented in white. In social network depicted in right-most figure, none of the stubborn agents, occupy the center. There, all regular agents' stationary beliefs coincide with the arithmetic average (represented in gray) of $x_{0}$ (white), and $x_{1}$ (black).}
\end{center}
\end{figure}
On its left-hand side, Figure \ref{linefig} reports the expected stationary beliefs and their variances for a social network with population size $n=5$, line (a special case of tree-like) topology: the two stubborn agents are positioned in the extremities, and plotted in white, and black, respectively, while regular agents are plotted in different shades of gray corresponding to their relative distance from the extremities, and hence to their expected stationary belief. In the right-hand side of Figure \ref{linefig}, a more complex tree-like topology is reported, again with two stubborn agents colored in white, and black respectively, and with regular agents colored by different shades of gray corresponding to their relative vicinity to the two stubborn agents. Figure \ref{starfig} reports two social networks with star topology (another special case of tree). In both cases there are two stubborn agents, colored in white, and black, respectively. In the left-most picture, the white stubborn agent occupies the center, so that all the rest of the population will eventually adopt his belief, and is therefore colored in white. In the right-most picture, none of the stubborn agents occupies the center, and hence all the regular agents, hence colored in gray, are equally influenced by the two stubborn agents.
\end{example}\medskip

\begin{example}(\textbf{Barbell})\label{examplebarbell}
For even $n\ge6$, consider a barbell-like topology consisting of two complete graphs with vertex sets $\mc V_0$, and $\mc V_1$, both of size $n/2$, and an extra link $\{a_0,a_1\}$ with $a_0\in\mc A_0$, and $a_1\in\mc A_1$ (see Figure \ref{figbarbell}). Let $\mc S=\{s_0,s_1\}$ with $s_0\ne a_0\in\mc V_0$ and $s_1\ne a_1\in\mc V_1$. Then, (\ref{harmonic3}) is satisfied by
$$\E[X_a]=\l\{\ba{lcl}
\ds\frac{4}{n+8}x_{s_0}+\frac{n+4}{n+8}x_{s_1}&\se&a=a_1\\[10pt]
\ds\frac{n+4}{n+8}x_{s_0}+\frac{4}{n+8}x_{s_1}&\se&a=a_0\\[10pt]
\ds\frac{2}{n+8}x_{s_0}+\frac{n+6}{n+8}x_{s_1}&\se&a\in\mc A_1\setminus\{a_1\}\\[10pt]
\ds\frac{n+6}{n+8}x_{s_0}+\frac{2}{n+8}x_{s_1}&\se&a\in\mc
A_0\setminus\{a_0\}\,. \ea\r.$$ In particular, observe that, as $n$
grows large, $\E[X_a]$ converges to $x_{s_0}$ for all $a\in\mc A_0$,
and $\E[X_a]$ converges to $x_{s_1}$ for all $a\in\mc A_1$. Hence,
the network polarizes around the opinions of the two stubborn
agents.
\begin{figure}\begin{center}
\includegraphics[height=6.4cm,width=12cm]{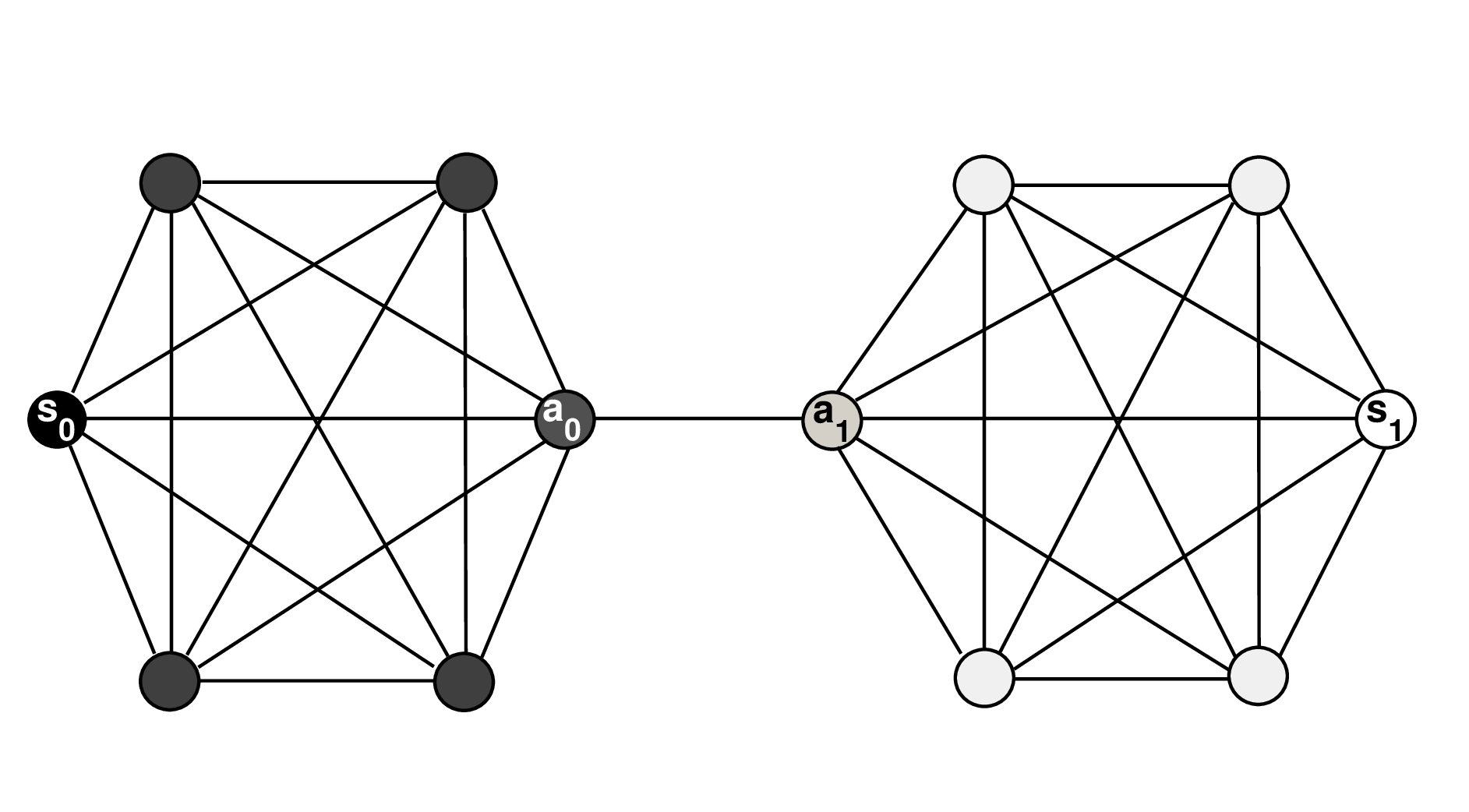}
\caption{\label{figbarbell}A social network with population size $n=12$, a barbell-like topology, and two stubborn agents. In each of the two halves of the graph the expected average beliefs concentrate around the beliefs of the stubborn agent in the respective half.}
\end{center}
\end{figure}
\end{example}\medskip

\begin{example}(\textbf{Abelian Cayley graph})\label{exampletorus}
Let us denote by $\Z_m$ the integers modulo $m$. Put $\mc V=\Z_m^d$, and
let $\Theta\subseteq \mc V\setminus\{0\}$ be a subset generating
$\mc V$ and such that if $x\in\Theta$, then also $-x\in \Theta$.
The Abelian Cayley graph associated with $\Theta$ is the graph
$\mc G=(\mc V,\mc E)$ where $\{v,w\}\in \mc E$ iff $v-w\in\Theta$.
Notice that Abelian Cayley graphs are always undirected and
regular, with $d_v=|\Theta|$ for any $v\in \mc V$. Denote by $e_i\in \mc
V$ the vector of all $0$'s but the $i$-th component equal to $1$.
If $\Theta=\{\pm e_1,\dots ,\pm e_d\}$, the corresponding $\mc G$
is  the classical $d$-dimensional torus of size $n=m^d$. In particular, for $d=1$, this is a cycle, while, for $d=2$, this is the torus (see Figure \ref{figcycletorus}).


Let the stubborn agent set consist of only two elements: $\mc S:=\{ s_0, s_1\}$.
Then the following formula holds (see \cite[Ch.~2, Corollary 10]{AldousFillbook}):
\be\label{meantime}\gamma^v_{s_0}=\P_v(T_{s_1}<T_{s_0})=\frac{E_{vs_0}-
E_{vs_1}+ E_{s_1 s_0}}{ E_{ s_0 s_1}+  E_{s_1s_0}}\ee where
$E_{vw}:=\E_v[T_{\{w\}}]$ denotes the expected time it takes to a Markov chain
started at $v$ to hit node $w$ for the first time. On the other hand,
average hitting times $E_{vw}$ can be expressed in terms of the Green
function of the graph, which is defined as the unique matrix $Z\in
\R^{\mc V\times \mc V}$ such that
$$Z\1=0\,,\quad (I-P)Z=I-n^{-1}\1\1^{\rm T}\,,$$
where $\1$ stands for the all-$1$ vector. The relation with the
hitting times is given by 
\be\label{green-mean}E_{vw}=n^{-1}(Z_{ww}-Z_{vw})\,.\ee
(See, e.g., \cite[Ch.~2, Lemma 12]{AldousFillbook}.)

Let $P$ be the stochastic matrix corresponding to the simple
random walk on $\mc G$. It is a standard fact that $P$ is
irreducible and its unique invariant probability is the uniform
one. (See, e.g., \cite[Chapter 15]{Behrends}.) Morever, there is an orthonormal basis of eigenvectors for $P$ good
for every $\Theta$: if $  l=(l_1,\dots l_d)\in\mc V$, define
$\Upsilon_{{  l}}\in \R^\mc V$ by
$$\Upsilon_{{  l}}({  k})=m^{-d/2}\exp\left(\frac{2\pi
i}{m}{  l}\cdot{  k}\right)\,,\quad {  k}=(k_1,\dots
,k_d)\in \mc V\,,$$ (where ${  l}\cdot{  k}:=\sum_il_ik_i$).
The corresponding eigenvalues can be expressed as follows
$$\lambda_{{  l}}=\frac{1}{|\Theta^+|}\sum\limits_{{  k}\in\Theta^+}\cos\left(\frac{2\pi}{m}{  l}\cdot {  k}\right)$$ where $\Theta^+$
 is any subset of $\Theta$ such that for all $x\in\Theta$,
 $|\{x,-x\}\cap\Theta^+|=1$. Hence,
\be\label{GreenCayley}Z_{vw}=m^{-d}\sum_{{  l}\neq
0}\frac{\exp\left(\frac{2\pi i}{m} {  l}\cdot({  v}-{
w})\right)}{1-\frac{1}{|\Theta^+|}\sum\limits_{{
k}\in\Theta^+}\cos\left(\frac{2\pi }{m}{  l}\cdot{
k}\right)}\,,\qquad v,w\in\mc V\,.\ee

From (\ref{meantime}), (\ref{green-mean}), and the fact that
$E_{s_0s_1}=E_{s_1s_0}$ by symmetry, one obtains \be\label{solutionCayley} \gamma^a_{s_1}=\frac{1}{2}+\ds\frac{m^{-d}\ds\sum_{{
l}\neq 0}\ds\frac{\exp\left(\frac{2\pi i}{m} {  l}\cdot({
a}-{  s_1})\right)-\exp\left(\frac{2\pi i}{m} {  l}\cdot({
a}-{  s_0})\right)}{1-\frac{1}{|\Theta^+|}\sum\limits_{{
k}\in\Theta^+}\cos\left(\frac{2\pi }{m}{  l}\cdot{
k}\right)}}{2m^{-d}\ds\sum_{{  l}\neq
0}\ds\frac{1-\cos\left(\frac{2\pi }{m}{  l}\cdot({  s_0}-{
s_1})\right)}{1-\frac{1}{|\Theta^+|}\sum\limits_{{
k}\in\Theta^+}\cos\left(\frac{2\pi }{m}{  l}\cdot{
k}\right)}}\,,\qquad a\in\mc A\,.\ee
The expected stationary beliefs can then be computed using Theorem \ref{theoasymptoticmeanvariance} and (\ref{solutionCayley}).

\begin{figure}\begin{center}
\subfigure{\includegraphics[height=5cm,width=5cm]{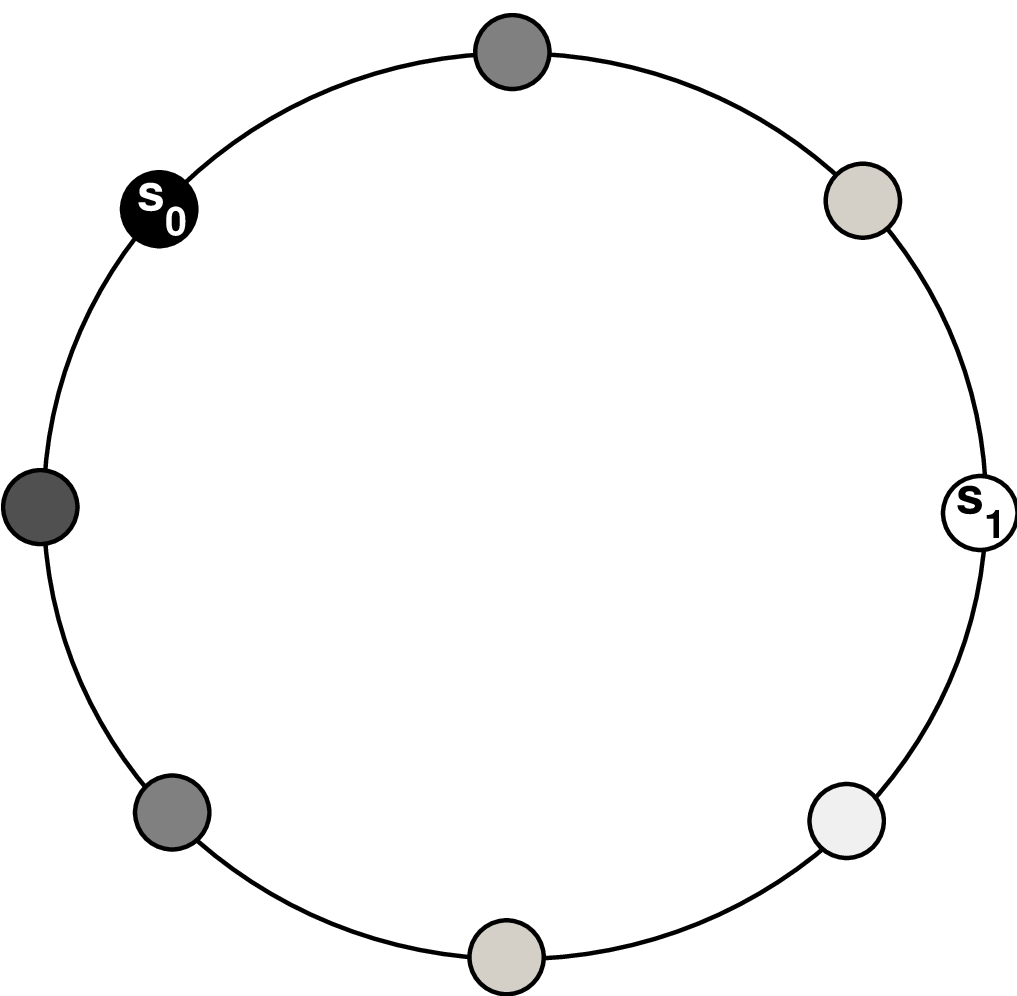}}\hspace{0.5cm}
\subfigure{\includegraphics[height=5cm,width=6.5cm]{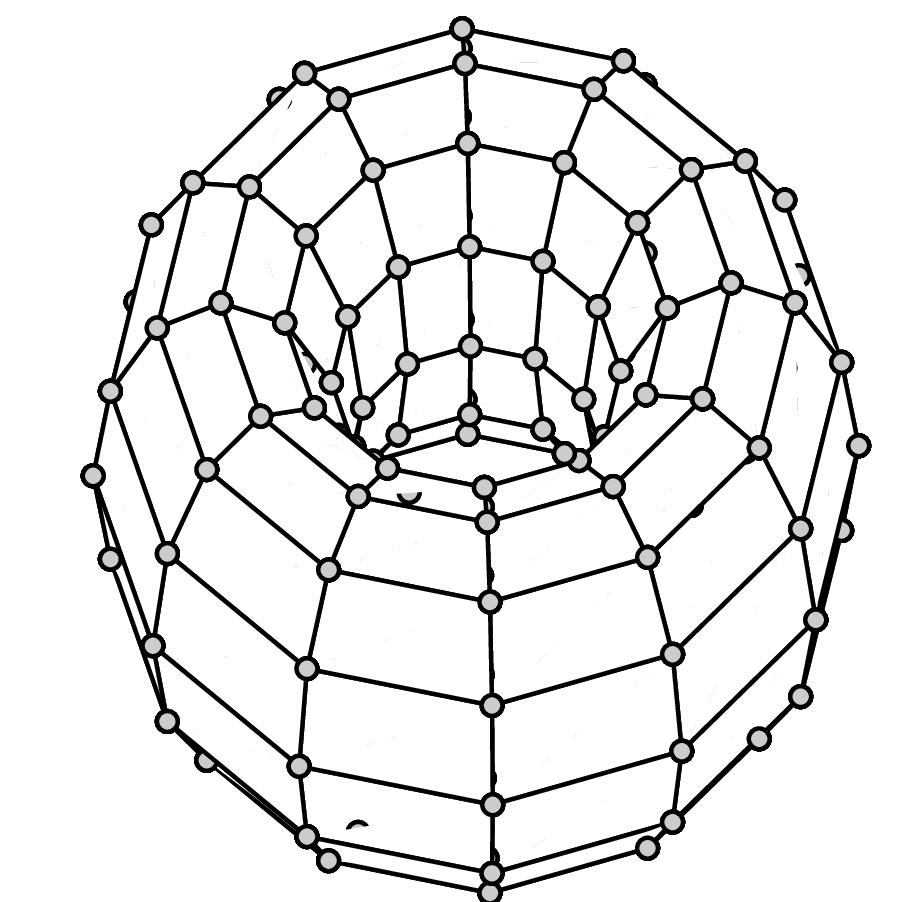}}
\caption{\label{figcycletorus}Two social networks with cycle and $2$-dimensional toroidal topology, respectively.}
\end{center}
\end{figure}

\end{example}

\section{Homogeneous influence in highly fluid social networks}\label{sectestimates}
In this section, we present estimates for the expected stationary beliefs and belief variances as a function of the underlying social network. First, we will introduce the notion of fluidity of a social network, a quantity which depends only on the geometry of the network and on the size of the stubborn agent set. Then, we will prove that, when the social network is highly fluid, the influence of the stubborn agents on the rest of the society is homogeneous, meaning that most of the regular agents have approximately the same stationary expected belief and (in the special case when $\theta_e=1$ for all $e\in\ora{\mc E}$)  belief variance. 

\subsection{Network fluidity and homogeneous influence}
Recall that Theorem \ref{theoasymptoticmeanvariance} allows one to express the first moment of the stationary beliefs in terms of the hitting probability distributions $\gamma^v$ on the stubborn agent set $\mc S$ of the continuous-time Markov chain $V(t)$ with state space $\mc V$ and transition rate matrix $Q$. It is a simple but key observation that such hitting probabilities only depend on the restriction of $Q$ to $\mc A\times\mc V$ (the other rows affecting only the behavior of $V(t)$ after hitting $\mc S$), and they do not change if any row of $Q$ is multiplied by some positive scalar (this multiplication having the only affect of speeding up or slowing down $V(t)$ without changing the jump probabilities). Formally, we have the following
\begin{lemma}\label{lemma:trivial}
Let $P\in\R^{\mc V\times\mc V}$ be stochastic and such that \be\label{Pdef} P_{av}=\alpha_aQ_{av}\,,\qquad \forall a\in\mc A\,,v\in\mc V\,,a\ne v\,,\ee 
for some $\alpha_a>0$. Let $W(k)$, for $k=0,1,\ldots$, be a discrete-time Markov chain with transition probability matrix $P$. Then, the hitting probability distributions of the continuous-time Markov chain $V(t)$ with transition rate matrix $Q$ coincide with those of $W(k)$, i.e., 
$$\gamma^v_s=\P_v(W(U_{\mc S})=s)\,,\qquad \forall v\in\mc V\,,\, s\in\mc S\,,$$
where $U_{\mc S}:=\min\{k\ge0:\,W(k)\in\mc S\}$. 
\end{lemma}

Lemma \ref{lemma:trivial} allows one to consider the discrete-time chain $W(k)$ with any transition probability matrix $P$ satisfying (\ref{Pdef}), in order to compute the hitting probability distributions $\gamma^v$. In fact, it is convenient to consider stochastic matrices $P\in\R^{\mc V\times\mc V}$ that, in addition to satisfying (\ref{Pdef}), are irreducible and aperiodic. Let us denote the set of all such matrices by $\mc P$. Observe that, provided that Assumption \ref{assumptionconnected} holds, the set $\mc P$ is non-empty, since it can be easily checked to contain, e.g., the matrix $P\in\R^{\mc V\times\mc V}$ with entries $P_{sv}=1/n$, $P_{aa}=0$, $P_{aw}=-Q_{aw}/Q_{aa}$ for all $s\in\mc S$, $v\in\mc V$, $a\in\mc A$, and $w\in\mc V\setminus\{a\}$. 
Observe that, for every $P\in\mc P$, irreducibility implies the existence of a unique invariant probability measure, and, together with aperiodicity, convergence of the time-$k$ distribution 
\be\label{pkdef}p^v(k)=\{p^v_w(k):\,w\in\mc V\}\,,\qquad p^v_w(k):=\P_v(W(k)=w)\,,\ee
irrespectively of the initial state $v\in\mc V$. We introduce the following notation. 
\begin{definition}
Given a social network satisfying Assumption \ref{assumptionconnected}, and $P\in\mc P$, let $\pi=P'\pi$ be the unique invariant probability measure of $P$. 
Let $$\pi(\mc S):=\sum_s\pi_s\,,\qquad\pi_*:=\min_{v}\pi_v\,,$$
be the size of the stubborn agents set and, respectively, the minimum weight of an agent, as measured by $\pi$. Moreover, let 
\begin{equation}\tau
:=\inf\l\{k\ge0:\,\max_{v}\l|\l|p^v(k)-\pi\r|\r|_{TV}\le
\frac1{2e}\r\}\,,\label{mixingtime}\end{equation}
where $p^v(k)$ are as in (\ref{pkdef}), denote the (variational distance) mixing time of the discrete-time Markov chain $W(k)$ with state space $\mc V$ and transition probability matrix $P$.
\end{definition}\medskip

For the canonical construction of a social network introduced in Example \ref{examplestandardRW}, the quantities above have a more explicit characterization which allows for a geometric interpretation. 
\begin{example}\label{examplestandardRWcont} Let us consider the
canonical construction of a social network from a given connected
multigraph $\mc G=(\mc V,\mc E)$, outlined in Example
\ref{examplestandardRW}. Define $\tilde P\in\R^{\mc V\times\mc V}$ by putting $\tilde P_{vw}=\kappa_{v,w}/d_v$, where $\kappa_{v,w}$ is the multiplicity of the link $\{v,w\}$ in $\mc E$, and $d_v=\sum_w\kappa_{v,w}$ is the degree of node $v$ in $\mc G$. Then, put $P=(I+\tilde P)/2$, where $I$ stands for the identity matrix on $\mc V$.  
In fact, $P$ defined as above is known in the literature as the transition matrix of the simple lazy random walk on $\mc G$ \cite[page 9]{LevinPeresWilmer}. 
Observe that (\ref{Pdef}) is satisfied, connectedness of $\mc G$ implies irreducibility of $\tilde P$, and hence of $P$, while aperiodicity of $P$ (not necessarily of $\tilde P$) is immediate. 
Hence, $P\in\mc P$. Moreover, the invariant measure $\pi$ (of both $P$ and $\tilde P$) is given by
$$\pi_v=d_v/(n\ov d)\,,\qquad\forall v\in\mc V\,,$$ 
where  $\ov d:=n^{-1}\sum_v d_v$ is the
average degree of $\mc G$. Observe that, in this construction,
\be\label{piS}\pi(\mc S)=\l(\sum\nolimits_{v}d_v\r)^{-1}\sum\nolimits_{s}d_s\ee
is the fraction of the total degree of the stubborn agents set $\mc S$ and the total degree of the whole agent set $\mc V$ in $\mc G$.
Moreover, the mixing time can be bounded in terms of the \emph{conductance} of $\mc G$, defined as 
\be\label{eq:conductance}\Phi:=\min\l\{\phi(\mc W):\,\mc W\subseteq\mc V\,,\ 0<\sum\nolimits_{w}d_{w}\le n\ov d/2\r\}\,,\ee
where 
$$\phi(\mc W):=\frac{\sum_{w\in\mc W}\sum_{v\in\mc V\setminus\mc W}\1_{\mc E}(\{v,w\})}{\sum_{w\in\mc W}d_{w}}$$
is the bottleneck ratio of the set $\mc W\subseteq\mc V$, i.e., the ratio between the number of links connecting $\mc W$ to the rest of the node set and its total degree. 
In particular, standard results imply that 
\be\label{mixing-conductance}
\frac1{4\Phi}\le\tau\le\frac2{\Phi^2}\log\frac{2e}{\pi_*}\,.\ee
(See, e.g., \cite[Theorem 7.3]{LevinPeresWilmer} for the lower bound, and combine \cite[Theorem 12.3]{LevinPeresWilmer} and \cite[Theorem 6.2.1]{Durrettbook} in order to get the upper bound.)

We conclude this example by observing that different choices of the multigraph $\mc G$ may result in the same social network, hence in particular in the same directed graph $\ora{\mc G}$. For example, $\mc G$ could have links between pairs of nodes both belonging to $\mc S$. While such links are irrelevant for the belief dynamics (they have no correspondence in the directed graph $\ora{\mc G}$), they do affect the stochastic matrix $P$, hence the quantities $\pi(\mc S)$ and $\tau$. In particular, while removing such links from $\mc G$ clearly has the effect of decreasing $\pi(\mc S)$, it also has the potential of increasing the mixing time $\tau$ (since less connected graphs tend to have larger mixing time). Hence, while the stationary belief distribution does not depend on the presence of links connecting pairs of stubborn agents in $\mc G$, the estimates of the stationary beliefs' moments derived in this section could potentially benefit from considering these links. 
\end{example}\medskip

We are now ready to introduce the notion of \emph{fluidity} of a social network. 

\begin{definition}\label{def:highlyfluid}
Let the social network satisfy Assumption \ref{assumptionconnected}. 
For every $P\in\mc P$ let 
\be\label{def:psi}\psi(P):=\frac{n\pi_*}{\tau\pi(\mc S)\log\l(e^2/(\tau\pi(\mc S))\r)}\,.\ee
The \emph{fluidity} of the social network is 
\be\label{fluiditydef}\Psi:=\sup\l\{\psi(P):\,P\in\mc P\r\}\,.\ee
A sequence of social networks (or, more briefly, a
social network) of increasing population size $n$ is {\it highly
fluid} if $\Psi$ diverges as $n$ grows large. \end{definition}\medskip

Our estimates will show that for large-scale highly fluid social networks, the first two moments of the stationary beliefs of most of the regular agents in the population can be approximated by those of a \emph{weighted-mean belief} $Z$, supported in the finite set $\mc X:=\{x_s:\,s\in\mc S\}$, and given by
\be\label{ovgammadef} \P(Z=z)=\sum\nolimits_{s}\ov\gamma_s\1_{\{z\}}(x_s)\,,\qquad z\in\mc X\,,\qquad\ov\gamma_s:=\sum\nolimits_v\pi_v\gamma_s^v\,,\qquad s\in\mc
S\,.\ee We refer to the probability distribution $\{\ov
\gamma_s:\,s\in \mc S\}$ as the {\it stationary stubborn agent
distribution}. Observe that $\ov\gamma_s=\P_{\pi}(W(U_{\mc S})=s)$
coincides with the probability that the Markov chain $W(k)$, started
from the stationary distribution $\pi$, hits the stubborn agent
$s$ before any other stubborn agent $s'\in\mc S$. In fact, as we
will  clarify below, one may interpret $\ov\gamma_s$ as a relative
measure of the influence of the stubborn agent $s$ on the society
compared to the rest of the stubborn agents $s'\in\mc S$.

More precisely, let us denote the expected value and variance of the
weighted-mean belief $Z$ by
\be\label{ovxsigma2def}\E[Z]:=\summ\nolimits_{s}\ov\gamma_sx_s\,,\qquad\sigma^2_Z:=\sum\nolimits_s\ov\gamma_s\l(x_s-\E[Z]\r)^2\,.\ee
Let $\sigma_v^2$ denote the variance of the stationary belief of agent
$v$,
$$\sigma_v^2:= \E[X_v^2] - \E[X_v]^2\,.$$
We also use the notation  $\Delta_*$ for the maximum
difference between stubborn agents' beliefs, i.e.,
\be\label{Delta*def}\Delta_*:=\max\l\{x_s-x_{s'}:\,s,s'\in\mc S\r\}.\ee

The next theorem presents the main result of this section.

\begin{theorem}\label{theoopinionconc}
Let Assumption \ref{assumptionconnected} hold. Then, for all $\eps>0$,
\be\label{expectedvalueconcentration}
\frac1n\l|\l\{v:\,\Big|\E[X_v]-\E[Z]
\Big|>\Delta_*\eps\r\}\r|\le\frac{1}{\eps\Psi}\,.\ee
Furthermore, if the trust parameters satisfy $\theta_{av}=1$ for all
$(a,v)\in\ora{\mc E}$, then
\be\label{varianceconcentration}\frac1n\l|\l\{v:\,\Big|\sigma_v^2-\sigma_Z^2
\Big|>\Delta_*^2\eps\r\}\r|\le\frac{1}{\eps\Psi}\,.\ee
\end{theorem}

\begin{figure}[t]\begin{center}
\subfigure[]{\includegraphics[height=4cm,width=5.2cm]{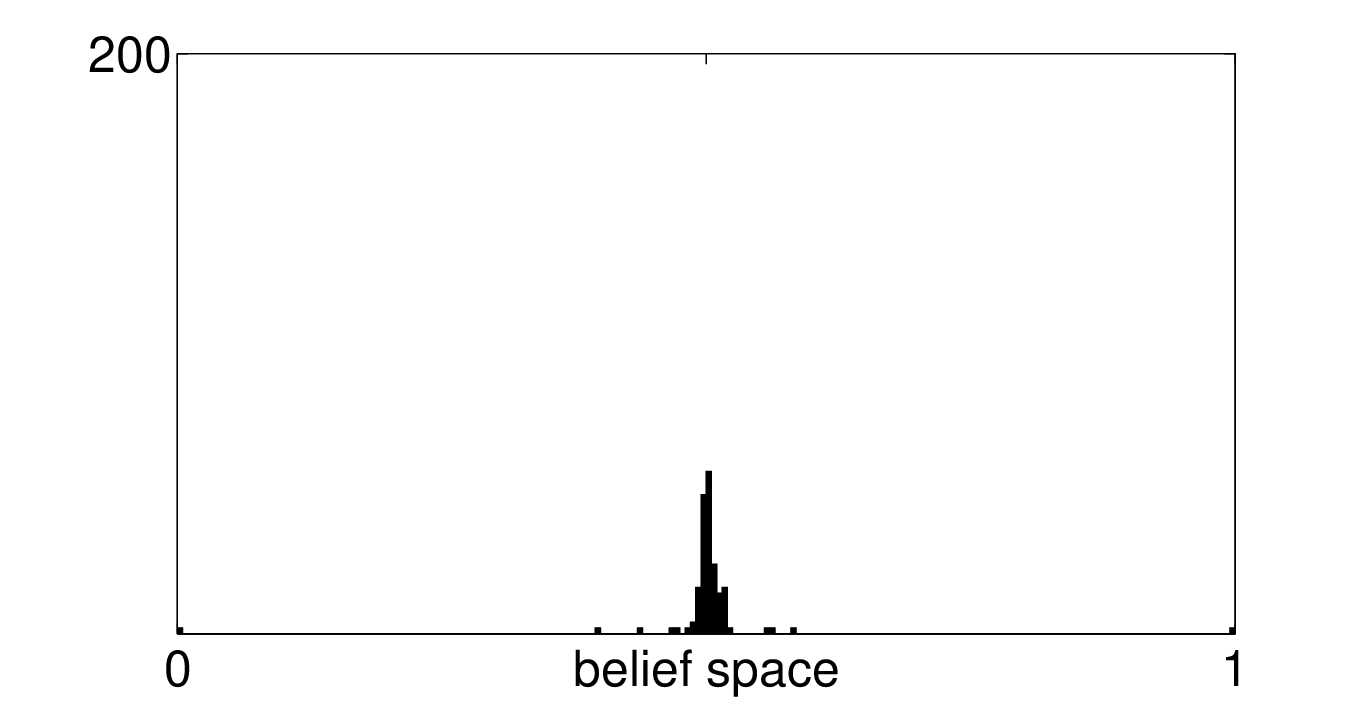}}
\subfigure[]{\includegraphics[height=4cm,width=5.2cm]{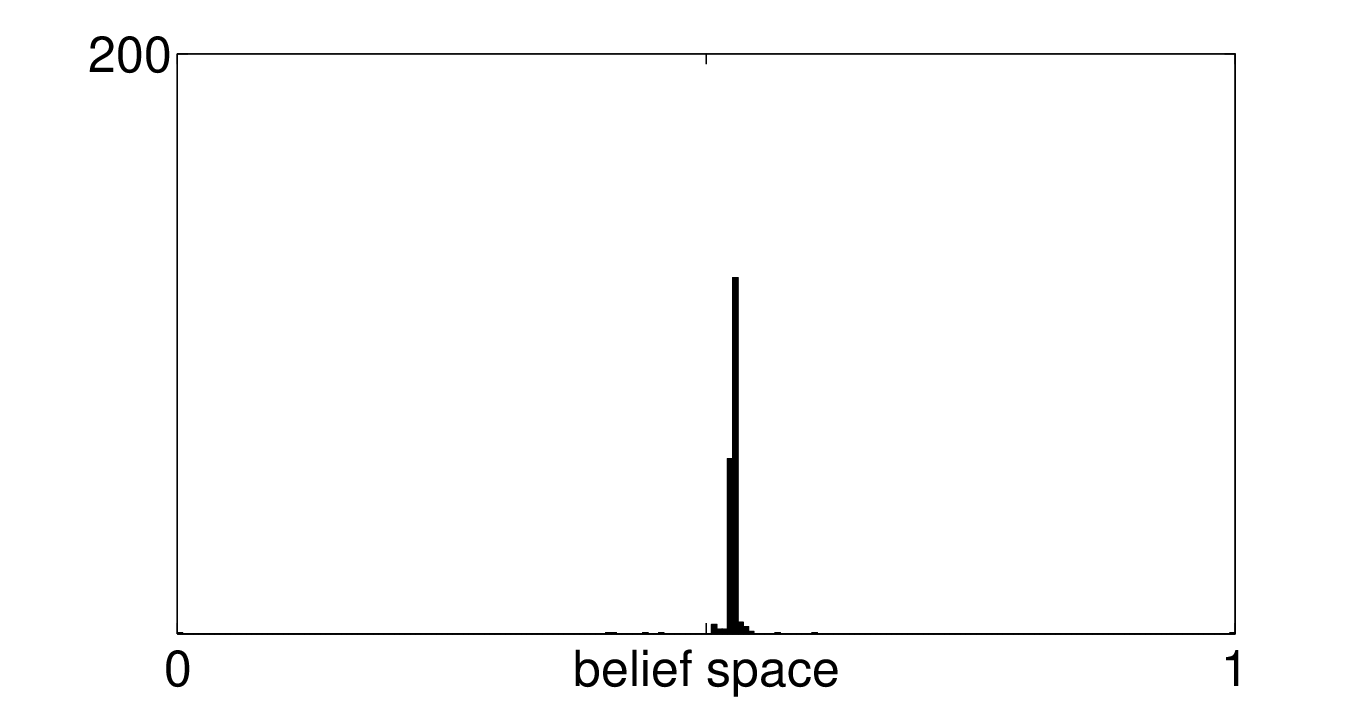}}
\subfigure[]{\includegraphics[height=4cm,width=5.2cm]{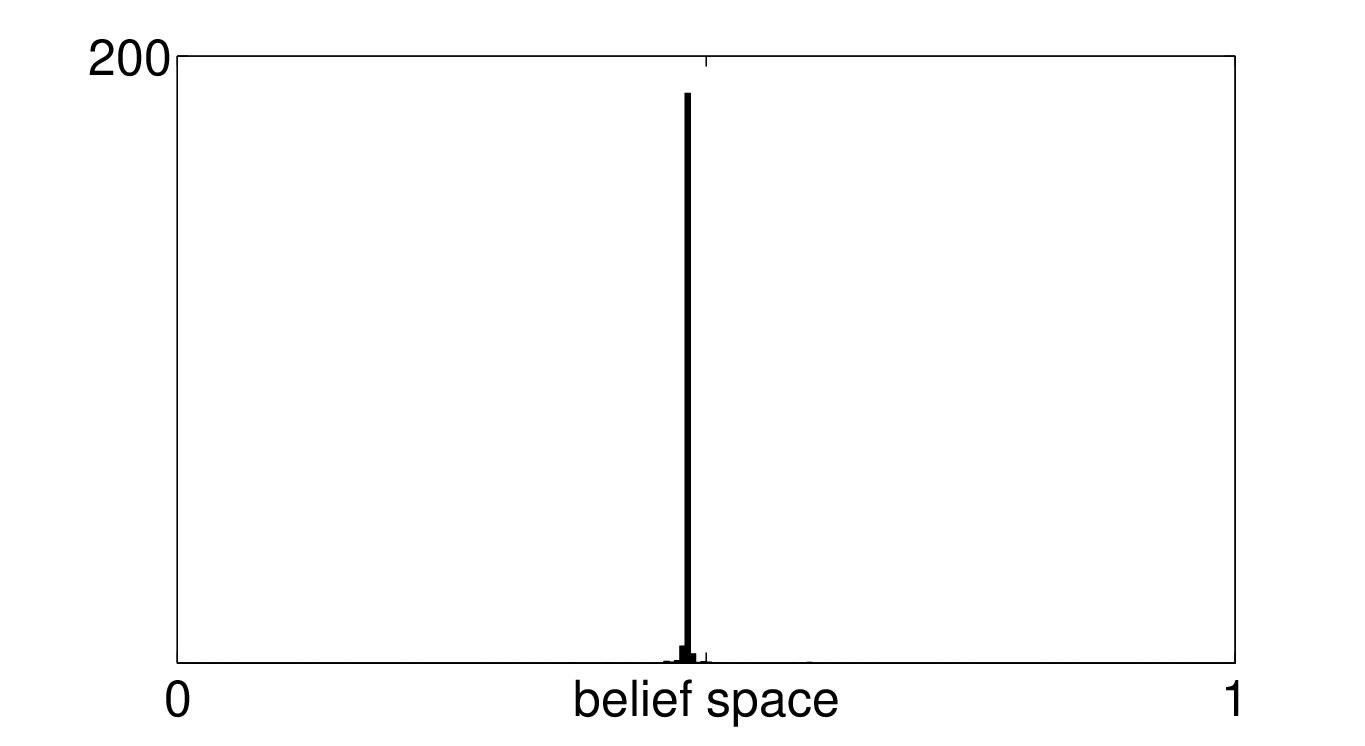}}
\subfigure[]{\includegraphics[height=4cm,width=5.2cm]{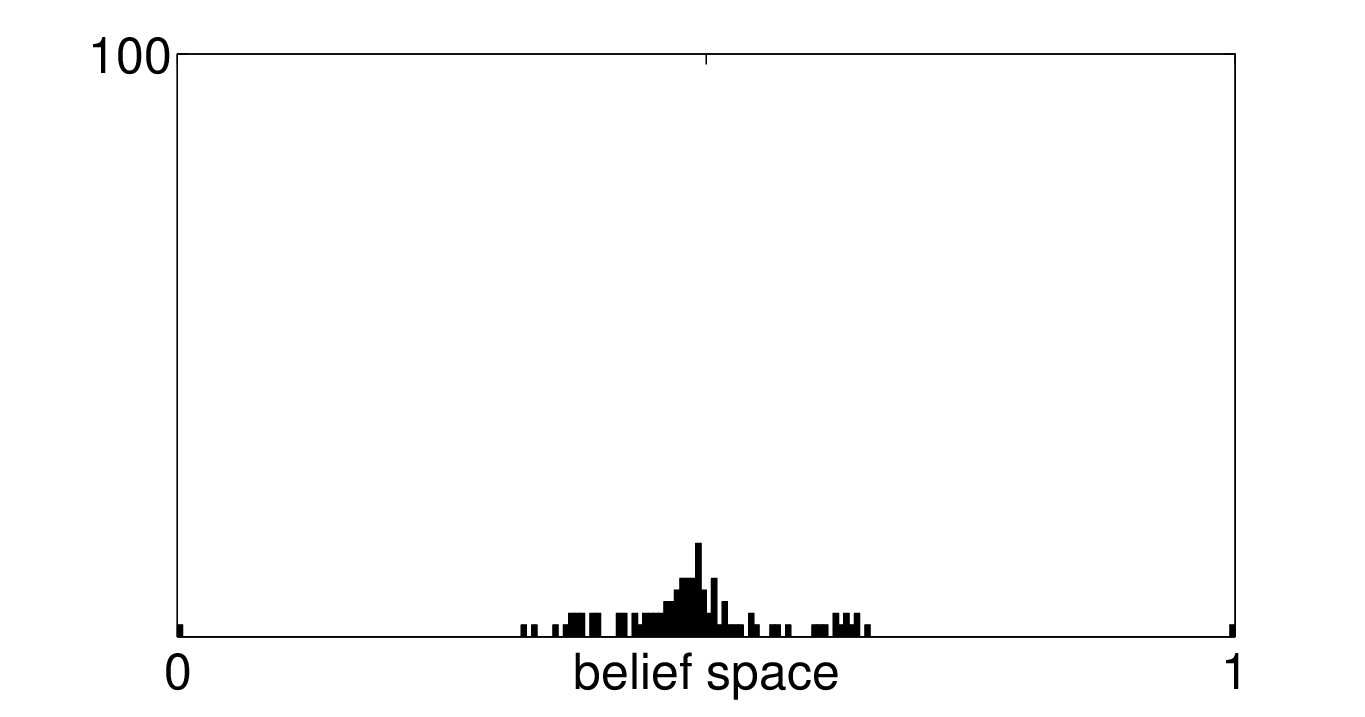}}
\subfigure[]{\includegraphics[height=4cm,width=5.2cm]{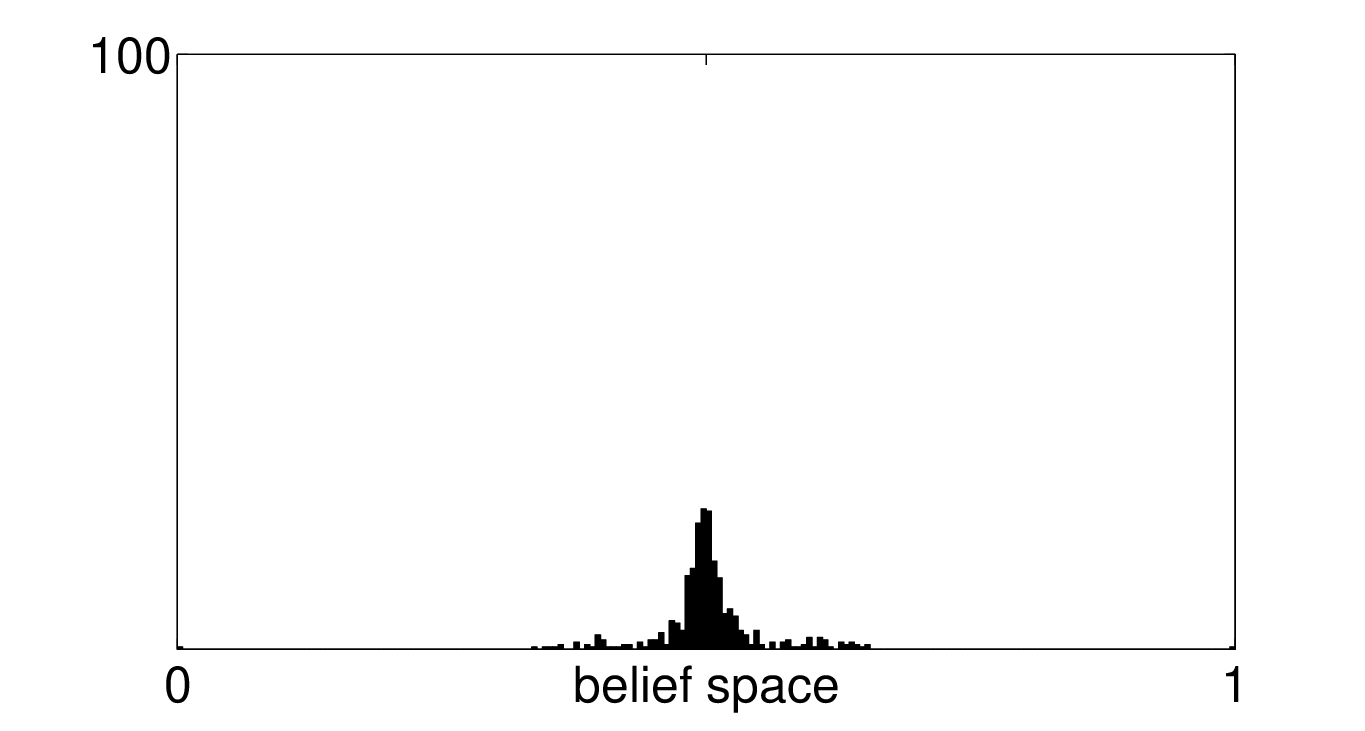}}
\subfigure[]{\includegraphics[height=4cm,width=5.2cm]{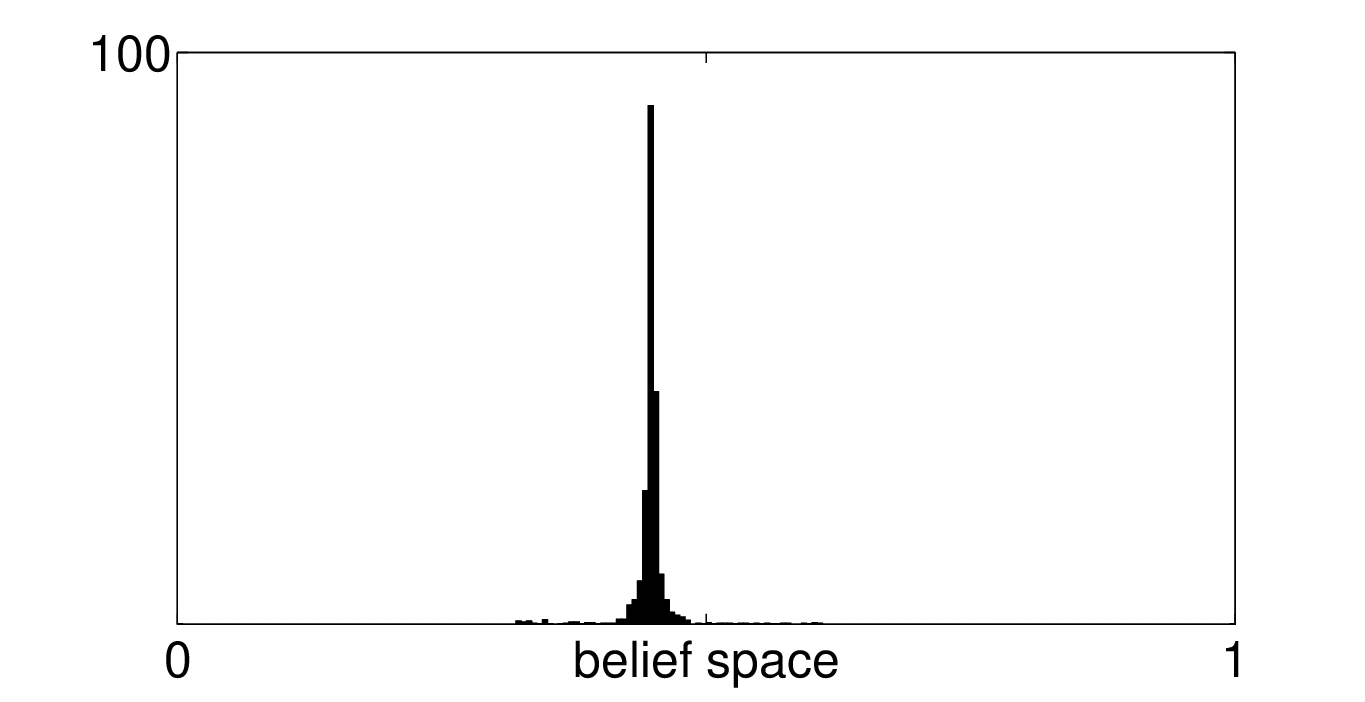}}
\caption{\label{fig:PA}
Homogeneous influence in preferential attachment networks of increasing sizes. The figures show the empirical density of the expected stationary beliefs, for typical realizations of such graphs. The population sizes are $n=100$ in (a) and (d), $n=500$ in (b) and (e), and $n=2000$ in (c) and (f), while $m=4$.  In each case, there are two stubborn agents holding beliefs equal to $0$, and $1$, respectively.  In (a), (b), and (c), the stubborn agents are chosen to coincide with the two latest attached nodes, and therefore tend to have the lowest degree. In contrast, in (d), (e), and (f), the stubborn agents are chosen to be the two initial nodes, and therefore tend to have the highest degrees. As predicted by Theorem \ref{theoopinionconc}, the empirical density of the expected stationary beliefs tends to concentrate around a single value as the population size grows large. The rate at which this concentration occurs is faster in the top three figures, where $\sum_sd_s$ is smaller, and slower in the bottom three figures, where $\sum_sd_s$ is larger. 
}
\end{center}
\end{figure}

This theorem implies that in large-scale highly fluid social networks, as the population size $n$ grows large, the stationary expected beliefs and variances of the regular agents concentrate around fixed values corresponding to the expected weighted-mean belief $\E[Z]$, and, respectively, its variance $\sigma^2_Z$ (see Figures \ref{fig:ER} and \ref{fig:PA}). We refer to this phenomenon as \emph{homogeneous influence} of the stubborn agents on the rest of the society---meaning that their influence on most of the agents in the society is approximately the same. Indeed, it amounts to homogeneous first and second moment of the agents' stationary beliefs. This shows that in highly fluid social networks, most of the regular agents are affected by the stubborn agents in approximately the same way.

Observe that, provided that $n\pi_*$ remains bounded from below by a positive constant, as we will prove to be the case in all the considered examples, a social network is highly fluid when the stationary measure $\pi(\mc S)$ of the stubborn agents set vanishes fast enough to compensate for the possible growth of the mixing time $\tau$, as the network size $n$ grows large. 
Hence, intuitively, Theorem \ref{theoopinionconc} states that, if the set $\mc S$ and the mixing time $\tau$ are both
small enough, then the influence of the stubborn agents will be felt by
most of the regular agents much later then the time it takes them to
influence each other, so that their beliefs' empirical averages and
variances will converge to values very close to each other. Theorem
\ref{theoopinionconc} is proved in Section \ref{secttheoopinionconcproof}. Its proof relies on the
characterization of the expected stationary beliefs and variances in terms of the hitting probabilities $\gamma^v_s$. The definition of highly
fluid network implies that the (expected) time it takes a Markov chain to
hit the set $\mc S$, when started from most of the nodes, is much
larger than the mixing time $\tau$. Hence, before hitting $\mc S$,
the chain looses memory of where it started from, and approaches $\mc
S$ almost as if started from the stationary distribution $\pi$.

It is worth stressing how the condition of homogeneous influence
may significantly differ from an approximate consensus. In fact,
the former only involves the  (first and second moments of) the
marginal distributions of the agents' stationary beliefs, and does
not have any implication for their joint probability law. A
distribution in which the agents' stationary beliefs are all
mutually independent would be compatible with the condition of
homogeneous influence, as well as an approximate consensus
condition, which would require the stationary beliefs of most of the
agents to be close to each other with high probability. We will study this topic in ongoing work.

Before proving Theorem \ref{theoopinionconc} in Section \ref{secttheoopinionconcproof}, 
we present some examples of highly fluid social networks in Section \ref{sectexamples}.

\subsection{Examples of large-scale social networks}\label{sectexamples}
We now present some examples of families of social networks that are
highly fluid in the limit of large population size $n$. All the
examples will follow the canonical social network construction of
Examples \ref{examplestandardRW} and \ref{examplestandardRWcont}, starting from an undirected graph
$\mc G$. 


We start with an example of a social network which is not highly
fluid.
\begin{example}(\textbf{Barbell}) 
\label{examplebarbell2} For even $n\ge6$,
consider the barbell-like topology introduced in Example
\ref{examplebarbell}. It is not hard to see that
the minimum in the right-hand side of (\ref{eq:conductance}) is achieved by $\mc W=\mc V_0$, so that the conductance satisfies 
$$\Phi=\l(\frac n2\l(\frac n2-1\r)+1\r)^{-1}\le\frac4{(n+1)^2}\,.$$
It then follows from (\ref{mixing-conductance}) that 
$\tau\ge(4\Phi)^{-1}\ge(n+1)^2/16$.
Since $d_v\ge n/2-1$ for all $v$, it follows that the barbell-like network is never highly fluid provided that $|\mc S|\ge1$.
In fact, we have already seen in Example \ref{examplebarbell} that the expected stationary beliefs polarize in this case, so that the influence of the stubborn agents on the rest of the society is not homogeneous.
\end{example}\medskip


Let us now consider a standard deterministic family of symmetric graphs.
\begin{figure}[t]\begin{center}
\subfigure[]{\includegraphics[height=3.8cm,width=5.2cm]{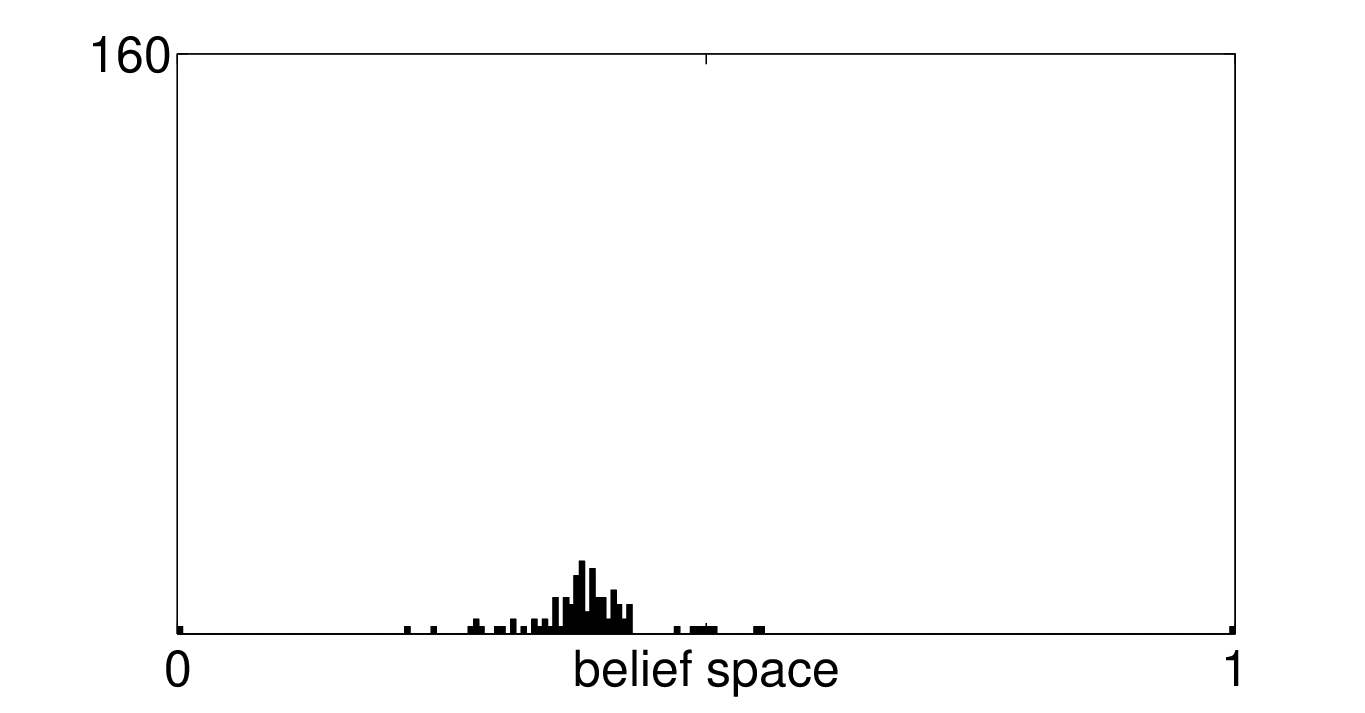}}
\subfigure[]{\includegraphics[height=3.8cm,width=5.2cm]{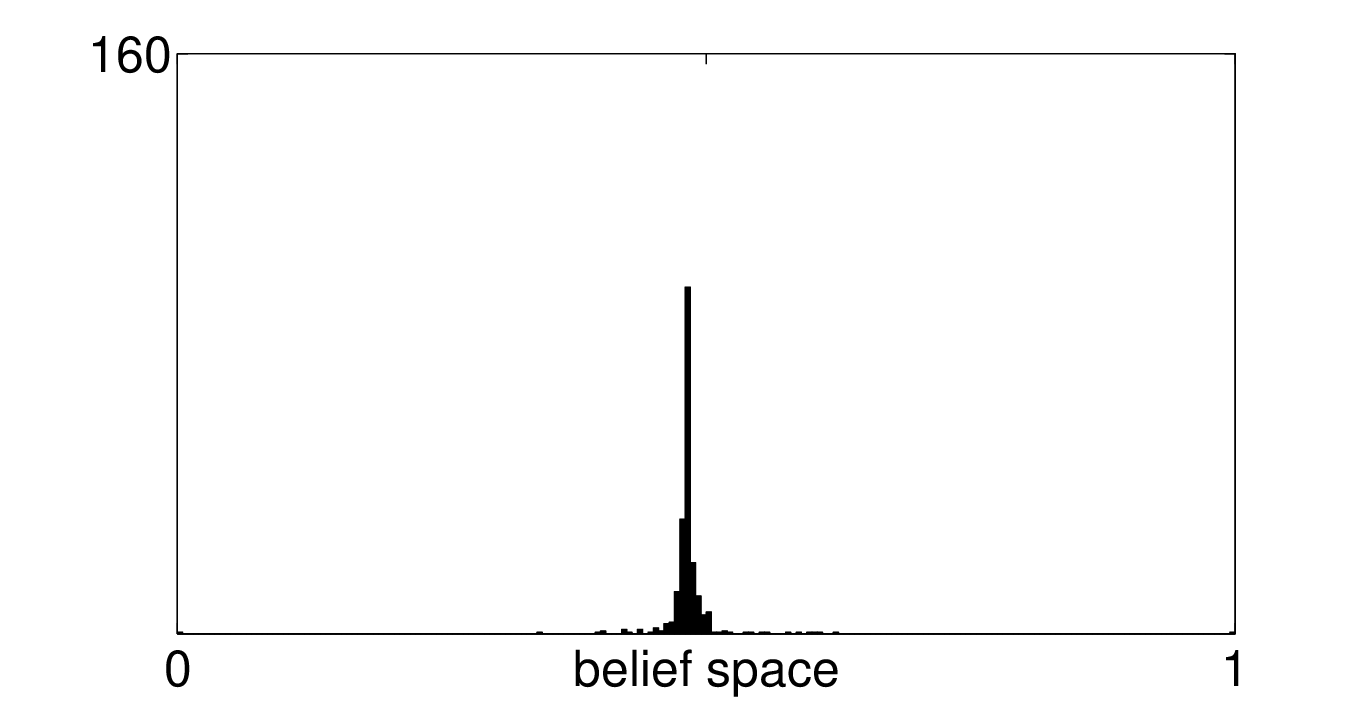}}
\subfigure[]{\includegraphics[height=3.8cm,width=5.2cm]{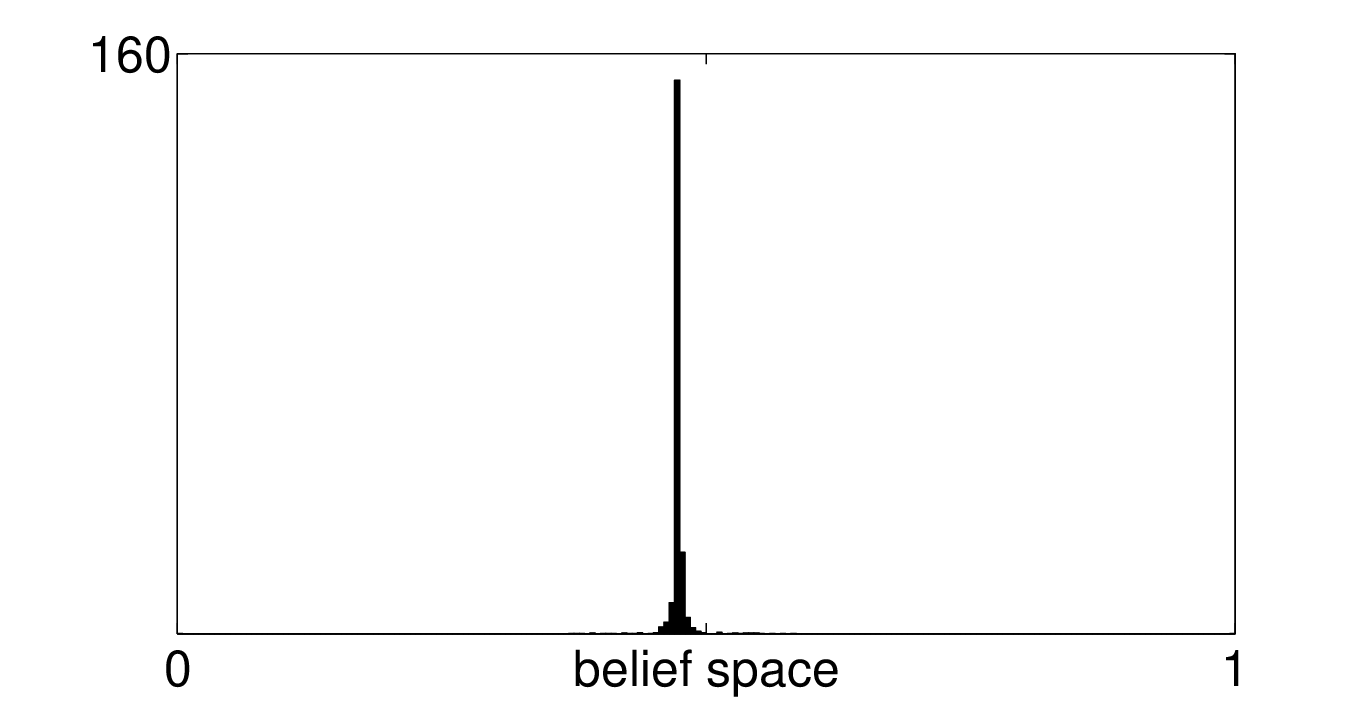}}
\caption{\label{fig:ER}Homogeneous influence in Erd\"os-R\'enyi graphs of increasing sizes. The population sizes are $n=100$ in (a), $n=500$ in (b), and $n=2000$ in (c), while $p=2n^{-1}\log n$. In each case, there are two stubborn agents randomly selected from the node set $\mc V$ with uniform probability and holding beliefs equal to $0$, and $1$, respectively. The figures show the empirical density of the expected stationary beliefs, for typical realizations of such graphs. As predicted by Theorem \ref{theoopinionconc}, such empirical density tends to concentrate around a single value as the population size grows large.}
\end{center}
\end{figure}
\begin{example}\label{exampletori}(\textbf{$d$-dimensional tori})
Let us consider the case of a $d$-dimensional torus of size $n=m^d$,
introduced in Example \ref{exampletorus}. Since this is a regular
graph, one has $\pi_*n=1$, and $\pi(\mc S)=|\mc S|/n$. Moreover, it is well known that (see, e.g., \cite[Theorem 5.5]{LevinPeresWilmer}) $\tau\le C_d n^{2/d}$, for
some constant $C_d$ depending on the dimension $d$ only. Then, $\tau
\pi(\mc S)\le C_d|\mc S|n^{2/d-1}$. For $d\ge3$, this implies that the
social network with toroidal topology is highly fluid, and hence homogeneous influence holds, provided that $|\mc S|=o(n^{1-2/d})$. 

In contrast, for $d\le2$, our arguments do not allow one to prove high fluidity of the social network. In fact, using the explicit calculations of Example \ref{exampletree}, one can see that the stubborn agents' influence is not homogeneous in the case $d=1$, since the expected stationary beliefs do not concentrate. 
On the other hand, in the case $d=2$, we conjecture that, using the explicit expression (\ref{solutionCayley}) and Fourier analysis, one should be able to show that the condition $|\mc S|=o(n^{1/2})$ would be sufficient for homogeneous influence. In fact, a more general conjecture is that $|\mc S|=o(n^{1-1/d})$ should suffice for homogeneous influence, when $d\ge2$. 
Proving this conjecture would require an analysis finer than the one developed in this section, possibly based on discrete Fourier transform techniques. 
The motivation behind our conjecture comes from thinking of a limit continuous model, which can be informally summarized as follows. First, recall that the expected stationary beliefs
vector solves the Laplace equation on $\mc G$ with boundary conditions assigned on the stubborn agent set $\mc S$. Now, consider
the Laplace equation on a $d$-dimensional manifold with boundary
conditions on a certain subset. Then, in order for the problem to be
well-posed, such subset should have dimension $d-1$.
Similarly, one should need $|\mc S|=\Theta(n^{1-1/d})=\Theta(m^{d-1})$
in order to guarantee that the expected stationary beliefs vector is
not almost constant in the limit of large $n$.

\end{example}\medskip


We now present four examples of random graph sequences which have
been the object of extensive research. Following a common
terminology, we say that some property of such graphs holds
with high probability, if the probability that it holds approaches
one in the limit of large population size $n$.

\begin{example}(\textbf{Connected Erd\"os-R\'enyi})\label{exampleER}
Consider the Erd\"os-R\'enyi random graph $\mc G=\mc {ER}(n,p)$, i.e.,
the random undirected graph with $n$ vertices, in which each pair of
distinct vertices is a link with probability $p$, independently
from the others. We focus on the regime $p=cn^{-1}\log n$,
with $c>1$, where the Erd\"os-R\'enyi graph is known to be connected
with high probability \cite[Thm.~2.8.2]{Durrettbook}. In this regime, results by Cooper and Frieze \cite{CooperFrieze06} ensure that, with high probability,
$\tau=O(\log n)$, and that there exists a positive constant $\delta$
such that $\delta c\log n\le d_v\le4c\log n$ for each node $v$
\cite[Lemma 6.5.2]{Durrettbook}. In particular, it follows that,
with high probability, $(\pi_*n)^{-1}\le4/\delta$. Hence, using (\ref{piS}), one finds that the resulting social network is highly fluid, provided that $|\mc S|=o(n/\log n)$, as $n$ grows large. Figure \ref{fig:ER} shows the empirical density of the expected stationary beliefs for typical realizations of Erd\"os-R\'enyi graphs of increasing size $n=100,500,2000$,  and constant stubborn agents number $|\mc S|=2$.

\end{example}\medskip

\begin{example}(\textbf{Fixed degree distribution})\label{exampleFD}
Consider a random graph $\mc G=\mc {FD}(n,\lambda)$, generated as follows. 
Fix $\mc V$ with $|\mc V|\ge2$, and let $\{d_v:\,v\in\mc V\}$ be a family of independent and identically
distributed random variables with $\P(d_v=k)=\lambda_k$, for
$k\in\N$. Assume that $\lambda_1=\lambda_2=0$, that
$\lambda_{2k}>0$ for some $k\ge2$, and that the first two moments
$\sum_k\lambda_kk$, and $\sum_k\lambda_kk^2$ are finite.
Then, let $\mc G=\mc {FD}(n,\lambda)$ be the multigraph of vertex set $\mc V$ generated by 
conditioning on the event $E_n:=\{\sum_vd_v\text{ is even}\}$ (whose  probability converges either to $1/2$ or to $1$ as $n$ grows large)
and matching the vertices uniformly at random given their degree. (See \cite[Ch.~3]{Durrettbook} for details on this construction)  
Then, results in \cite[Ch.~6.3]{Durrettbook} show that the mixing time of the lazy random walk on $\mc G$ satisfies $\tau=O(\log n)$ with high probability.
Therefore, using (\ref{piS}), one finds that the resulting social network is highly fluid with high
probability provided that $\sum_{s} d_s=o\big({n}/{\log n}\big)$.
\end{example}\medskip
\begin{example}(\textbf{Preferential attachment})\label{examplePA}
The preferential attachment model was introduced by Barabasi and
Albert \cite{BarabasiAlbert} to model real-world networks which
typically exhibit a power law degree distribution. We follow
\cite[Ch.~4]{Durrettbook} and consider the random multigraph $\mc
G=\mc{PA}(n,m)$ with $n$ vertices, generated by starting with two
vertices connected by $m$ parallel links, and then subsequently
adding a new vertex and connecting it to $m$ of the existing nodes
with probability proportional to their current degree. As shown in
\cite[Th.~4.1.4]{Durrettbook}, the degree distribution converges in
probability to the power law
$\lambda_k=2m(m+1)/k(k+1)(k+2)$, and the graph is
connected with high probability \cite[Th.~4.6.1]{Durrettbook}. In
particular, it follows that, with high probability, the average
degree $\ov d$ remains bounded, while the second moment of the
degree distribution diverges an $n$ grows large. On the other hand,
results by Mihail et al.~\cite{Mihail} (see also
\cite[Th.~6.4.2]{Durrettbook}) imply that the mixing time of the lazy random walk satisfies 
$\tau=O(\log n)$, with high probability. Therefore, thanks to (\ref{piS}), the resulting social network is highly
fluid with high probability if $\sum_{s} d_s=o\big(n/{\log n}\big)$.
\end{example}\medskip

\begin{example}(\textbf{Watts \& Strogatz's small world})\label{exampleSW}
Watts and Strogatz \cite{WattsStrogatz}, and then Newman and Watts
\cite{NewmanWatts} proposed simple models of random graphs to
explain the empirical evidence that most social networks contain a
large number of triangles and have a small diameter (the latter has
become known as the small-world phenomenon). We consider
Newman and Watts' model, which is a random graph $\quad\mc G=\mc
N\mc W(n,k,p)$, with $n$ vertices, obtained starting from a Cayley
graph on the ring $\Z_n$ with generator
$\{-k,-k+1,\ldots,-1,1,\ldots,k-1,k\}$, and adding to it a Poisson
number of shortcuts with mean $pkn$, and attaching them to randomly
chosen vertices. In this case, the average degree remains bounded
with high probability as $n$ grows large, while results by Durrett
\cite[Th.~6.6.1]{Durrettbook} show that the mixing time
$\tau=O(\log^3 n)$. This, and (\ref{piS}) imply that the network is highly fluid with high probability provided that $\sum_{s} d_s=o\big(n/{\log^3 n}\big)$.
\end{example}



\subsection{Proof of Theorem \ref{theoopinionconc}}\label{secttheoopinionconcproof}
In order to prove Theorem \ref{theoopinionconc}, we will  obtain
estimates on the hitting probability distributions $\gamma^v$. 

The following result provides a useful estimate on the total variation
distance between the hitting probability distribution $\gamma^v$
over $\mc S$ and the stationary stubborn agent distribution
$\ov\gamma$.

\begin{lemma}\label{lemmamixingS} Let the social network satisfy Assumption \ref{assumptionconnected} and $P\in\mc P$. 
Then, for all $P\in\mc P$, and $k\ge0$, 
\be\label{opinionconc1}
||\gamma^v-\ov\gamma||_{TV}
\le \P_v(U_{\mc S}<k)+\exp(-\l\lfloor k/\tau \r\rfloor)\,,\qquad v\in\mc V\,,\ee
where $\tau$ is the mixing time of the discrete-time Markov chain $W(k)$ with transition probability matrix $P$ (cf.~(\ref{mixingtime})), and $U_{\mc S}:=\min\{k\ge0:\,W(k)\in\mc S\}$ is the hitting time of such chain on $\mc S$. 
\end{lemma}
\proof{Proof.}
Notice that (\ref{opinionconc1}) is trivial when $k=0$ or $v\in\mc S$. For $k\ge1$ and $a\in\mc A$, one can reason as follows.
Let $\tilde U_{\mc S}:=\inf\{k'\ge k:\,W(k')\in\mc S\}$.
Thanks to Lemma \ref{lemma:trivial}, one has that the distributions of $W(U_{\mc S})$ and $W(\tilde U_{\mc S})$, conditioned on $W(0)=a$, are given by $\gamma^a$, and $\sum_vp^a_v(k)\gamma^v$, respectively. Using the identity $$||\mu-\nu||_{TV}=\frac12\sup_{f\in[-1,1]^{\mc V}}\l\{\sum\nolimits_v(\mu_v-\nu_v)f_v\r\}$$ (see, e.g., \cite[Prop.~4.5]{LevinPeresWilmer}), and observing that the event $\{U_{\mc S}\ge k\}$ implies $\{W(U_{\mc S})=W(\tilde U_{\mc S})\}$, one gets that
$$\ba{rcl}\ds\l|\l|\gamma^a-\sum\nolimits_vp^a_v(k)\gamma^v\r|\r|_{TV}
&=&\ds\frac12\sup_{f\in[-1,1]^{\mc V}}\l\{\E_a\l[f(W(U_{\mc S}))-f(W(\tilde U_{\mc S}))\r]\r\}\\[10pt]
&=&\ds\frac12\sup_{f\in[-1,1]^{\mc V}}\l\{\E_a\l[\1_{\{U_{\mc S}<k\}}\l(f(W(U_{\mc S}))-f(W(\tilde U_{\mc S}))\r)\r]\r\}\\[12pt]
&\le&\ds\P_a\l(U_{\mc S}<k\r)\,.
\ea
$$
On the other hand, since the Markov kernel $v\mapsto \gamma^v$ is contractive in total variation distance, one has that
$$\l|\l|\sum\nolimits_vp^a_v(k)\gamma^v-\ov\gamma\r|\r|_{TV}
=\l|\l|\sum\nolimits_v(p^a_v(k)-\pi_v)\gamma^v\r|\r|_{TV}
\le||p^a(k)-\pi||_{TV}
\,.$$
Finally, submultiplicativity of the maximal total variation distance from the stationary distribution (see, e.g., \cite[Lemma 4.12]{LevinPeresWilmer}) implies that 
$$||p^a(k)-\pi||_{TV}\le\exp(-\lfloor k/\tau\rfloor)\,.$$
By applying the triangle inequality and the three bounds above, one gets that
$$
||\gamma^a-\ov\gamma||_{TV}\le\l|\l|\gamma^a-\sum\nolimits_v\gamma^a_v(t)\gamma^v\r|\r|_{TV}+\l|\l|\sum\nolimits_v\gamma^a_v(t)\gamma^v-\ov\gamma\r|\r|_{TV}
\le\P_a\l(U_{\mc S}<k\r)+\exp(-\lfloor k/\tau\rfloor)\,,$$
thus proving the claim.
\qed\endproof

Lemma \ref{lemmaopinionconc}, stated below, is the main technical result of this section. 
\begin{lemma}\label{lemmaopinionconc}
Consider a social network satisfying Assumption \ref{assumptionconnected}. Then, 
$$\frac1n\l|\l\{v\in\mc V:\,||\gamma^v-\ov\gamma||_{TV}\ge\eps\r\}|\r|\le\frac{1}{\Psi\eps}\,,$$
for every $\eps>0$. 
\end{lemma}
\proof{Proof.} 
Fix an arbitrary $P\in\mc P$, and let $\pi=P'\pi$ be its invariant measure. 
Let $W(k)$ be a discrete-time Markov chain with transition probability matrix $P$. 
For every nonnegative integer $k$, stationarity of $\pi$ and the union bound yield 
\be\label{UB}\P_{\pi}(U_{\mc S}<k)=\P_{\pi}\l(\bigcup\nolimits_{0\le j<k}\l\{W(j)\in\mc S\r\}\r)\le\sum_{0\le j<k}\P_{\pi}(W(j)\in\mc S)= k\pi(\mc S)\,,\ee
Combining (\ref{UB}) with Lemma \ref{lemmamixingS}, one gets that 
$$\sum_v\pi_v||\gamma^v-\ov\gamma||_{TV}\le\P_{\pi}(U_{\mc S}<k)+\exp(-\lfloor k/\tau\rfloor)\le k\pi(\mc S)+\exp(-\lfloor k/\tau\rfloor)\,.$$
Choosing $k=\tau\lfloor\log(e/(\tau\pi(\mc S))\rfloor$ yields  
$$\frac1n\sum_v||\gamma^v-\ov\gamma||_{TV}\le\sum_v\frac{\pi_v}{n\pi_*}||\gamma^v-\ov\gamma||_{TV}\le\frac1{n\pi_*}\l({\tau\pi(\mc S)}\log\frac{e}{\tau\pi(\mc S)}+\exp\l(-\log\frac{e}{\tau\pi(\mc S)}+1\r)\r)=\frac1{\psi(P)}\,.$$
Then, 
$$\frac{\eps}n\l|\l\{v\in\mc V:\,||\gamma^v-\ov\gamma||_{TV}\ge\eps\r\}\r|\le\frac1n\sum_v||\gamma^v-\ov\gamma||_{TV}\le\frac{1}{\psi(P)}\,.$$
The claim now follows from the arbitrarinessss of $P\in\mc P$. 
\qed\endproof

\proof{Proof of Theorem \ref{theoopinionconc}.}
 Let
$y_s:=x_s+\Delta_*/2-\max\{x_{s'}:\,s'\in\mc S\}$ for all $s\in\mc
S$. Clearly $|y_s|\le\Delta_*/2$. Then, it follows from Theorem
\ref{theoasymptoticmeanvariance} that
$$\Big|\E[X_v]-\E[Z]\Big|=\big|\sum\nolimits_s\gamma^v_sx_s-\sum\nolimits_s\ov\gamma_sx_s\Big|=\big|\sum\nolimits_s\gamma^v_sy_s-\sum\nolimits_s\ov\gamma_sy_s\Big|\le\Delta_*||\gamma^v-\ov\gamma||_{TV}\,,$$
so that (\ref{expectedvalueconcentration}) immediately follows from
Lemma \ref{lemmaopinionconc}.

On the other hand, in order to show (\ref{varianceconcentration}),
first recall that, if $\theta_e=1$ for all $e\in\ora{\mc E}$, then
Eq.\ (\ref{Mdef}) provides the transition rates of coalescing Markov chains. In particular, if $V(0)=V'(0)$, then $V(T_{\mc S})=V'(T'_{\mc
S})$, so that  $\eta^{vv}_{ss'}=\gamma^v_s$ if $s=s'$, and
$\eta^{vv}_{ss'}=0$ otherwise. Then, it follows from Theorem \ref{theoasymptoticmeanvariance} that
$$\ba{rcl}\sigma^2_v
&=&\E[X_v^2]-\E[X_v]^2\\[7pt]
&=&\ds\sum\nolimits_{s,s'}\eta^{vv}_{ss'}x_sx_{s'}-\l(\ds\sum\nolimits_s\gamma^v_sx_s\r)^2\\[10pt]
&=&\ds\sum\nolimits_s\gamma^v_sx_s^2-\l(\ds\sum\nolimits_s\gamma^v_sx_s\r)^2\\[10pt]
&=&\ds\frac12\sum\nolimits_s\ds\sum\nolimits_{s'}\gamma^v_s\gamma^v_{s'}(x_s-x_{s'})^2\,.
\ea$$
Similarly,
$$\sigma^2_Z=\frac12\sum\nolimits_{s,s'}\ov\gamma_s\ov\gamma_{s'}(x_s-x_{s'})^2\,,$$
so that
$$\ba{rcl}|\sigma^2_v-\sigma^2_Z|
&\le&\ds\frac12\sum\nolimits_{s,s'}\l|\gamma^v_s\gamma^v_{s'}-\ov\gamma_s\ov\gamma_{s'}\r|(x_s-x_{s'})^2\\[10pt]
&\le&\ds\frac12\sum\nolimits_{s,s'}\l|\gamma^v_s\gamma^v_{s'}-\ov\gamma_s\ov\gamma_{s'}\r|\Delta_*^2\\[10pt]
&\le&\ds\frac12\sum\nolimits_{s,s'}\l(\gamma^v_s\l|\gamma^v_{s'}-\ov\gamma_{s'}\r|+\ov\gamma_{s'}\l|\gamma^v_{s}-\ov\gamma_{s}\r|\r)\Delta_*^2\\[10pt]
&=&\l|\l|\gamma^v-\ov\gamma\r|\r|_{TV}\Delta_*^2\,.
\ea$$
Now, (\ref{varianceconcentration}) follows again from a direct application of Lemma \ref{lemmaopinionconc}.\qed\endproof

\begin{remark}\label{remarkTS}
For a  stochastic matrix $P\in\mc P$ which is reversible, i.e., such that $\pi_vP_{vw}=\pi_wP_{wv}$ for all $v,w\in\mc V$, (observe that this additional property is enjoyed by the matrix $P$ considered in Example \ref{examplestandardRWcont} for the canonical construction of a social network from an undirected graph $\mc G$) one can potentially obtain tighter estimates on the homogeneity of the agents' influence. In fact, one could use the results on the approximate exponentiality of hitting times (i.e., the property that the distribution of  $T_{\mc S}/\E_{\pi}[T_{\mc S}]$ is close to a rate-$1$ exponential distribution, see, e.g., \cite[Ch.~3.5]{AldousFillbook}) in order to show that, for a continuous-time Markov chain with transition rate matrix $P-I$, one has $\P_{\pi}(T_{\mc S}\ge t)\le(t+\tau)/\E_{\pi}[T_{\mc S}]$ for all $t\ge0$. Using this bound in place of (\ref{UB}), arguments analogous to those developed in this section imply that  $\tau/\E_{\pi}[T_{\mc S}]=o(1)$ is a sufficient condition for homogeneous influence. Observe that, using Markov's inequality and (\ref{UB}) with $k=\lfloor1/(2\pi(\mc S))\rfloor$, gives
$$\label{ETS}\E_{\pi}[T_{\mc S}]=\E_{\pi}[U_{\mc S}]\ge\left\lfloor\frac{1}{2\pi(\mc S)}\right\rfloor{\P_{\pi}\left(U_{\mc S}\ge\left\lfloor\frac{1}{2\pi(\mc S)}\right\rfloor\right)}\ge
\left\lfloor\frac{1}{2\pi(\mc S)}\right\rfloor\l(1-\pi(\mc S)\left\lfloor\frac{1}{2\pi(\mc S)}\right\rfloor\r)\ge\frac{1-4\pi^2(\mc S)}{4\pi(\mc S)}\,.$$
Hence, this argument would provide potentially a weaker sufficient condition for homogenous influence in situations where $\pi(\mc S)=o(1/\E_{\pi}[T_{\mc S}])$. 
\end{remark}

\section{Conclusion}\label{conclusions}
In this paper, we have studied a possible mechanism explaining
persistent disagreement and opinion fluctuations in social networks. We have considered an inhomogeneous 
stochastic gossip model of continuous opinion dynamics, whereby some stubborn agents in the
network never change their opinions. We have shown that the
presence of these stubborn agents leads to persistent fluctuations
and disagreements among the rest of the society: the beliefs of regular agents
do not converge almost surely, and keep on fluctuating
in an ergodic fashion. A duality argument allows for characterizing expected stationary beliefs in terms of the hitting probabilities of a Markov chain on the graph describing the social network, while the correlation between the stationary beliefs of any pair of regular agents can be characterized in terms of the hitting probabilities of a pair of coupled Markov chains. We have shown that in highly fluid social networks, whose associated Markov chains have mixing times which are sufficiently smaller than the inverse of the stubborn agents' set size, the vectors of the stationary expected beliefs and variances are almost constant, so that the stubborn agents have homogeneous influence on the rest of the society. We wish to emphasize that homogeneous influence in highly fluid societies needs not imply approximate consensus among the agents, whose beliefs may well fluctuate in an almost uncorrelated way. A deeper understanding of this topic is ongoing work.

\section*{Acknowledgments.} The authors would like to thank an anonymous Referee for many detailed comments which significantly helped in improving the presentation. This research was partially supported by the NSF grant SES-0729361, the AFOSR grant FA9550-09-1-0420, the ARO grant  911NF-09-1-0556, the Draper UR\&D program, and the AFOSR MURI R6756-G2. The work of the second author was partially supported by the Swedish Research Council through the LCCC Linnaeus Center and the junior research grant `Information dynamics over large-scale networks'. 
\bibliographystyle{amsplain}
\bibliography{influence}

%
%

\end{document}